\documentclass[12pt]{article}

\usepackage{newtxtext,newtxmath}

\usepackage{graphicx}
\usepackage{enumitem}

\usepackage{rotating}
\usepackage{float}
\usepackage[fleqn]{cases}
\usepackage{subcaption}
\usepackage{caption}
\usepackage{subcaption}
\usepackage[letterpaper,margin=1in]{geometry}

\renewenvironment{abstract}
	{\quotation}
	{\endquotation}

\date{}

\makeatletter
\renewcommand{\fnum@figure}{\textbf{Figure \thefigure}}
\renewcommand{\fnum@table}{\textbf{Table \thetable}}
\makeatother

\usepackage{scicite}

\usepackage{url}

\def\scititle{
Universal Scaling Laws in Freeway Traffic
}
\title{\bfseries \boldmath \scititle}

\author{
	Garyoung~Lee$^{1}$,
	Aryaman~Jha$^{2}$,
	Kurt~Wiesenfeld$^{2}$,
    Jorge~Laval$^{1\ast}$\and
	\small$^{1}$School of Civil and Environmental Engineering, Georgia Institute of Technology, Atlanta \& 30332, United States.\and
	\small$^{2}$School of Physics, Georgia Institute of Technology, Atlanta \& 30332, United States.\and
	\small$^\ast$Corresponding author. Email: jorge.laval@ce.gatech.edu\and
}

\begin{document} 

\maketitle

\begin{abstract} \bfseries \boldmath

Traffic congestion, a daily frustration for millions and a multi-billion dollar drain on economies, has long resisted deep physical understanding. 
While simple theoretical models of traffic flow have suggested connections to critical phenomena and non-equilibrium universality, direct empirical validation is lacking. Using extensive, high-resolution vehicle trajectory data from the I-24 MOTION testbed, we show that traffic flow exhibits both a percolation phase transition that is self-organized critical and fluctuations consistent with the Kardar-Parisi-Zhang universality in 1+1 dimensions.
This suggests that the complex and seemingly chaotic formation of traffic jams has predictable statistical properties, which opens new avenues in traffic science for developing advanced forecasting and management strategies grounded in universal scaling laws.

\end{abstract}

\noindent

Understanding traffic congestion has been a long-standing puzzle for scientists from diverse disciplines. Potentially consequential clues have come in the form of conjectures regarding the universal behavior of traffic flow. It has long been conjectured that traffic belongs in the same category as systems driven by equilibrium critical behavior, such as fluids undergoing a phase transition \cite{herman1979two, nagel1995emergent,   paczuskitextordfeminine1996self,nagel2003still, laval2024traffic}. This critical state is right on the edge between free-flow and widespread jamming, it is chaotic, \textit{i.e.,} highly sensitive to small perturbations, and exhibits scale-free jam clusters that would exhibit power-law statistics. 
While there is ample empirical evidence of a critical percolation phase transition at a coarse link level in urban networks and highways \cite{zhang2019scale}, at the more fundamental microscopic-trajectory level, devoid of network effects, this is not the case. In addition, there is still no consensus in the literature on the matter as the Nagel-Schreckenberg (NaSch) model \cite{nagel1992cellular}, 
\textit{i.e.}, a simple but influential microscopic stochastic traffic model,  has been shown not to exhibit critical behavior, although the debate is ongoing   \cite{roters1999critical, chowdhury2000comment, roters2000reply, schadschneider2010stochastic,jha2025evidence}.
This paper closes this gap by showing that microscopic-level traffic does exhibit self-organized criticality \cite{bak1987self} under the lens of percolation phase transitions.

More recently, it has been also conjectured that traffic falls within the   Kardar-Parisi-Zhang (KPZ) universality class \cite{kardar1986dynamic}, which describes a broad set of non-equilibrium systems where interfaces grow randomly but with specific statistical fingerprints regardless of microscopic details, ranging from bacterial colonies to turbulent flows \cite{halpin2015kpz,odor2004universality, johansson2003discrete, corwin2015renormalization}.
The first indication that traffic flow belongs to the KPZ came in the early 2000s when the influential Totally Asymmetric Simple Exclusion Process (TASEP) \cite{spitzer1991interaction}  was rigorously shown to exhibit KPZ universality \cite{johansson2000shape,johansson2003discrete}. TASEP describes particles hopping stochastically in one direction with hard-core exclusion on a discrete lattice, and its parallel-update variant is considered a foundational model of traffic flow on a single lane. It has been extended to incorporate bounded vehicle accelerations and an arbitrary maximum speed in the NaSch model, and has been shown to exhibit KPZ universality even on multi-lane roads with lane-changing behavior, which speaks to the robustness of this universality class \cite{de2019kardar}.

Despite the pivotal nature of these conjectures, to date, empirical validation has been missing, possibly owing to the fine-grain and large-scale spatial and temporal extent of the empirical data required for the task.   
Leveraging the unprecedented spatial extent of the recently released I-24 MOTION trajectory dataset offers a rare opportunity to address this gap. It 
provides vehicle positions at 25 Hz for four morning peak hours along a 4.2 mi segment of Interstate 24 in Tennessee, USA in 2022, covering both eastbound and westbound traffic in four lanes \cite{gloudemans202324}. 

\section*{Criticality of traffic flow}

\begin{figure}[h]
    \centering
    \includegraphics[width=\linewidth]{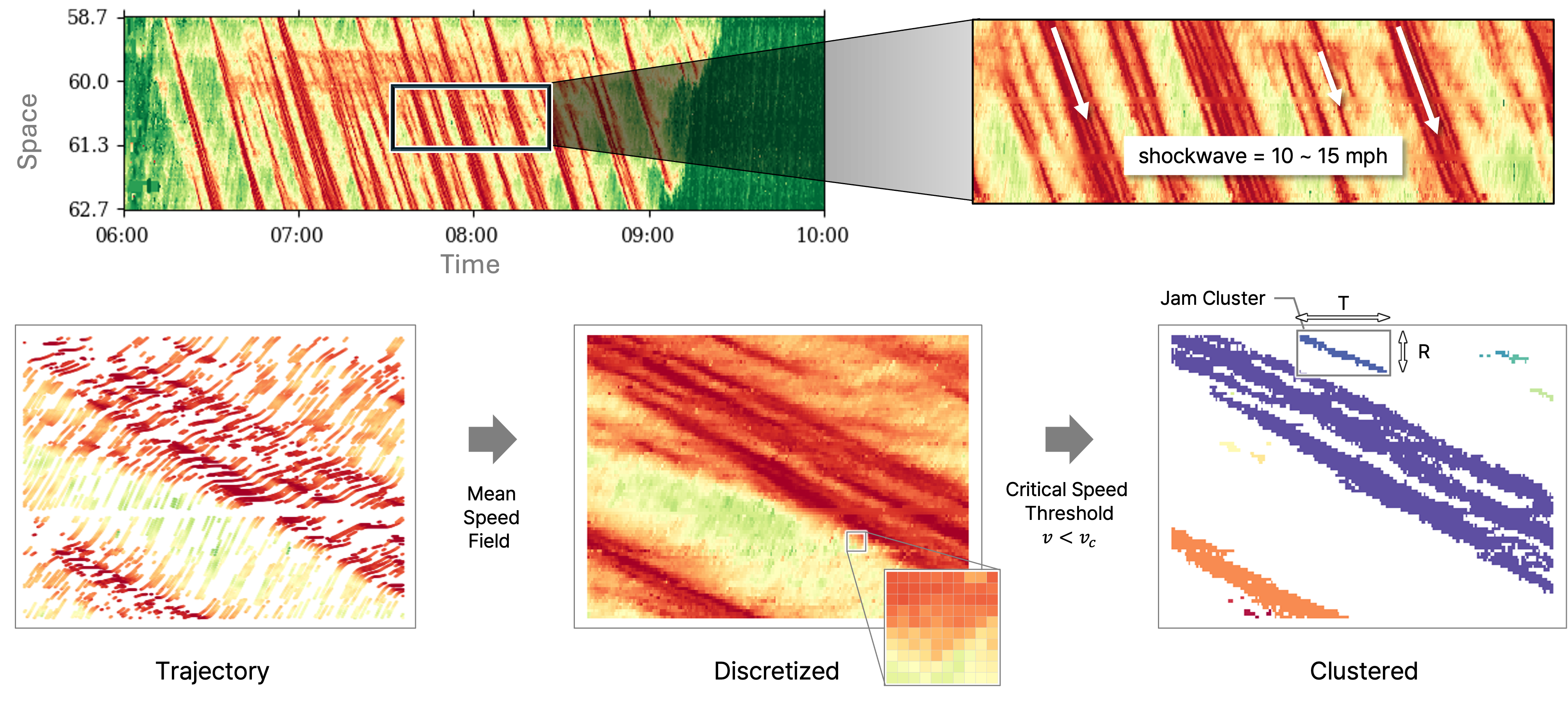}
    \caption{\textbf{Traffic jam clusters extracted from trajectory data.} We begin by discretizing the trajectory data into a time-space speed field 
    of  resolution $0.02$ miles by  $6$ seconds. 
    %
    A critical speed threshold $v_c$ is applied to binarize this field: cells with speeds below $v_c$ are labeled as congested (jammed), and those above $v_c$ as free-flow. Cells are subsequently grouped into clusters based on nearest-neighbor adjacency, where connectivity is defined by shared edges in either space or time. A set of adjacent congested cells constitutes a single \textit{jam cluster} representing localized connected space-time regions of low speed. Their size, shape, and distribution reflect the emergent macroscopic patterns of congestion and serve as the primary objects of our scaling analysis.
}
    \label{fig:clustering}
\end{figure}

The analogy between simple fluids and traffic flow on a single lane has long been suspected  \cite {nagel2003still}, but the obstacle has been the interpretation of the thermal energy. Inspired by the analogy between the flow-density diagram in traffic flow and the coexistence curve in gas-liquid phase transitions, it was recently proposed the flow of cars as a viable interpretation \cite{laval2024traffic}, enabling the analysis of single-lane traffic flow criticality through the lens of percolation theory thanks to the Fortuin–Kasteleyn representation of the Ising model \cite{fortuin1972random}. This representation allows one to study criticality through the geometry of fluid particle clusters due to attraction; in our case these are \textit{jam clusters} of stopped vehicles in the time-space plane. Our conjecture is that if traffic constitutes a critical phenomenon, then there should exist a critical threshold at which a percolation phase transition occurs, where jam clusters shift from being localized and disconnected to spanning the entire length of the freeway. This is indeed what we observe, along with an unexpected additional finding: rather than exhibiting a single critical threshold as typically seen in percolation phenomena, the system displays a \textit{wide range} of threshold values where critical behavior emerges. This extended critical range provides strong evidence for self-organized criticality (SOC), which characterizes systems that naturally evolve toward a critical state without requiring precise tuning of control parameters. This behavior was previously conjectured for simple traffic flow models \cite{laval2023self}. It arises as a consequence of drivers' tendency to travel as fast as possible, resulting in outflows from jam clusters operating in the critical state.

To construct the percolation representation of traffic in each lane, the trajectory data is converted into a discrete time-space field of speed data.
Each cell in this field is either congested or not, depending on a \textit{critical speed threshold}, $v_c$, and the resulting set of adjacent congested cells constitutes a single \textit{jam cluster}; see   Figure~\ref{fig:clustering}. These clusters are commonly known as traffic shockwaves, which propagate against the direction of traffic at an approximately constant wave speed, producing the elongated shapes in the figure. 

\begin{figure}[htbp!]
    \centering
    \includegraphics[width=\linewidth]{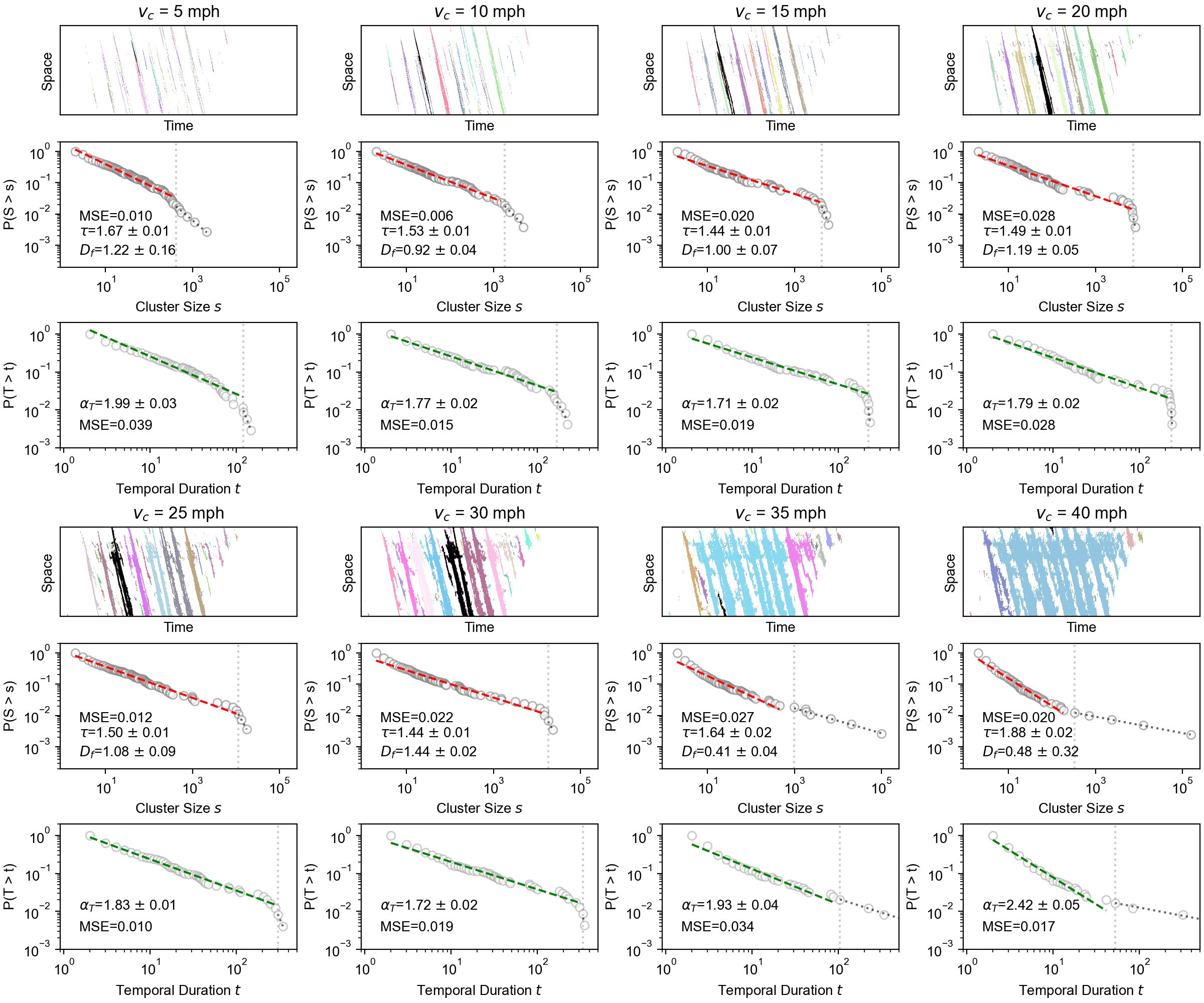}
    \caption{\textbf{Cluster structures and survival functions  $P(S>s) $ and $P(T>t)$ of cluster sizes and temporal durations under varying critical speed thresholds on Nov.  22.} Top panels: time-space diagrams of traffic jam clusters at different speed thresholds. The cutoff cluster is colored in black. Bottom panels: corresponding log-log plots of $P(S>s) $ and $P(T>t)$. Estimates are estimated through piecewise linear, forcing the cutoff to happen at the last 10\% of the points. The reported mean squared error corresponds to the first linear segment. All other days in our sample show very similar results; see Methods and Supplementary Figures.
    %
    }
    \label{fig:tau_1122}
\end{figure}

To identify the critical threshold, we leverage the scaling hypotheses in percolation theory based on the self-similarity exhibited by the geometric properties of jam clusters at criticality: their size $S$, spatial extent $R$, and temporal duration $T$; see Figure~\ref{fig:clustering}. In the limit of infinitely large systems, their marginal distribution follows power-laws of the form:
\begin{equation}\label{tau}
P(S) \sim S^{- \tau}, \quad P(R) \sim R^{-\alpha_R}, \quad P(T) \sim T^{-\alpha_T},
\end{equation}
where $ \tau $ is Fisher’s critical exponent, while
$\alpha_R$ and $\alpha_T$ are the scaling exponents for the spatial extent and temporal duration distributions, respectively. 
In finite systems of size $L$, these power-laws hold only up to a characteristic cutoff size $ s_c $, which scales with the system size $L$ as
\begin{equation} \label{sc}
    s_c \sim L^{D_f},
\end{equation} where $D_f $ denotes the fractal dimension of the cutoff cluster size. Beyond this cutoff, the frequency of large clusters drops sharply due to finite-size constraints. Figure \ref{fig:tau_1122} illustrates this behavior for lane 1 of Nov. 22, where the expected power-law behavior prevails for a wide range of critical speed thresholds. It is important to note that Eqs.~\ref{tau} and~\ref{sc} exhibit the expected finite-size scaling behavior; see Supplementary Text.

We also have
the fractal dimension $D_R$ and $D_T$, relating the size of a cluster to its spatial extent and temporal duration, $S \sim R^{D_R}$ and $S \sim T^{D_T}$, and the dynamic exponent of the percolation process ${z_P = D_R/ D_T}$, relating the spatial extent of a cluster to its temporal duration, $T \sim R^{z_P}$.
These exponents are linked by universal relationships  \cite{zhang2019scale}: 
\begin{align} \label{eq:alphaR} \alpha_R &= D_R(\tau - 1) + 1 \\ \label{eq:alphaT} \alpha_T &= D_T(\tau - 1) + 1 \end{align}


\begin{figure}[h!]
    \centering
    \includegraphics[width=\textwidth]{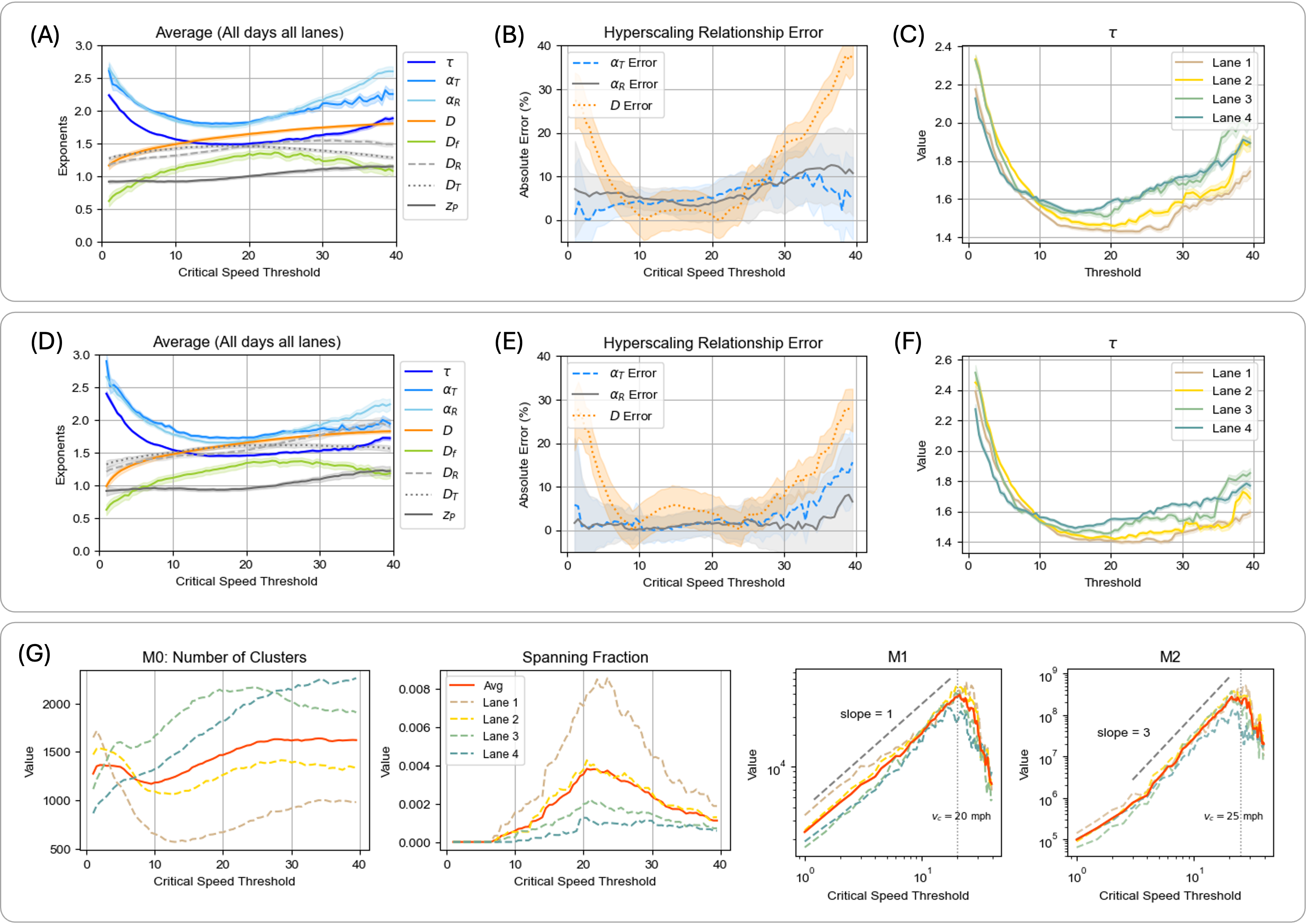}
\caption{\textbf{Empirically measured critical exponents and derived metrics as a function of the speed threshold $v_c$.}
\textbf{(A–C)} Critical exponents computed with clusters of size $\geq 2$ cells. \textbf{(D–F)} Critical exponents with clusters of size $\geq 10$ cells.
\textbf{(A, D)} Average critical exponents across all days and lanes as a function of the critical speed threshold $v_c$ (lane-specific results are provided in the Supplementary Text).
\textbf{(B, E)} Errors in hyperscaling relationships from Eqs.~\ref{eq:alphaR}, \ref{eq:alphaT}, and \ref{eq:D}, showing the deviation between left-hand side and right-hand side expressions using independently estimated exponents. Detailed comparisons are available in the Supplementary Text.
\textbf{(C, F)} Fisher exponent $\tau$ averaged across all days for each lane. Outer lanes exhibit higher estimates and shorter plateau ranges, which become more stable when clusters smaller than 10 cells are excluded.
\textbf{(G)} Derived metrics used to identify critical transitions, including $\mathcal{M}_0$, $\mathcal{M}_1$, $\mathcal{M}_2$, and the fraction of spanning clusters.  } 
    \label{fig:exponents_vs_vc}
\end{figure}

All critical exponents (except $z_P$) were independently extracted from the empirical data by fitting power-laws to the respective distributions and scaling relationships as a function of the threshold speed $v_c$. The dynamic exponent $z_P$ was not directly fitted, but computed as the ratio $z_P = D_R/ D_T$, using the separately estimated values of $D_R$ and $D_T$; see Methods~\cite{methods}.
Figure~\ref{fig:exponents_vs_vc} presents the average results across all days and lanes for two settings: subfigures A–C use clusters $\geq 2$ cells (corresponding to 12 seconds or 0.04 miles), while D–F use clusters  $\geq 10$ cells. 
The presence of a plateau region for exponents $\tau$, $\alpha_R$, and $\alpha_T$ within the range $10 \lesssim v_c \lesssim 35$ indicates a robust critical regime regardless of the precise definition of a \textit{congested} cluster. This robust regime supports the presence of SOC, contrasting sharply with traditional percolation transitions, which typically occur at a single critical point. 
This persistent critical behavior is further validated in B and E, which show that the average error in Eqs.~\ref{eq:alphaR} and~\ref{eq:alphaT} across the plateau remains below 10\%, and decreases further when small clusters are excluded. 

At the individual lane level, we observe 
that outer lanes (3 and 4) show higher values of  $\tau$  and narrower plateau ranges, while the HOV lane (lane 1) exhibits a more stable trend. These discrepancies are reduced when small clusters are excluded in the analysis, suggesting that outer lanes contain more small and fragmented clusters possibly due to frequent lane-changing activity near on-/off-ramps. This leads to an artificial inflation of $\tau$ in lanes 3 and 4, particularly at higher $v_c$ values, where the cutoff occurs early (see power-laws in Figure~\ref{fig:tau_1122}). 
Differences in other exponents by lane are detailed in the Supplementary Text.

%
%

Different cluster dynamics in $10 \lesssim v_c \lesssim 35$  suggest a crossover between two distinct transitions, as shown in Figure~\ref{fig:tau_1122}. At $v_c \approx 10$ mph, the first elongated clusters (shockwaves) appear, spanning the full freeway length. This \textit{shockwave critical point} corresponds to a minimum in the total number of clusters $\mathcal{M}_0$; see Figure~\ref{fig:exponents_vs_vc}G. As $v_c$ increases, spanning clusters become more frequent until around $v_c \approx 20$ mph when lateral cluster merging due to lane-changing activity becoming dominant. By $v_c \approx 30$ mph, a single giant cluster typically forms, which defines the \textit{lateral-merging critical point}. This second transition is also reflected in the sharp drops of the first- and second-order moments of the cluster size distribution, denoted $\mathcal{M}_1$ and $\mathcal{M}_2$. 
Notably, unlike $\mathcal{M}_0$ and the fraction of spanning clusters, $\mathcal{M}_1$ and $\mathcal{M}_2$ are relatively insensitive to lane differences, which makes them promising indicators for identifying the lateral-merging critical point. 

Now that criticality has been established, it can be leveraged to estimate the total cluster mass, $M$, which represents the total congestion delay in the given time-space region \cite{laval2023self}, and satisfies:
\begin{equation} \label{eq:MLz}
M\sim L^D
\end{equation}
with $D$ the delay fractal dimension. 
Figure \ref{fig:exponents_vs_vc} shows that the delay fractal dimension $D$ remains near 1.5 to 1.75 across the plateau critical region; see Methods.
To connect  $D$ to other exponents, we note 
that the typical delay $M$ can  be expressed as the total number of clusters, $N$, times the expected  cluster size:
\begin{align}  \label{eq:M}
M = N \int_{s_{\text{min}}}^{s_c}  s P(s) ds 
&\sim L^d \int_{s_{\text{min}}}^{L^{D_f}} s^{1 - \tau} ds 
\sim 
L^{d + D_f(2 - \tau)},       ~~~\tau \neq 2.             
\end{align}
Note that $N \sim L^d$ should only be valid near the shockwave critical point since it is determined by the initial conditions given the directed percolation-like nature of shockwaves; see Figure~\ref{fig:tau_1122}. 
Comparing Eqs.~\ref{eq:MLz} and \ref{eq:M} we obtain the hyper-scaling relation: 
\begin{equation} \label{eq:D}
D = d + D_f(2 - \tau),
\end{equation}
which provides an alternative method for measuring $D$. As a consistency check, evaluating the right-hand side of Eq.~\ref{eq:D} using $d = 1$, $\tau \approx 1.5$, and $D_f \approx 1$ at $v_c = 10$ mph gives $D \approx 1.5$, consistent with our direct measurements.
Figure~\ref{fig:exponents_vs_vc}B and E show that the hyperscaling error for Eq.~\ref{eq:D} is very low in $10 \lesssim v_c \lesssim 20$, but increases thereafter, as expected. 

\section*{KPZ universality signatures in traffic flow}

The central quantity describing systems that exhibit KPZ universality is a growing interface, represented by its height at position $x$ and time $t$, $h(x,t)$, whose fluctuations follow predictable scaling laws. 
In the context of traffic flow, $h(x,t)$ is interpreted as the cumulative count curve of vehicles up to spatial position $x$ at a fixed time $t$ \cite{laval2024traffic}.
Members of the KPZ universality class converge upon renormalization to the KPZ fixed point \cite{matetski2021kpz}, characterized by three critical exponents: the roughness exponent $\alpha$, growth exponent $\beta$, and dynamic exponent $z$, which satisfy the hyper-scaling relations
\begin{equation}\label{KPZhyperScale}
z = {\alpha}/{\beta}\qquad \mbox{and} \qquad \alpha+z=2
\end{equation}
and take the universal values $\alpha = 1/2$, $\beta = 1/3$, and $z = 3/2$ in 1+1 dimensions  (\textit{i.e.}, one spatial, one time).   The exponent $\alpha$ describes the spatial roughness of these curves, $\beta$ quantifies how their fluctuations evolve over time, and $z$ relates the spatial and temporal scales.

To construct the cumulative count curves $h(x,t)$ per lane from the raw trajectory data, we discretize it spatially into segments the size of a vehicle length, approximately 7~m. For a given time slice $t$, a one-dimensional cumulative count curve is generated by assigning a value of $+1$ to each spatial segment containing a vehicle and $0$ otherwise. These values are then summed sequentially along the spatial axis to produce a cumulative count curve that will consequently display an average increasing spatial trend given by the vehicle density in the segment, $\rho$. To remove this trend, we use the detrended height function $h'(x,t)=h(x,t)-\rho x$ in what follows.
Armed with this data, we proceed to the estimation of the KPZ exponents, which we tackle by analyzing three different qualities: the interface width, dynamic correlations, and the Hurst exponent.

\begin{figure}
    \centering
    \includegraphics[width=\linewidth]{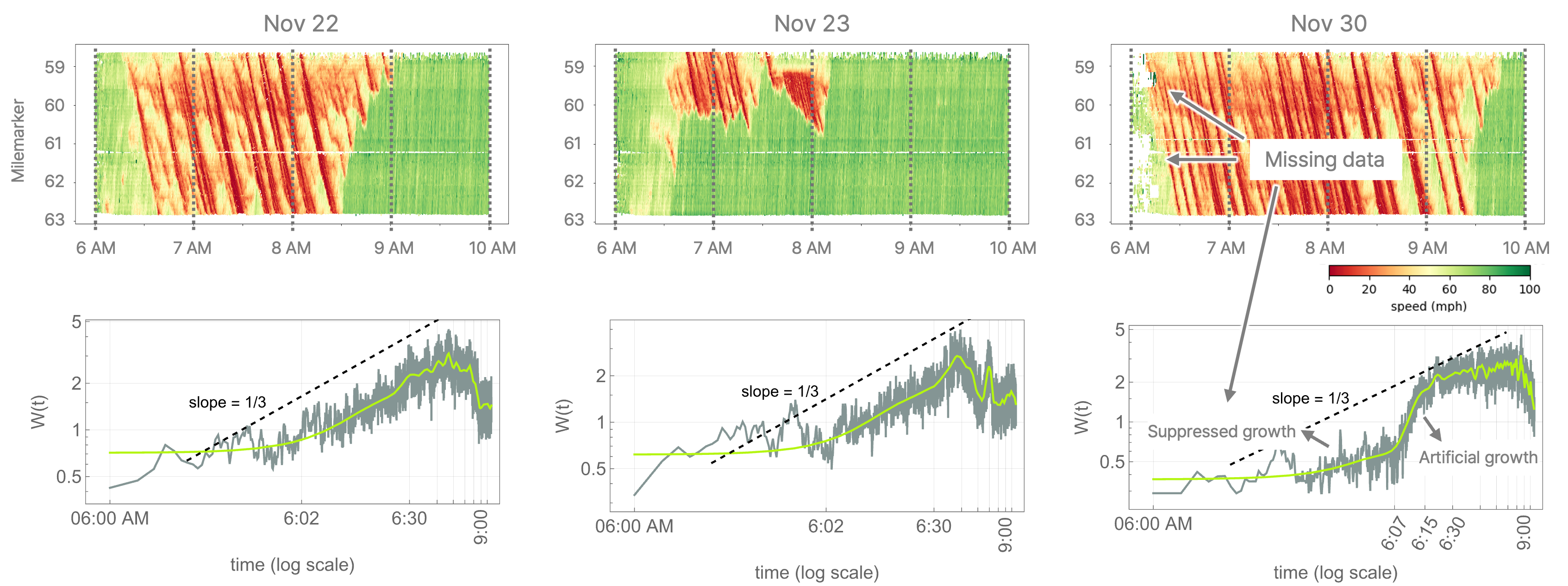}
    \caption{\textbf{Time evolution of the interface width $W(t)$ 
    for three representative days, averaged across all four lanes, one second resolution.} Nov. 22 and 23 are the only days in our sample that have complete trajectory coverage during the early free-flow period, while all other days report missing data similar to Nov. 30.
    It becomes clear that in days with complete coverage
    $W(t)$ grows as a power-law with slope consistent with the KPZ prediction $\beta = 1/3$, before saturating. In contrast, days with missing data exhibit \textit{{suppressed growth}} in the early stages due to incomplete sensor coverage, leading to a lower estimate of the interface width $W(t)$ during the initial period. Once full coverage resumes, $W(t)$ appears to {grow rapidly}, but this apparent burst is \textit{artificial}—it simply reflects the recovery of previously unrecorded fluctuations. As a result, the slope of $ W(t)$ in the plot is {temporarily overestimated}, 
    and therefore a {data artifact}, not a change in the underlying dynamics. The time ticks in the width evolution plot after 6:30 a.m. are shown in 30-minute intervals. A similar pattern is observed in the eastbound direction, which features free-flow traffic only (see Supplementary Text). 
    }

    \label{fig:beta_width}
\end{figure}

A key quantity for assessing KPZ behavior is the interface width, defined as the standard deviation of the height function at time $t$, $W(t)={ \left\langle \left[ h'(x,t) - \left\langle h'(x,t) \right\rangle \right]^2 \right\rangle}^{1/2} $, where $\langle\cdot\rangle$ represents averaging over space. It captures the typical fluctuation size of the interface and is expected to follow KPZ scaling: initially growing as a power-law, $W(t) \sim t^\beta$, with $\beta = 1/3$, before saturating once the correlation length reaches the system size. 
Figure~\ref{fig:beta_width} illustrates this behavior for
Nov.  22 and 23, which unfortunately are the only two days in our sample that have complete trajectory coverage during the early free-flow period. Luckily, for these days $W(t)$ grows cleanly with a slope consistent with $\beta = 1/3$ before saturating shortly after 6:30 am, after the appearance of the first shock waves and congestion spans the entire segment.


To further examine KPZ universality, we analyze the dynamic scaling behavior of both the interface \textit{local} width $W(\ell,t)$ and the height-height correlation function $C(r,t)$. The local width is defined as
$
W(\ell,t) = \left\langle \left[ h'_\ell(x,t) - \left\langle h'_\ell(x,t) \right\rangle \right]^2 \right\rangle^{1/2},
$ where $\ell$ denotes the size of a window over which the height fluctuations are measured. The correlation function $C(r,t) = \langle [h'(x+r, t) - h'(x, t)]^2 \rangle$  quantifies the spatial correlation of height fluctuations at a separation $r$. Although $\ell$ and $r$ serve distinct roles, the former as a window size and the latter as a separation length, both probe spatial fluctuation structures and follow analogous dynamic scaling laws.

In systems described by the KPZ, both $W(\ell,t)$ and $C(r,t)$ are expected to obey dynamic scaling forms in the large-scale limit:
\begin{equation}
    W(\ell,t) \sim t^{\beta} F_w(\ell/t^{1/z}) ~~\text{and}~~ \sqrt{C(r,t)} \sim t^{\beta} F_c(r/t^{1/z}),
\end{equation} where $F_w(u_\ell)$ and $F_c(u_r)$ are universal scaling functions, and the rescaled variables $u_\ell = \ell / t^{1/z}$ and $u_r = r / t^{1/z}$ respectively characterize the two observables.  Despite the different meanings, the asymptotic behaviors of these two functions are structurally similar. For small arguments $u \ll 1$, they grow as power-laws: $F_{\_}(u) \sim u^\alpha$, reflecting correlated fluctuations with self-affine roughness for distances below the correlation length $\xi(t) \sim t^{1/z}$; for large arguments $u \gg 1$, they reach a plateau indicating independent and growing fluctuations for distances  beyond $\xi(t)$ \cite{barabasi1995fractal}. This crossover behavior encodes the dynamic growth of the correlation length over time and serves as a hallmark of KPZ universality.

\begin{figure}[htbp!]
    \centering

     \begin{subfigure}[b]{.475\textwidth}
        \includegraphics[width=\textwidth]{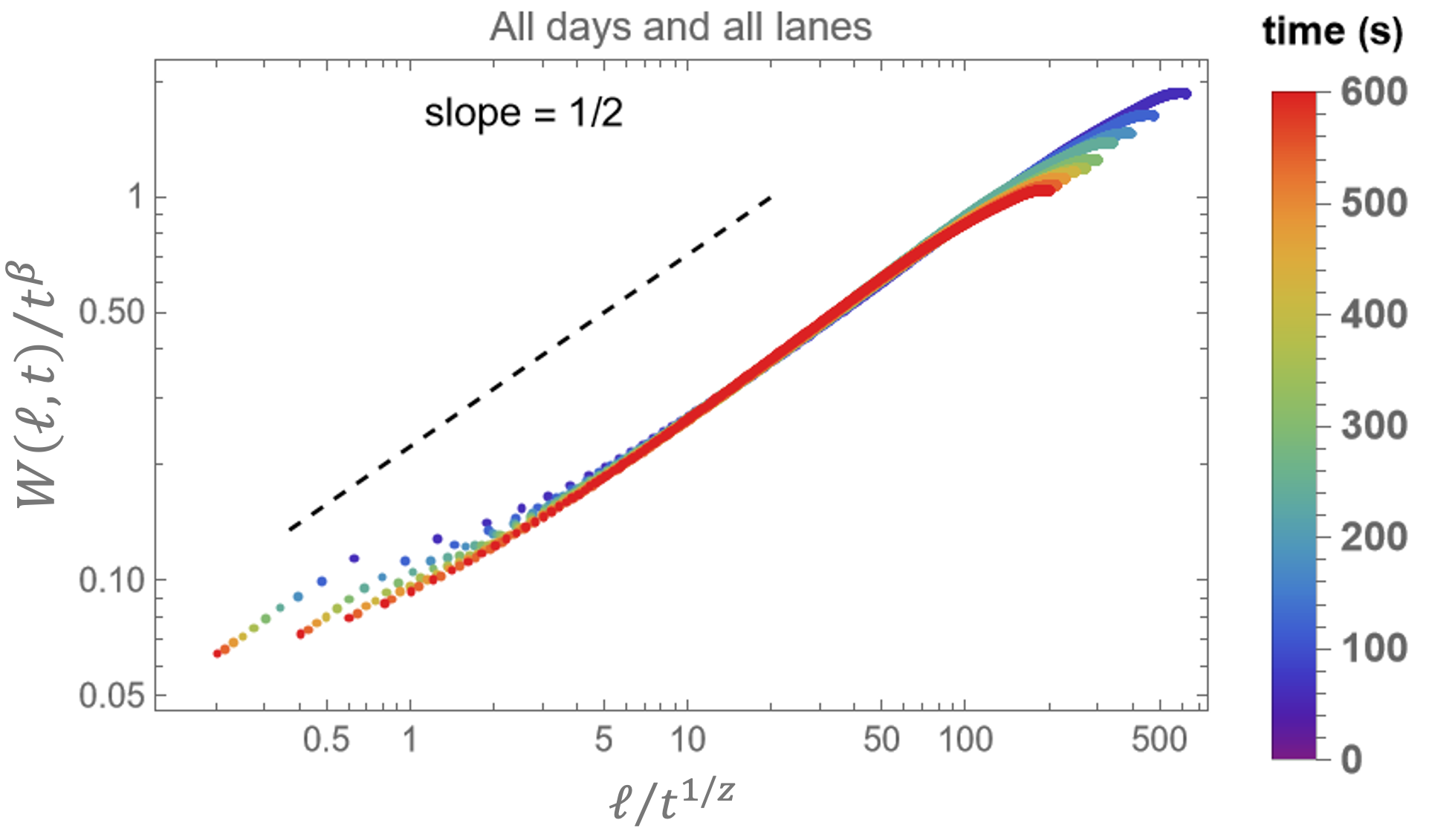}
         \caption{} 
         \label{fig:l3_H}
     \end{subfigure} 
     \begin{subfigure}[b]{0.475\textwidth}
         \includegraphics[width=\textwidth]{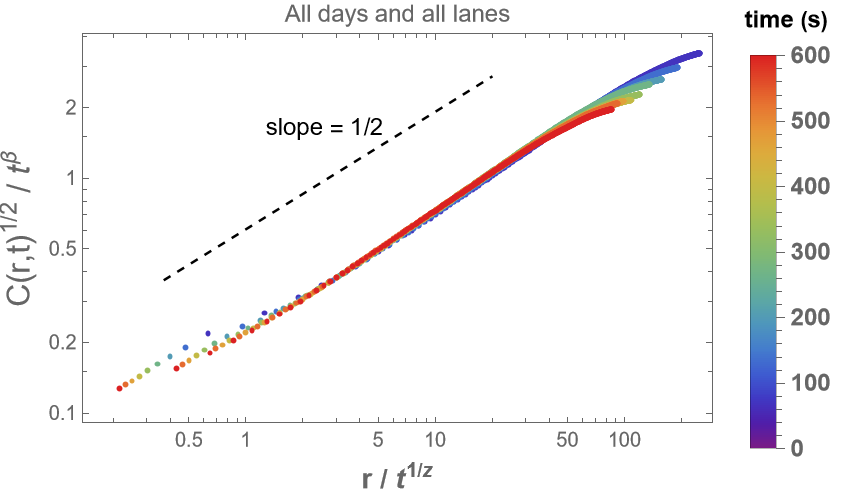}
         \caption{} 
         \label{fig:l2_H}
     \end{subfigure} 
    \begin{subfigure}[b]{0.475\textwidth}
        \includegraphics[width=.95\textwidth]{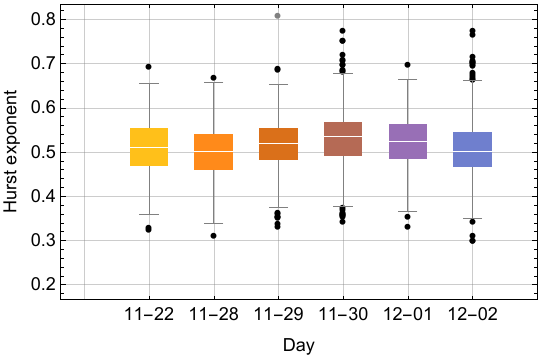}
                  \caption{} 
         \label{fig:H}
     \end{subfigure} 
    \begin{subfigure}[b]{0.475\textwidth}
        \includegraphics[width=.95\textwidth]{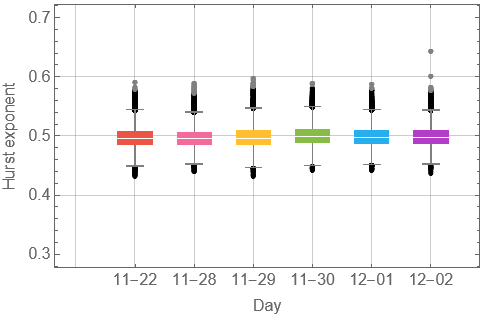}
                  \caption{} 
         \label{fig:H_mle}
     \end{subfigure}

    \caption{\textbf{Evidence of KPZ universality in traffic flow.}
%
%
    \textbf{(A)}  Log-Log plot of rescaled interface width $W(\ell,t)/t^\beta$ as a function of $\ell/t^{1/z}$, with  $\beta=1/3$ and $z=3/2$, for $1\le \ell\le 800$ vehicle lengths ($L=1000$) averaged across all days and lanes. 
    Interface widths are averaged for window size $\ell$ or a given day at time $t$. This is repeated every minute, yielding approximately 180 time slices. Then divide the full window into 18 sequential groups, each containing 10 consecutive time slices. For each of the slices within these groups (\textit{i.e.}, the 1st, 2nd, ..., 10th slice), we average the corresponding slices across all 18 groups, resulting in 10 different lines.
    \textbf{(B)} Same as (A) but for rescaled height-height correlation functions $\sqrt{C(r,t)}/t^{\beta}$ versus $r/t^{1/z}$, and for $1\le r\le 400$. 
    \textbf{(C)} Box plots of Hurst exponents estimated using the R/S method across 1,440 spatial series per day, consistently yielding values near $\alpha = 0.5$.
   \textbf{(D)} Box plots of Hurst exponents obtained by fitting spatial cumulative profiles to fractional Brownian motion via maximum likelihood estimation. The estimates are more tightly clustered around $\alpha = 0.5$ compared to (C).    
}
    \label{fig:corrlength0}
\end{figure}

Therefore, plotting $W(\ell,t)/t^\beta$ or $\sqrt{C(r,t)}/t^\beta$ against their respective rescaled variables $\ell / t^{1/z}$ and $r / t^{1/z}$ should reveal a data collapse onto the scaling functions $F_{\_}(u)$ if the hypotheses is to hold. 
And this is indeed what the data suggests. Figure~\ref{fig:corrlength0}A-B demonstrates that using $\beta=1/3$ and $z=3/2$ produces a notable data collapse onto the expected scaling functions. In both cases, the collapsed curves exhibit a power-law regime with a slope consistent with $\alpha = 1/2$, verifying the self-affine nature of fluctuations predicted by KPZ theory. The absence of the plateau regime in these figures suggests that most of our observations fall in the saturated regime, possibly due to the missing data problem at the beginning of the rush, as mentioned earlier.

To further support these findings, we estimate the roughness exponent $\alpha$ as the Hurst exponent of the detrended height function $h'(x,t)$. We employ two distinct methods to estimate the Hurst exponent:  the rescaled range (R/S) method \cite{hurst1951long}, a model-free approach used to quantify the scaling behavior observed in our data, and maximum likelihood estimation (MLE) under the assumption that the data follows a fractional Brownian motion (fBM), which provides a parametric framework for estimating $H$. 
While both methods indicate that the theoretical value $\alpha = 1/2$ holds, the MLE-based estimates are considerably more tightly centered around this value across all lanes, compared to the R/S method. However, it is important to note that the Brownian motion assumption holds in KPZ-class systems only at a \textit{local} scale, where fluctuations are approximately Gaussian and exhibit short-range correlations; at larger scales, non-Gaussian effects and long-range correlations emerge, which may limit the validity of parametric models such as fBM for capturing the full scaling behavior of the interface.
Estimation details and sensitivity of both methods are provided in the Methods \cite{methods} and the Supplementary Text, respectively.

\subsection*{Discussion}

This study provides the first empirical support that real-world freeway traffic dynamics exhibit self-organized critical percolation, with fluctuations consistent with the KPZ universality class in 1+1 dimensions.
Our cluster analysis revealed that stop-and-go traffic is a critical state that the system naturally evolves to, with critical exponents largely independent of the speed threshold defining congestion. The interface defined by the cumulative vehicle count roughens and grows over time in a manner that is statistically identical to a wide range of other physical systems, from growing bacterial colonies to the burning edge of papers. This establishes a crucial link between traffic and fundamental physics and mathematics, which may open powerful new avenues for traffic science. Key considerations are outlined below.


\textit{Traffic criticality and implications on traffic flow models.} 
We found that real traffic exhibits critical behavior, as jam clusters 
exhibit robust power-law distributions in both size and duration across a broad threshold range ($10 \lesssim v_c \lesssim 35$). The presence of a broad plateau in scaling exponents, rather than a sharp critical point, suggests that real-world freeway traffic does not exhibit criticality in the classical sense, but instead SOC.  

Our results raise two important considerations for traffic flow models. First, near the shockwave critical point, the observed exponents remain stable with $\tau \approx 1.5$, $D_f \approx 1$, $z_P \approx 1$,  and $D \approx 1.5$.  Rather surprisingly, these exponents exactly match the exponent recently found for the \textit{deterministic} version of TASEP \cite{jha2025evidence}, where particles advance with probability $p = 1$ if allowed, but which is not in KPZ. In these systems, the exponents $\tau = 1.5$ and $z_P = 1$ are known exactly and arise from mappings to random walk return-time statistics \cite{laval2023self}. This suggests that in real traffic, the stochastic noise is vanishingly low, but not so low as to lose the KPZ membership. This warrants the careful study of TASEP in the limit as $p\rightarrow1$. 
This can also settle the contentious question in the literature whether or not  TASEP  exhibits critical behavior when $p<1$, as well as what observables and regimes are appropriate for identifying critical phenomena   \cite{roters1999critical, chowdhury2000comment, roters2000reply, schadschneider2010stochastic}.  We suspect that the percolation approach proposed here, not used so far in the literature, could be one of the potential remedies. 

Second, the measured exponent 
are notably distinct from those in the Directed Percolation (DP) universality class, which has a well-established dynamic exponent $z_P \approx 1.58$, but a less universally agreed-upon value for the Fisher exponent $\tau$, which is typically reported around $\tau \approx 1.277$ depending on the observable considered \cite{hinrichsen2000non}. Although traffic jam clusters exhibit anisotropic growth reminiscent of DP, the underlying system lacks key DP features such as irreversible absorbing states. Instead, traffic operates in a quasi-stationary, bidirectional regime where congestion can spontaneously dissipate and re-form. This difference suggests that, despite similarities in growth morphology, freeway traffic belongs to a distinct universality class. 

\textit{Comparison with existing literature.} Our results are generally consistent with the findings of \cite{zhang2019scale}, which remains the only empirical study to investigate traffic criticality on freeways. While their analysis used a fixed critical threshold of 25 mph without exploring sensitivity to this choice, they reported $\tau$ and $\alpha_T$ in the range of 1.7–2 and $D_T$ from 0.91 to 1.25.  In comparison, our study estimates $\tau \approx 1.5$, $\alpha_T \approx 1.75$, and $D_T \approx 1.5$. Note that both pairs of estimates satisfy the hyper-scaling relation of Eq.~\ref{eq:alphaT}. 

These variations likely stem from methodological and contextual differences: (i) our data is collected at finer resolution of cell, whereas their analysis used link-level data; (ii) differences in lane-changing behavior between the two freeways can lead to differing degrees of cluster fragmentation and lateral merging; and (iii) we use the survival function $P(S > s)$ to estimate $\tau$, while their analysis applies logarithmic binning to $P(s)$, a method that can bias exponent estimation in finite systems depending on bin size and range.



\textit{Improved congestion management.} Our results present an opportunity to rethink traditional congestion management strategies, which typically rely on first-order indicators, such as mean speed, flow, or density, to diagnose and control traffic states.  This is especially relevant given the non-Gaussian distributions observed in SOC/KPZ systems, 
which challenge classical assumptions of normality. 
A real-time traffic management system could take advantage of both the SOC nature and the quantifiable scaling behavior characteristic of KPZ universality. By continuously monitoring traffic fluctuations (\textit{e.g.}, speed or headway variance), such a system could estimate scaling exponents and detect early signs of phase transitions, much like how fluctuation statistics are used to forecast rogue waves or financial crashes. The observed plateau in the critical threshold range supports this idea which indicates that traffic operates in a phase-sensitive regime where small shifts in system parameters can avert large-scale breakdowns. Rather than relying on static thresholds, the emergence of power-law scaling across observables may serve as a dynamic early warning signal for impending congestion.

This dual perspective encourages a rethinking of control strategies by not simply preventing threshold violations, but actively steering the system away from self-organized criticality. Traditional approaches like ramp metering or variable speed limits may be enhanced by incorporating not just mean traffic conditions, but also the magnitude and spatiotemporal structure of fluctuations. Detecting KPZ-like scaling patterns provides actionable metrics, while the SOC framework reminds us that localized interventions may only offer temporary relief unless they are designed to disrupt the feedback loops that drive the system toward criticality. Taken together, SOC and KPZ offer a complementary foundation for phase-aware traffic control: one that is both predictive and adaptive, capable of minimizing the formation and propagation of large congestion clusters before they fully emerge.

\textit{The effects of on/off-ramps. }
The direct impact of ramps can be observed in how they modify the shape and the distribution of clusters. Comparing Figures~\ref{fig:tau_1122} and \ref{fig:tau_all_lane}, we find that in outer lanes, clusters tend to become thicker and more laterally spread rather than long and narrow. This effect is especially pronounced in lane 4, which lies directly adjacent to ramps. We observe frequent lateral merging of clusters around on-ramps, while off-ramps often coincide with cluster break points. These ramp-induced disruptions fragment otherwise coherent clusters and create short, wide congestion patterns, inflating the count of small clusters and increasing the estimated value of $\tau$. Future research may aim to systematically explore how ramp-induced flow variations affect scaling laws, both through theoretical extensions that relax the conservation constraint and empirical analyses on datasets with detailed ramp-level annotations.

\clearpage 

%

%
%


\section*{Acknowledgments}
We sincerely thank Dr. Daniel Remenik for his valuable feedback and insightful discussions. We also gratefully acknowledge the I-24 MOTION team for making their trajectory data openly available, which made this study possible.

\paragraph*{Funding:}
This research was supported by the US National Science Foundation through grant CIS-2311159.
\paragraph*{Author contributions:}
The authors confirm their contribution to the paper as follows: study design: G. Lee, and J. Laval; analysis: G. Lee and J. Laval; interpretation of results: G. Lee, J. Laval, A. Jha, and K. Wiesenfeld; manuscript preparation: G. Lee, A. Jha, K. Wiesenfeld, and J. Laval. All authors reviewed the results and approved the final version of the manuscript.
\paragraph*{Competing interests:}
There are no competing interests to declare.
\paragraph*{Data and materials availability:}
The raw I-24 MOTION dataset and its associated publication are available at \url{https://i24motion.org/} and \url{https://doi.org/10.1016/j.trc.2023.104311}, respectively.
The processed data and source code used for analysis and figure generation can be accessed at \url{https://github.com/GaryoungLee/kpz_i24}.


\subsection*{Supplementary materials}
Materials and Methods\\
Supplementary Text\\
Figs. S1 to S18\\
References \textit{(30-\arabic{enumiv})}\\ 

\newpage

\renewcommand{\thefigure}{S\arabic{figure}}
\renewcommand{\thetable}{S\arabic{table}}
\renewcommand{\theequation}{S\arabic{equation}}
\renewcommand{\thepage}{S\arabic{page}}
\setcounter{figure}{0}
\setcounter{table}{0}
\setcounter{equation}{0}
\setcounter{page}{1} 


\begin{center}
\section*{Supplementary Materials for\\ \scititle}


	Garyoung~Lee$^{1}$,
	Aryaman~Jha$^{2}$,
	Kurt~Wiesenfeld$^{2}$,
    Jorge~Laval$^{1\ast}$\and \\
	\small$^{1}$School of Civil and Environmental Engineering, Georgia Institute of Technology, Atlanta \& 30332, United States.\\
	\small$^{2}$School of Physics, Georgia Institute of Technology, Atlanta \& 30332, United States.\\
	\small$^\ast$Corresponding author. Email: jorge.laval@ce.gatech.edu\and
\end{center}

\subsubsection*{This PDF file includes:}
Materials and Methods\\
Supplementary Text\\
Figures S1 to S18\\
References \textit{(30-\arabic{enumiv})}

\newpage


\subsection*{Materials and Methods}

\subsubsection*{Dataset}
The rich dataset from I-24 MOTION supports our empirical investigation into the critical scaling laws of traffic flow. The testbed is designed to provide a real-world freeway setting for experimenting with advanced traffic management and automated vehicle technologies. 

Previous studies have leveraged this testbed for various experiments.  For example, \cite{jang2024reinforcement} and \cite{lee2024traffic} used it to experiment with 100 reinforcement learning control vehicles, called MegaVanderTest. Additionally, the effects of variable speed limit control strategies were explored using this testbed \cite{zhang2023marvel}. The currently available dataset, INCEPTION v1.0.0, includes natural vehicle trajectory data without any control interventions, representing the first version of this dataset. Vehicle trajectories in this dataset are processed using computer vision algorithms applied to video footage \cite{gloudemans2021vehicle, gloudemans2024so}. Despite comprehensive processing efforts, some challenges remain. These include reconciling the same vehicles captured by different cameras \cite{wang2024automatic, wang2023online} and reconstructing trajectories to eliminate false positive collisions and impute missing parts caused by overpasses and tracking errors \cite{ji2023platoon, ji2024virtual}.  Data artifacts due to hardware and software issues including missing poles, overpass, and network bandwidth limitations were reported. See Supplementary Text for details \cite{gloudemans202324}.

\begin{figure}[htbp!]
\centering
     \begin{subfigure}[b]{0.31\textwidth}
         \includegraphics[width=\textwidth]{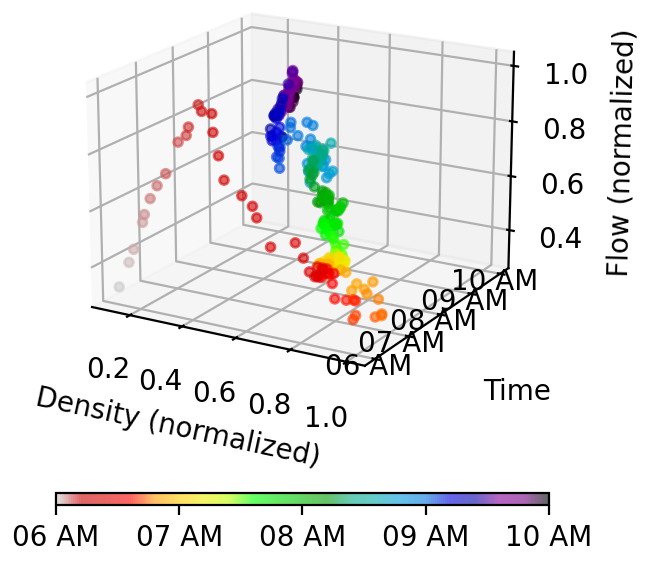} 
         \vskip -0.06 in
         \includegraphics[width=0.97\textwidth]{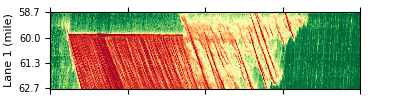}
         \caption{Nov. 21st} 
     \end{subfigure} 
    \begin{subfigure}[b]{0.31\textwidth}
         \includegraphics[width=\textwidth]{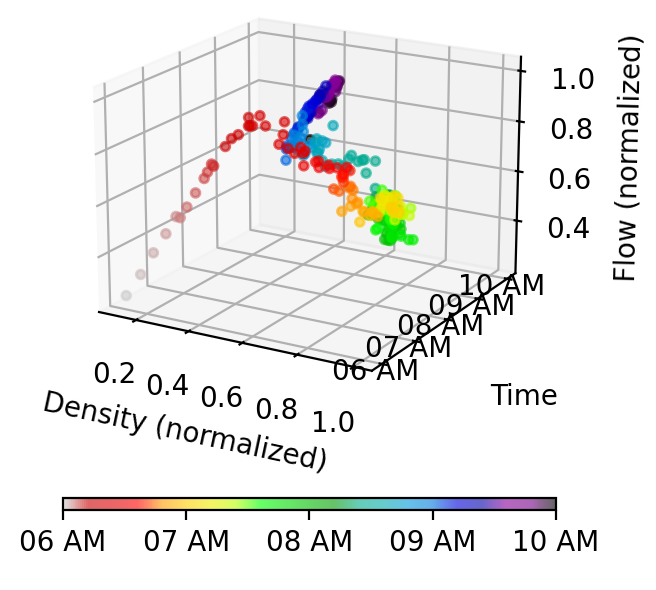} 
         \vskip -0.06 in
         \includegraphics[width=0.97\textwidth]{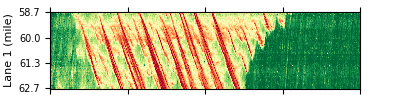}
         \caption{Nov. 22nd} 
     \end{subfigure} 
    \begin{subfigure}[b]{0.31\textwidth}
         \includegraphics[width=\textwidth]{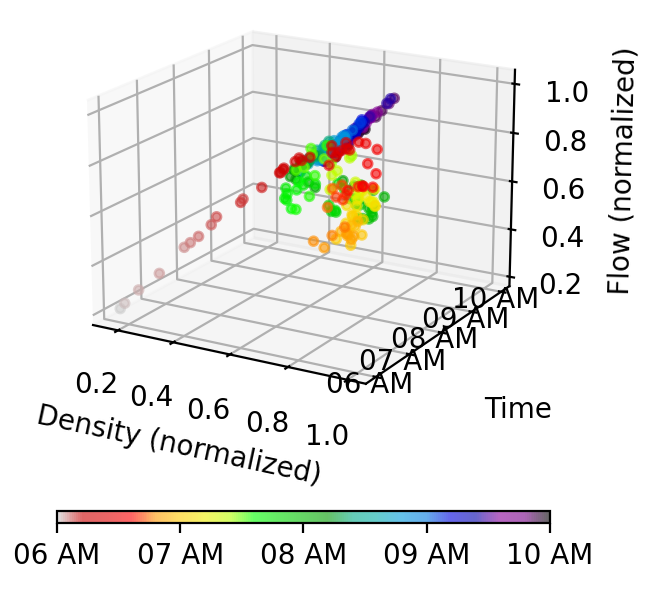} 
         \vskip -0.06 in
         \includegraphics[width=0.97\textwidth]{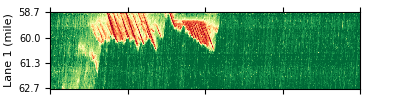}
         \caption{Nov. 23rd} 
    \end{subfigure} \\
 
    \begin{subfigure}[b]{0.31\textwidth}
         \includegraphics[width=\textwidth]{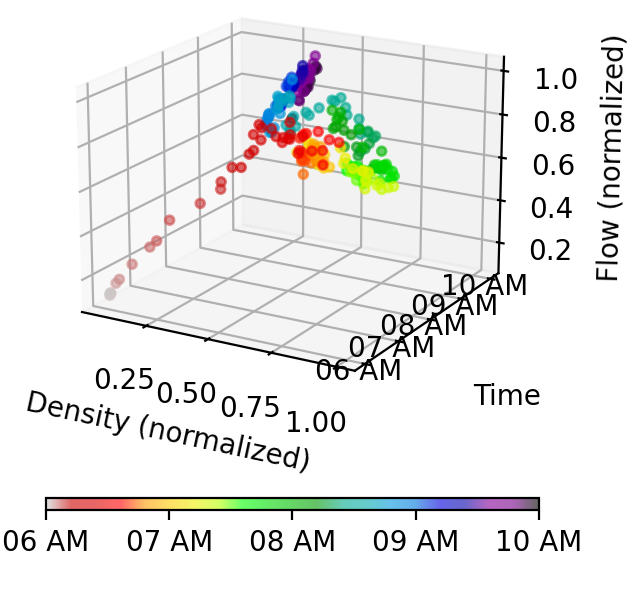} 
         \vskip -0.06 in
         \includegraphics[width=0.97\textwidth]{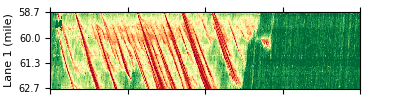}
         \caption{Nov. 28th} 
     \end{subfigure} 
          \begin{subfigure}[b]{0.31\textwidth}
         \includegraphics[width=\textwidth]{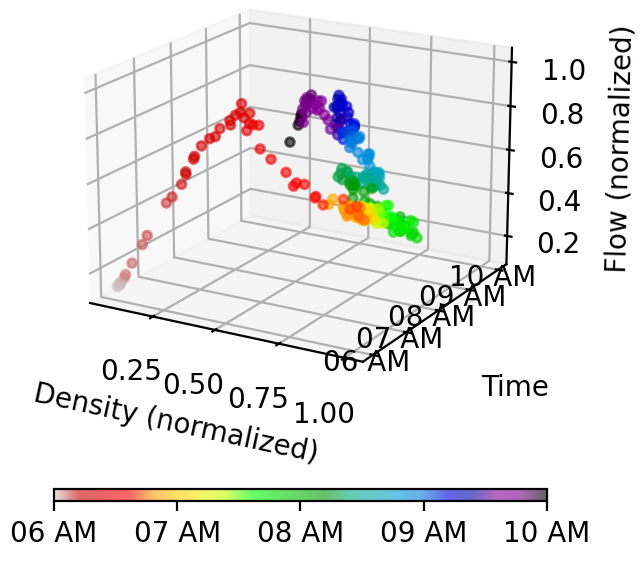} 
         \vskip -0.06 in
         \includegraphics[width=0.97\textwidth]{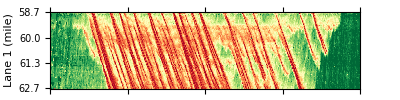}
         \caption{Nov. 29th} 
     \end{subfigure} 
          \begin{subfigure}[b]{0.31\textwidth}
         \includegraphics[width=\textwidth]{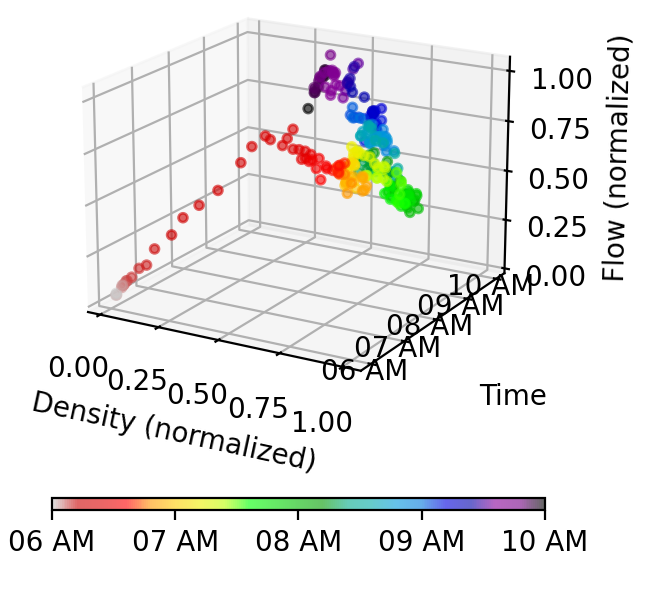} 
         \vskip -0.06 in
         \includegraphics[width=0.97\textwidth]{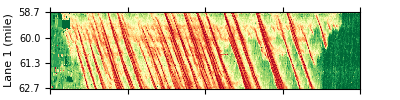}
         \caption{Nov. 30th} 
         \end{subfigure} \\
      
              \begin{subfigure}[b]{0.31\textwidth}
         \includegraphics[width=\textwidth]{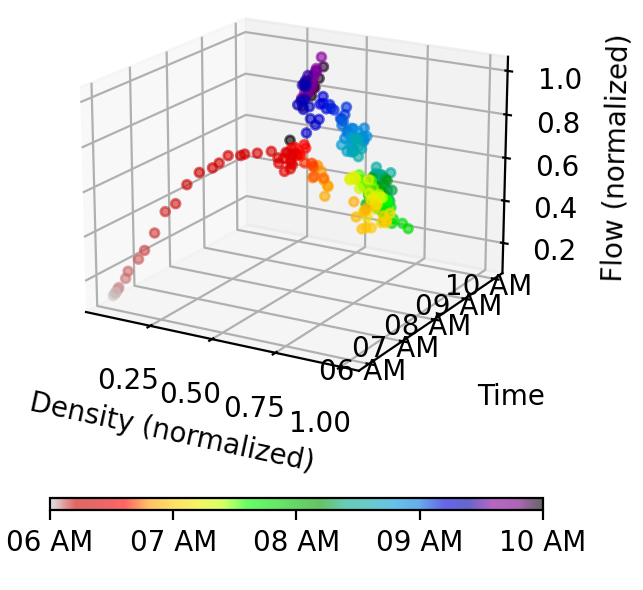} 
         \vskip -0.06 in
         \includegraphics[width=0.97\textwidth]{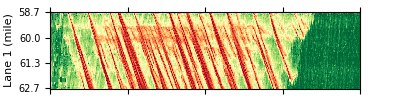}
         \caption{Dec. 1st} 
     \end{subfigure} 
          \begin{subfigure}[b]{0.31\textwidth}
         \includegraphics[width=\textwidth]{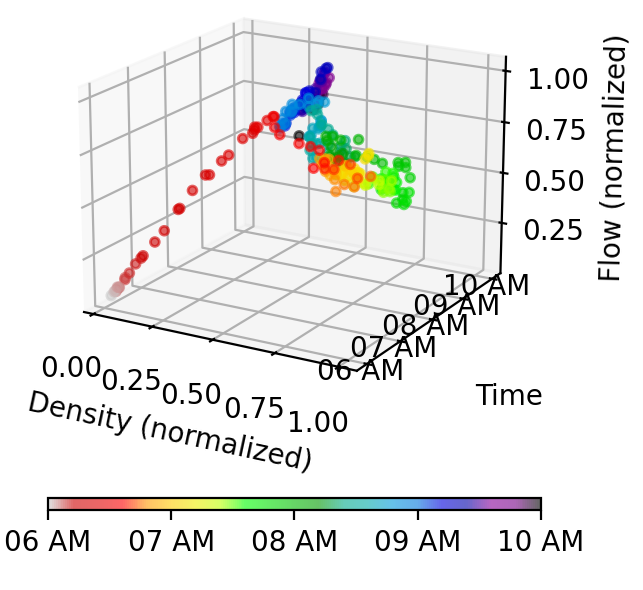} 
         \vskip -0.06 in
         \includegraphics[width=0.97\textwidth]{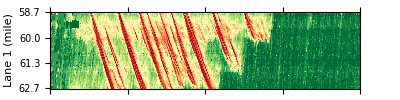}
         \caption{Dec. 2nd} 
     \end{subfigure} 
    \caption{\textbf{Fundamental diagram and time–space diagram for eight days of westbound traffic.} 
Top panel: one-minute aggregated fundamental diagram (flow–density), with color-coded trajectories indicating congestion dynamics. 
Bottom panel: time–space diagram for lane 1, where red denotes congestion and green indicates free flow.} \label{fig:data}
\end{figure}

Figure \ref{fig:data} shows eight days of westbound traffic having instances of observable congestion. Note that the morning rush is absent in the eastbound direction as it leads out of central Nashville. In each subfigure, the top panel presents a one-minute aggregated fundamental diagram, with the flow–density trajectory color-coded to highlight when congestion appears and dissipates. Directly below, the bottom panel depicts a lane 1 (leftmost) time–space diagram in which red shading indicates high congestion and green denotes free flow. Although the diagrams for lanes 2 to 4 (left to right) are omitted for brevity, it was indicated that shock waves are more frequently interrupted in outer lanes while inner lanes maintain more coherent wave patterns \cite{gloudemans202324}: see figure \ref{fig:tau_all_lane}. This is due to on-/off-ramps in the freeway. The studied segment includes five merging/diverging zones. These geometric features may influence wave formation and clustering behavior.

Traffic patterns across most days follow a consistent structure: free-flow conditions persist until approximately 6:30 AM, after which traffic transitions into a congested regime due to morning peak demand, and eventually recovers to free-flow by around 10 AM. This pattern, characterized by parallel congestion waves in the time-space diagram, reflects recurrent congestion, \textit{i.e.}, predictable congestion that results from routine commuting behavior. However, certain days deviate from this typical pattern. On November 21, an incident caused severe non-recurrent congestion, evident from disrupted wave patterns and a large, irregular congestion patch early in the observation window. Conversely, on November 23, the day before Thanksgiving, traffic demand was significantly lower. A minor incident did occur, but the resulting congestion was minimal and quickly resolved. 

Because our scaling analysis requires traffic conditions near the critical density--neither far above nor well below--it excludes those two anomalous days. The remaining six days thus provide a representative sample of near‐critical congestion for all subsequent exponent estimation.  As more data become publicly available, future analysis can expand beyond the Thanksgiving week to further test the consistency of the estimated scaling exponents.

\subsubsection*{Theory: KPZ Universality Class}

This section introduces the conceptual and theoretical foundations of the KPZ universality class and explains how its three critical exponents attain the universal values.

Dynamic systems across distinct physical phenomena often exhibit \textit{universal} patterns that can be described by similar mathematical frameworks. One such universal behavior is captured by the KPZ universality class, which describes how stochastic interfaces grow and fluctuate in a subset of non-equilibrium systems characterized by local interactions and lateral spreading \cite{kardar1986dynamic}. These interfaces can be found in many natural and man-made processes, 
\textit{e.g.,} the infiltration of fluids in porous media, the roughening of crystal surfaces \cite{takeuchi2010universal}, or the growth of bacterial colonies \cite{allen2018bacterial}. Specifically, consider a sheet of paper burning from one edge. As the flame spreads, it advances the boundary between the burned and unburned regions. The growth of this boundary is influenced by both deterministic spreading (\textit{i.e.}, the steady combustion of material) and random fluctuations (\textit{i.e.}, local variations in paper density or air flow that cause uneven burning). Over time, these effects produce an irregular, rough front whose fluctuations have been experimentally shown to follow the KPZ class \cite{myllys2001kinetic}.

The KPZ equation is the prototypical continuum model for the evolution of such an interface height \(h(x,t)\) at position \(x\) and time \(t\) as:
\begin{equation}
\frac{\partial h(x,t)}{\partial t} = \frac{\lambda}{2} (\nabla h(x,t))^2 +\nu \nabla^2 h(x,t) +  \eta(x,t)
\end{equation}

\noindent where $\lambda$ controls the strength of nonlinear lateral growth,  $\nu$ is the surface tension term promoting smoothness, and $\eta(x,t)$ is a noise term. The equation describes the stochastic evolution of an interface, incorporating three main effects: (i) lateral growth due to local slope-dependent effects ($\frac{\lambda}{2} (\nabla h)^2$), (ii) smoothing due to surface tension ($\nu \nabla^2 h$), and (iii) stochastic roughening due to random noise ($\eta$). In the case of traffic flow, this equation has been derived from the existence of the density-flux relationship  through the scaling limit of deterministic Hamilton-Jacobi equations supplemented with viscosity and stochastic perturbations \cite{laval2024traffic}.

A hallmark of KPZ in 1+1 dimensions is the 1:2:3 scaling law, which governs how time, space, and fluctuations scale near the KPZ fixed point \cite{matetski2021kpz}. Explicitly, the interface height $h(x,t)$ is rescaled as:
\begin{equation}\label{he}
h_\epsilon(x,t) := \epsilon^{1/2} h(\epsilon^{-1}x, \epsilon^{-3/2}t) - C_{\epsilon} t
\end{equation}
\noindent where $C_\epsilon$ is a constant capturing the average growth rate of the interface, and in the limit $\epsilon \rightarrow 0$, this rescaled solution approaches a universal distribution known as the KPZ fixed point. 
Under Eq.~\ref{he} time scales as $\epsilon^{-3/2} \sim t$, space scales as $\epsilon^{-1} \sim t^{2/3}$, and fluctuations grow as $\epsilon^{-1/2} \sim t^{1/3}$. Hence, the 1:2:3 scaling ratio is featured by explicitly indicating the fluctuations, space, and time scale as $t^{1/3}:t^{2/3}:t$, respectively, from which 
the KPZ scaling exponents follow: \begin{itemize}
\setlength\itemsep{0em}
\setlength{\itemindent}{-1em}
    \item[-] Roughness exponent $ \alpha = 1/2 $, since fluctuation scales with space as $ t^{1/3} \sim (t^{2/3})^\alpha $
    \item[-] Growth exponent $ \beta = 1/3 $, since fluctuation scales with time as $ t^{1/3} \sim (t)^\beta $
    \item[-] Dynamic exponent $ z = 3/2 $, since time scales with space as $ t \sim (t^{2/3})^z $
\end{itemize}

\subsubsection*{Percolation Theory}

Percolation theory investigates how large-scale connectivity emerges from local randomness in a system \cite{christensen2002percolation}. A classical model is site percolation, where each site on a lattice is independently occupied with probability $p$ and vacant with probability $1-p$. As $p$ increases, small groups of occupied sites merge to form larger connected clusters, and at a critical value $p_c$, a system-spanning cluster appears which marks a phase transition. In standard models, the critical threshold $p_c$ is identified by observing how the sizes of the largest and second-largest clusters evolve with increasing $p$. Below $p_c$, clusters remain small; at $p_c$, the second-largest cluster typically peaks before merging into a giant component. This probabilistic framework enables precise detection of criticality. 

However, our setting departs from this traditional formulation. In empirical systems like traffic, each site does not have a probability of being active, but instead possesses a measured value (e.g., speed $v$). Thus, percolation is induced by applying a deterministic threshold to these values: sites are considered percolating if they fall below a threshold speed $v_c$. This thresholding replaces the stochastic occupation probability $p$ in classical theory.

It is important to note that our system features a clear directional structure, implying that connectivity unfolds along a preferred direction. As such, our framework is more accurately described by concepts from directed percolation, which governs systems where activity propagates anisotropically or under temporal causality. While we do not fully implement a DP model here, the analogy provides theoretical grounding for our interpretation of the critical threshold and associated scaling behaviors. To identify the critical threshold, we do not rely on spanning clusters but instead search for power-law behavior in the empirical cluster size distribution. The emergence of a scale-free distribution suggests proximity to criticality and aligns with the universal behavior expected in percolation transitions.

\subsubsection*{Critical Exponent Estimation}
We estimate the critical exponents $\tau$, $\alpha_R$, and $\alpha_T$ using piecewise linear regression on the log-log survival functions of cluster size $S$, spatial extent $R$, and temporal duration $T$, respectively. The cutoff point is constrained to occur within the last 10\% of the data to ensure it reflects the finite size effects and is chosen to minimize the total mean squared error (MSE) of the two fitted linear segments. For $D_R$ and $D_T$, we directly estimate the exponents by performing linear regression on the relevant log-transformed variables. The dynamic exponent $z_P$ is not independently estimated but is obtained by taking the ratio of the estimated $D_R$ and $D_T$ values.

The fractal dimensions $D$ and $D_f$ are estimated via the box-counting method, as described in the next section. In particular, when the cutoff point is selected based on the piecewise fit for $\tau$, the corresponding cutoff fractal dimension $D_f$ is computed by averaging the fractal dimensions of five clusters around the cutoff size, \textit{i.e.}, the two clusters just smaller and the two just larger than the cutoff cluster.

\subsubsection*{Box-counting Method}

\begin{figure}[htbp!]
    \centering
    \includegraphics[width=\linewidth]{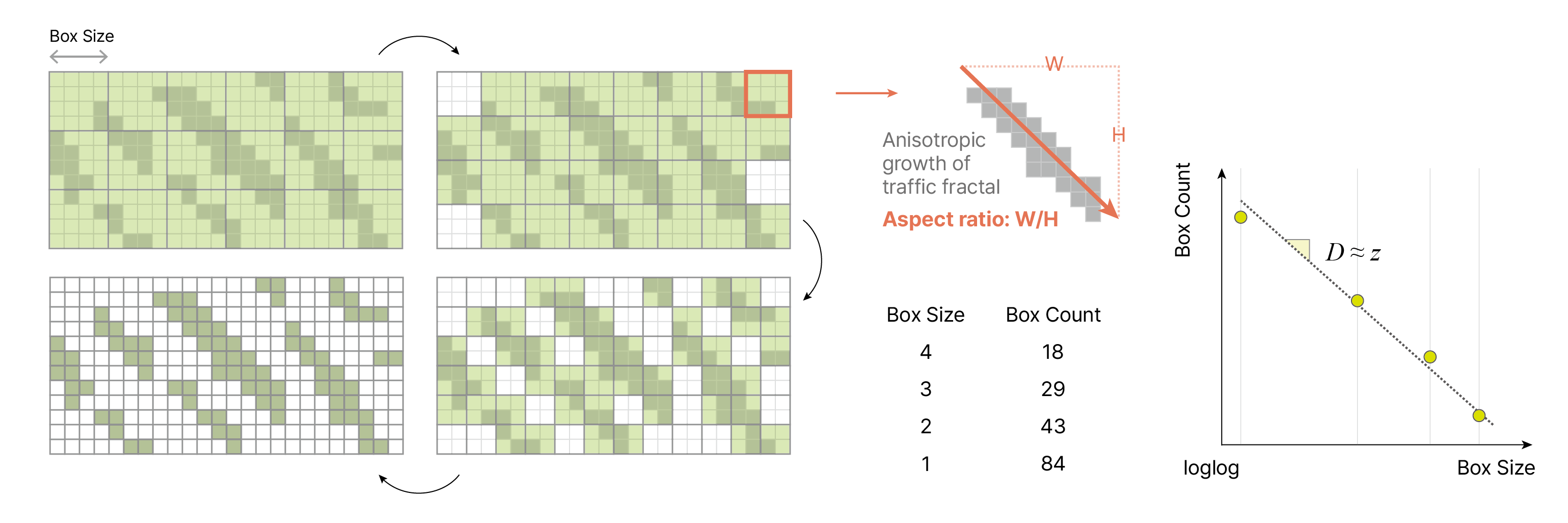}
    \caption{\textbf{Schematic illustration of the box-counting method applied to a simplified traffic congestion pattern.} This toy example demonstrates how varying box sizes are used to cover jammed regions and estimate fractal dimensions. The aspect ratio is set to reflect the anisotropic nature of traffic wave propagation, simply set to 1 here.}
    \label{fig:boxcounting}
\end{figure}

Box-counting is a widely used technique for estimating the fractal dimension, which characterizes the scaling behavior of spatial structures \cite{mandelbrot1982fractal}. The standard procedure involves superimposing a fixed grid of square boxes over the binary cluster field and counting how many boxes contain any part of the cluster structure at different scales. The log-log slope of box size versus count yields the estimated dimension. Figure \ref{fig:boxcounting} illustrates this process.

To reduce boundary effect and alignment sensitivity, we adopt a sliding-grid variant of the method, which averages counts over multiple grid placements with small shifts of 50\% overlap. This increases the robustness, particularly in systems with anisotropic or fragmented structures. In addition, we adjust the box aspect ratio to reflect the anisotropic nature of traffic wave propagation, aligning the counting process with the directional growth observed in real congestion patterns (Eq.~\ref{eq:hw}).

To estimate $D$ and $D_f$ from our data, we use the sliding-grid averaging version of the box-counting, which enhances the traditional approach by reducing edge effects and noise sensitivity \cite{cheng1997multifractal, reiss2015noise, mandelbrot1982fractal}.
A key consideration in the box-counting method is the choice of the box height $H$ and width $W$. While square boxes are commonly used, they may be inadequate for anisotropic structures like traffic congestion, which propagates upstream at an approximately constant wave speed in the 10-15 mph range \cite{treiber2010three, chen2012microscopic, gloudemans202324}.
We therefore conjecture that the box aspect ratio $W/H$ should align with the characteristic anisotropy of traffic congestion, as: 
\begin{equation} \label{eq:hw}
\text{wave~speed} = \frac{H}{W} \cdot \frac{\Delta x}{ \Delta t}  = \text{10~-~14~mph},
\end{equation}
which corresponds to an aspect ratio close to 1. Our choice of $\Delta t = 6$ seconds was made precisely to ensure that an aspect ratio of 1 would align with the empirically observed shockwave speed. However, if the resolution of the time-space grid were to change, the aspect ratio must be adjusted accordingly to maintain consistency with the anisotropic structure of traffic jams. 

To reduce sensitivity to the choice of box sizes, we average the fractal dimensions estimated using three subsets: \{5, 10, 20, 40, 80, 160\}, \{10, 20, 40, 80, 160\}, and \{20, 40, 80, 160\}. A comparison of both aspect ratio and box size range, presented in the Supplementary Text, confirms that deviations from the optimal aspect ratio lead to slightly different estimates of $D$ and $D_f$, highlighting the importance of anisotropy-aware box scaling.

\subsubsection*{Hurst Exponent Estimation}

To quantify the spatial roughness of the detrended cumulative vehicle count profiles $ h'(x, t) $, we estimate the Hurst exponent $ H $, which characterizes the long-term memory and fractal scaling of a series. Specifically, the Hurst exponent describes how variability grows with the observation window size: values $ H < 0.5 $ indicate anti-persistent and smoother behavior; $ H = 0.5 $ corresponds to uncorrelated, random walk–like roughness; and $ H > 0.5 $ reflects persistent and rougher fluctuations. Under KPZ scaling, the expected roughness exponent is $ \alpha = H = 1/2 $.

Two complementary methods were employed to estimate $ H $: the rescaled range (R/S) method and maximum likelihood estimation (MLE) assuming a fractional Brownian motion (fBM) model.

The rescaled range (R/S) method \cite{hurst1951long} provides a robust, model-free approach to estimating the roughness of a data series. The method calculates the cumulative deviation from the local mean within subintervals of length $ n $, computes the range of this cumulative deviation, and rescales it by the standard deviation within the same interval. The central idea is to analyze how the average rescaled range $ R/S(n) $ grows with the subinterval size $ n $. If a power-law relationship $ R/S(n) \sim n^H $ holds, the exponent $ H $ offers a measure of the signal's roughness. Since the R/S method does not assume stationarity, Gaussianity, or a specific generative process, it serves as a versatile tool for detecting long-range dependence and characterizing the roughness of a wide class of time series.
  For each subinterval of length $n$:

\begin{enumerate}
\setlength\itemsep{0em}
    \item Make sure the data $(x_1,\ldots x_m)$ exhibits no deterministic trends
    \item Compute the first differences (increments) of the data: $\Delta x_i = x_{i+1} - x_i$.
    \item  Subtract the mean of the segment to obtain centered deviations: $d_i =\Delta  x_i - \bar{\Delta  x}$.
    \item Construct the cumulative deviation series: $Z_k = \sum_{i=1}^{k} d_i$ for $k = 1, \ldots, n-1$.
    \item Compute the range of cumulative deviations: $R = \max(Z_k) - \min(Z_k)$.
    \item Compute the standard deviation $S$ of the segment $x_i$.
    \item The rescaled range is defined as $R/S$. This process is repeated for varying subinterval lengths $n$. 
    The slope of the best-fit line on 
    a log-log plot of $R/S$ versus $n$ 
    corresponds to the Hurst exponent $H$.
\end{enumerate}

Step 1 above introduces an additional parameter, the window size parameter $r$, needed to eliminate deterministic trends in the data. In this paper, we use a Gaussian filter for the task, which can be thought of as a weighted moving average with weights given by a Gaussian $(0,\sigma)$ kernel, with $\sigma=r/2$. In the main text, we chose $r=L/3$ as it implies that $6\sigma=L$, ensuring that the smoothing window spans the full scale of analysis; see Supplementary Text for the limited impact of this parameter on our results.

The second method is a parametric approach based on maximum likelihood estimation under the assumption that the spatial signal follows a fractional Brownian motion (fBM). This model assumes the process is Gaussian with stationary increments and long-range dependence. The MLE procedure jointly estimates $ H $, drift, and volatility parameters. 

However, the fBM assumption does not generally hold for systems governed by KPZ dynamics, which are inherently non-Gaussian and nonlinear. The spatial traffic count profiles can resemble Brownian motion of $H = 1/2$ only under certain conditions: (i) locally during transient evolution, or (ii) globally once the system reaches equilibrium at the KPZ fixed point. While verifying either condition requires further analysis, the fBM-based MLE still serves as a useful benchmark. It strengthens the case for KPZ scaling without requiring additional tuning parameters beyond those internally estimated by the model.


\subsection*{Supplementary Text}

\subsubsection*{Finite-Size Scaling and Evidence for the Critical Speed Threshold}   
The result of finite-size scaling (FSS) provides supporting evidence for the existence of a critical speed threshold in traffic congestion dynamics. It may serve as an additional validation tool to determine $v_c$, reinforcing findings derived from independent percolation-based analyses in the main text.

According to percolation theory, the survival function of cluster sizes is expected to follow the scaling form:
\begin{equation}
P(S > s) \sim s^{-(\tau-1)} f\left(\frac{s}{L^{D_f}}\right),
\end{equation}
\noindent where  $f(\cdot)$ is a universal scaling function. At criticality, plotting $ P(S > s) s^{\tau - 1}$ against $  s / L^{D_f}$ should yield a nice collapse of curves from different system sizes $L$.

To systematically identify the optimal $v_c$, we follow the method to quantify the quality of collapse, $C$, elaborated on \cite{serafino2021true}. For a range of candidate threshold values, we compute each of the rescaled coordinates $(s / L^{D_f}, P(S > s) s^{\tau - 1})$ across system sizes and construct the master curve $Y$ by locally fitting straight lines through pooled values of the rescaled survival functions. The quality of the collapse is quantified using a reduced–$\chi^2$-like statistic:

\begin{equation}
C = \frac{1}{3 |I|} \sum_{(L,s) \in I} \frac{\left[ P(S > s)s^{\tau - 1} - Y\left(\frac{s}{L^{D_f}}\right) \right]^2}{\sigma_{L,s}^2 + \Sigma\left(\frac{s}{L^{D_f}}\right)^2},
\end{equation}

\noindent where $\sigma_{L,s}$ and $\Sigma(\cdot)$ are the standard errors of the observed values and the fitted master curve, respectively, and $|I|$ is the total number of data points contributing to the collapse. The denominator includes a factor of 3 to account for degrees of freedom from the local linear fit and error propagation. A collapse metric $C \approx 1$ indicates high-quality alignment and near-critical behavior.

\begin{figure}[H]
    \centering
        \begin{subfigure}[b]{\textwidth}
         \includegraphics[width=\textwidth]{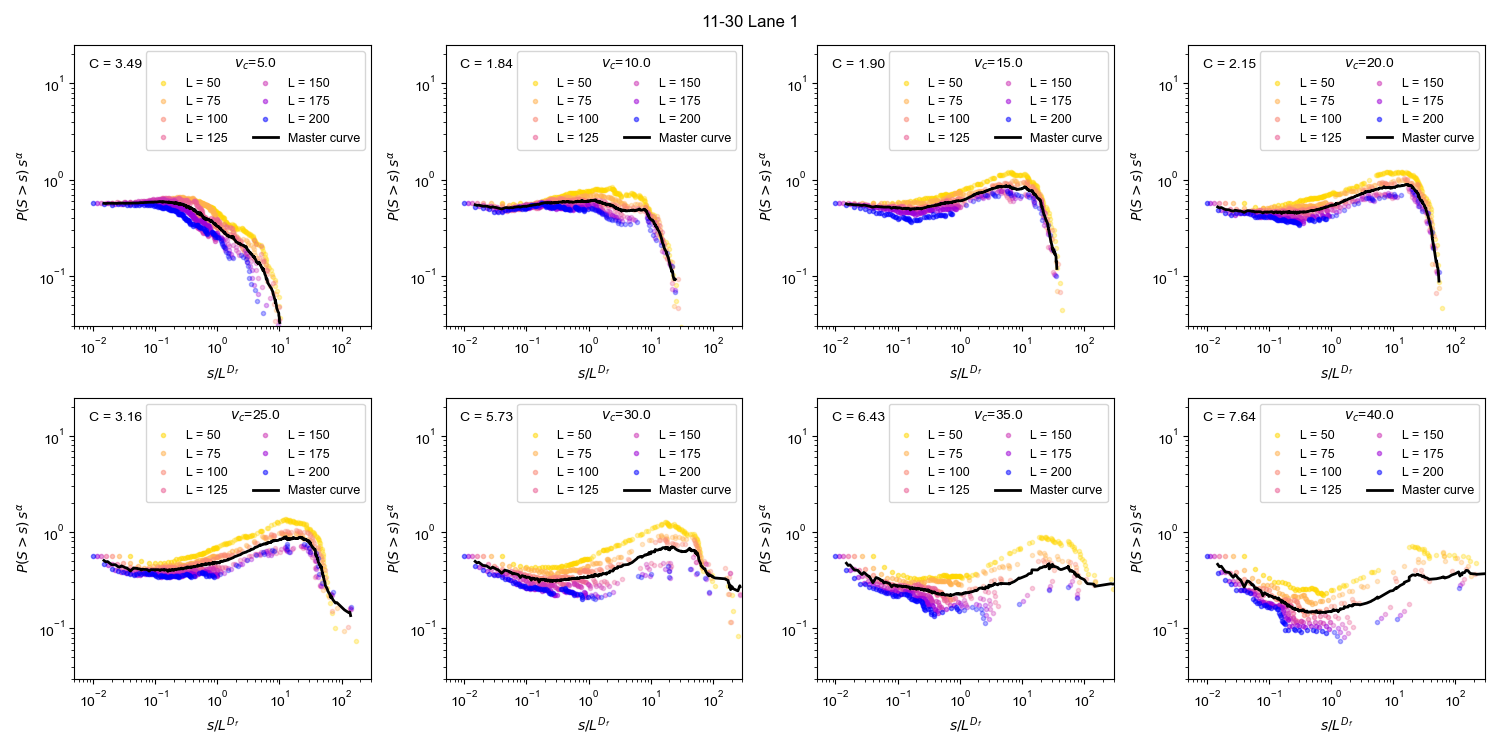}
         \caption{} \label{fig:fss_1130}
     \end{subfigure} 
     \begin{subfigure}[b]{0.49\textwidth}
         \includegraphics[width=\textwidth]{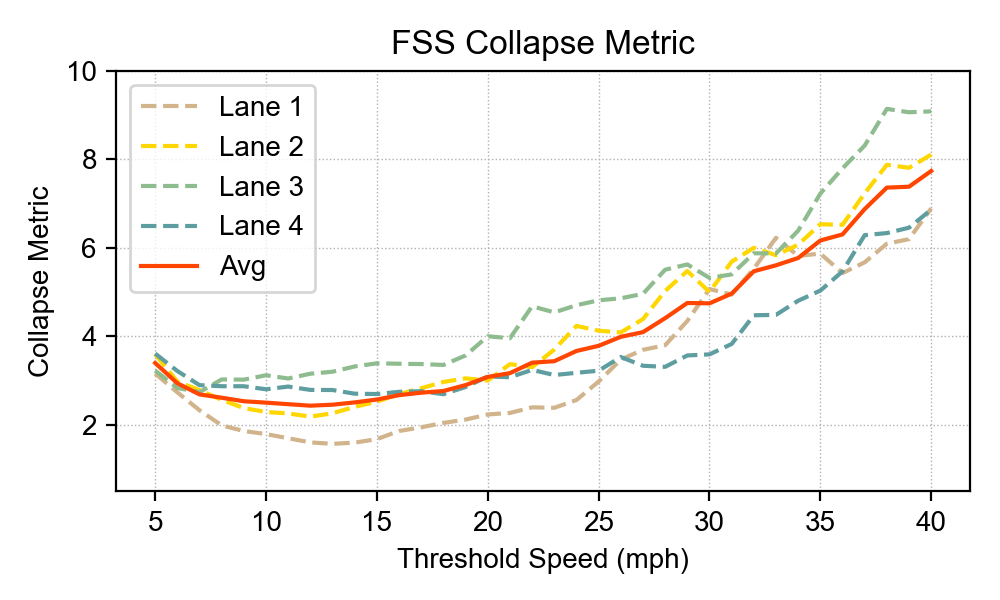}
         \caption{} \label{fig:fss_comp} 
     \end{subfigure} 
          \begin{subfigure}[b]{0.49\textwidth}
         \includegraphics[width=\textwidth]{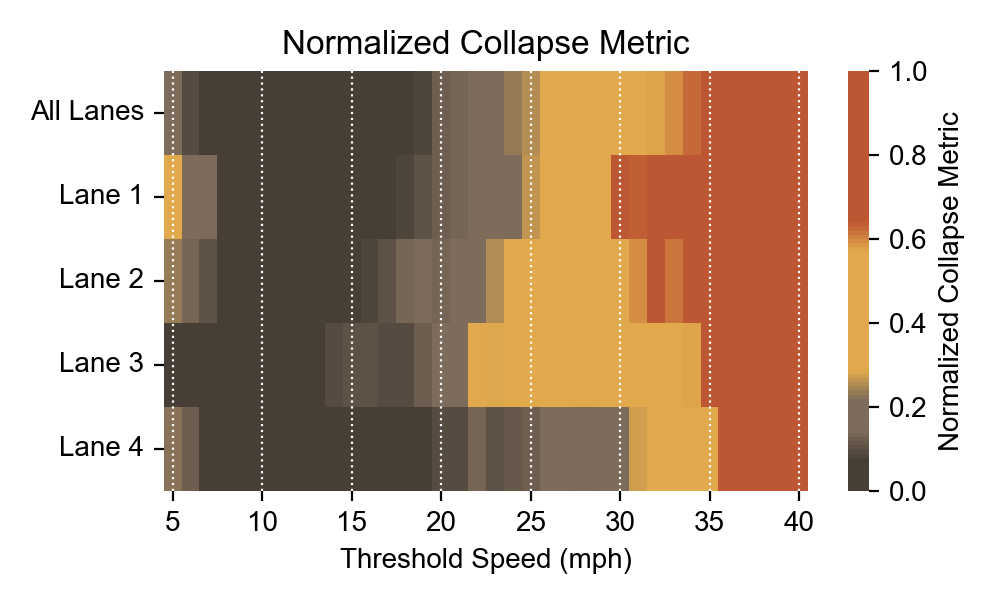}
         \caption{} \label{fig:heat}
     \end{subfigure} \quad
 \caption{\textbf{Finite-size scaling analysis as supporting evidence for critical speed threshold.} 
\textbf{(A)} Data collapse across system sizes for November 30 lane 1 under various candidate thresholds, with the best alignment observed near 10 - 20 mph, both visually and based on the collapse quality metric. 
\textbf{(B)} Collapse quality $C$ evaluated across candidate thresholds, shown as an average over all dates and all lanes as well as averages per lane. 
\textbf{(C)} Heatmap of the normalized collapse quality metric $C$ across thresholds and lanes, averaged over all dates.}
    \label{fig:fss_comparison}
\end{figure}

The speed threshold $v_c$ that minimizes $C$ can be regarded as the \textit{optimal critical speed threshold}. Figure~\ref{fig:fss_1130} shows the exemplary collapse results for November 30 lane 1, where 10 - 20 mph yields the best alignment among tested thresholds. Although the collapse is not perfect likely due to real-world effects such as noise and finite sample sizes, it still exhibits sufficient structure to support percolation-based interpretation. A notable feature observed in the rescaled survival curves is the bump just before the curves drop sharply. This bump becomes more pronounced as the threshold speed increases. It reflects the finite-size limitation of the system: near the upper tail of the distribution, only a few very large clusters exist which leads to statistical fluctuations in $P(S > s)$. Moreover, when the threshold is too high, many smaller jammed regions artificially merge into fewer but disproportionately large clusters. This effect reflects the overflow point of congestion and distorts the ideal power-law behavior thereby causing the observed bump.

To systematically evaluate threshold sensitivity, we compute the collapse metric $C$ over a dense threshold grid (5 to 40 mph in 1 mph increments) for all dates and lanes. Figure~\ref{fig:fss_comp} and Figure~\ref{fig:heat} summarize the average collapse metrics across all days and lanes, as well as the lane-specific averages. This aggregation enables robust identification of threshold values consistently exhibiting high-quality collapse across different spatial configurations.

Unlike the broader 10–30 mph range suggested in the main text, the FSS results do not exhibit widespread robustness across this interval. However, a pronounced local minimum in the collapse metric appears near 10–15 mph, providing additional support for the empirically identified critical threshold range.

\subsubsection*{Hyper-scaling Relationship}

\begin{figure}[hptb!]
    \centering
    \includegraphics[width=\linewidth]{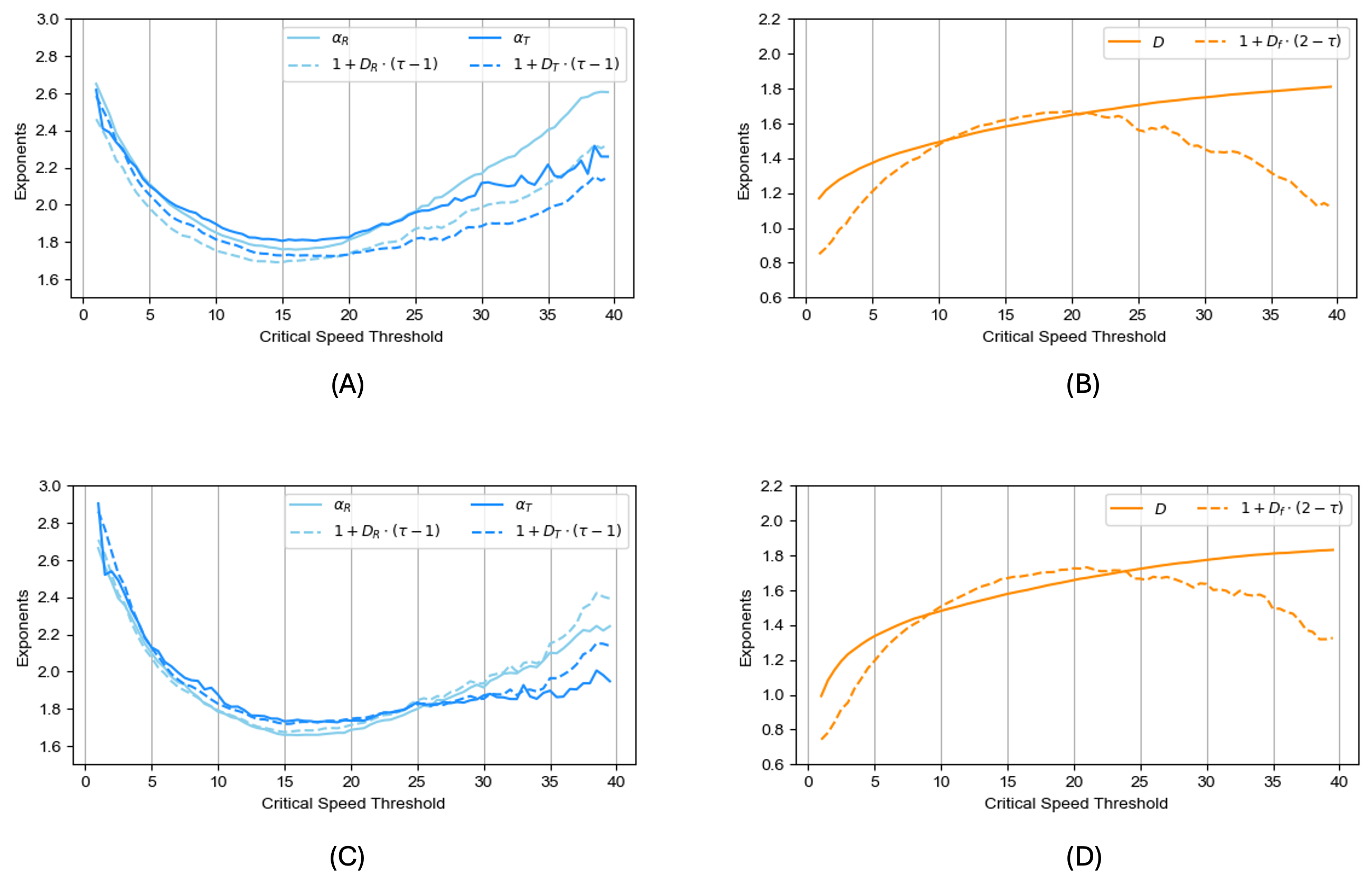}
    \caption{\textbf{Direct comparison of hyper-scaling relationships. }  $\alpha_R$, $\alpha_T$, $D$ compared with theoretical predictions using Eqs.~\ref{eq:alphaR}, ~\ref{eq:alphaT}, and \ref{eq:D}. \textbf{(A-B)} All clusters included. \textbf{(C-D)} Clusters with size $\ge10$. }
    \label{fig:hyper}
\end{figure}

As shown in Figures~\ref{fig:exponents_vs_vc}B and E, the average absolute error across the plateau remains below 10\%, supporting the consistency of the hyper-scaling relationships. Figure~\ref{fig:hyper} directly compares both sides of the hyper-scaling equations.

The top row includes all clusters, while the bottom row excludes clusters with sizes smaller than 10. We observe that the hyper-scaling relationships of Eqs.~\ref{eq:alphaR} and~\ref{eq:alphaT} show improved agreement in the plateau region when small clusters are excluded. In contrast, the hyper-scaling relationship for the delay fractal dimension (Eq.~\ref{eq:D}) exhibits higher errors within the plateau range but holds over a broader threshold range. It seems the delay fractal dimension hyper-scaling is more sensitive to deviations from criticality. 

\subsubsection*{Sensitivity of the Box-counting Method of Fractal Dimension Estimation}

In this section, we examine how the estimated fractal dimensions depend on the choices made in the box-counting method: the range of box sizes used and the aspect ratio of the boxes. These factors influence the accuracy and stability of both the delay fractal dimension $D$ and the cutoff cluster dimension $D_f$.

\begin{figure}[htbp!]
    \centering
    \includegraphics[width=\textwidth]{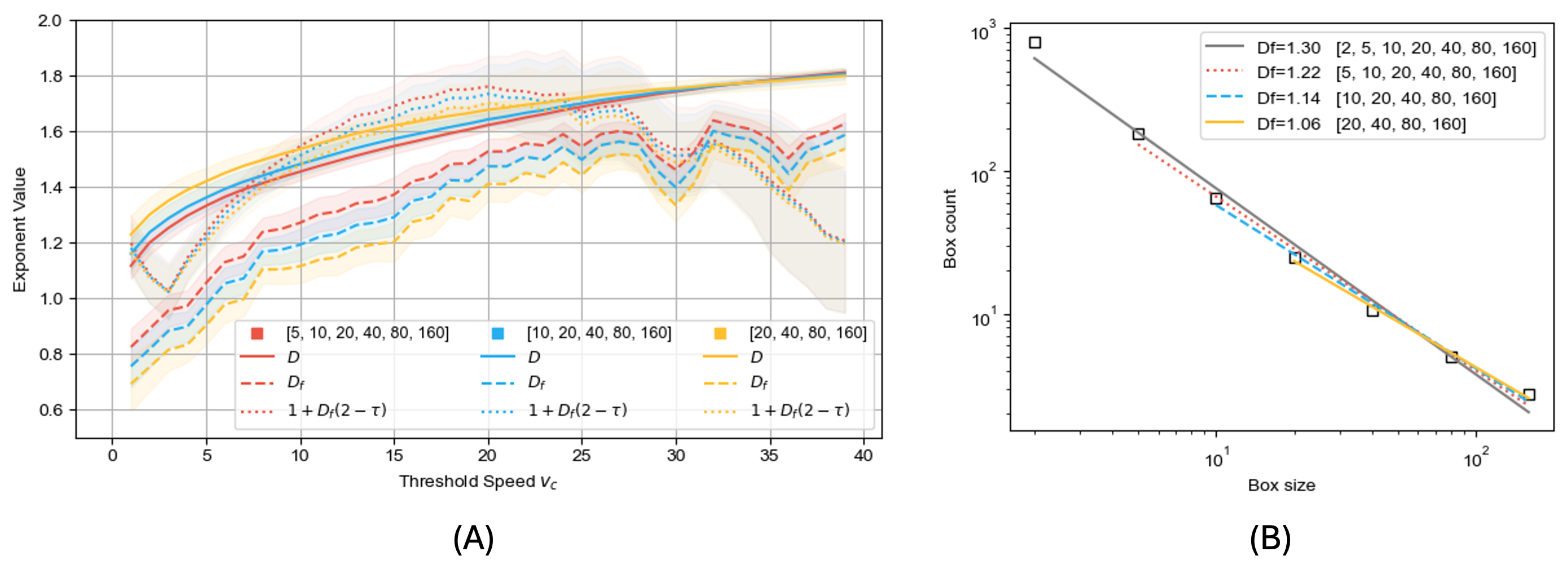}
    \caption{\textbf{Sensitivity of estimated fractal dimensions to the choice of box sizes.}
    \textbf{(A)} Estimates of $D$, $D_f$, and the hyper-scaling relationship of Eq.~\ref{eq:D} using different subsets of box sizes at fixed aspect ratio 1. While $D$ remains relatively stable, $D_f$ is highly sensitive to the box size range used in regression, highlighting its structural variability at different scales. \textbf{(B)} Example of box-count regression of $D_f$ from a single day (Nov. 22, lane 2 at $v_c = 10$ mph)  }
    \label{fig:boxsize_sensitivity}
\end{figure}

Figure~\ref{fig:boxsize_sensitivity}A shows the average (for all days and lanes) values of the delay fractal dimension $D$, the cutoff cluster dimension $D_f$, and the corresponding hyper-scaling estimate $1 + D_f(2 - \tau)$ across different box size subsets: \{5, 10, 20, 40, 80, 160\}, \{10, 20, 40, 80, 160\}, and \{20, 40, 80, 160\}. While $D$ is relatively stable, $D_f$ shows significant variation. Including smaller box sizes yields higher $D_f$ estimates, likely due to data noise, which can artificially increase box counts at small scales. In contrast, starting from larger boxes may smooth over these fine details of fractal structure and result in underestimation. Figure~\ref{fig:boxsize_sensitivity}B illustrates one realization of the cutoff fractal dimension $D_f$ in Nov. 22 lane 2 at $v_c = 10$ mph, where lower estimates are observed when small box sizes are excluded. Based on the selection of box sizes, the error in the hyper-scaling relation also varies. These results suggest that $D_f$ is more sensitive than $D$ to the scale of observation, and imply the importance of exploring a representative range of box sizes for robust estimation.

\begin{figure}[H]
    \centering
    \includegraphics[width=\linewidth]{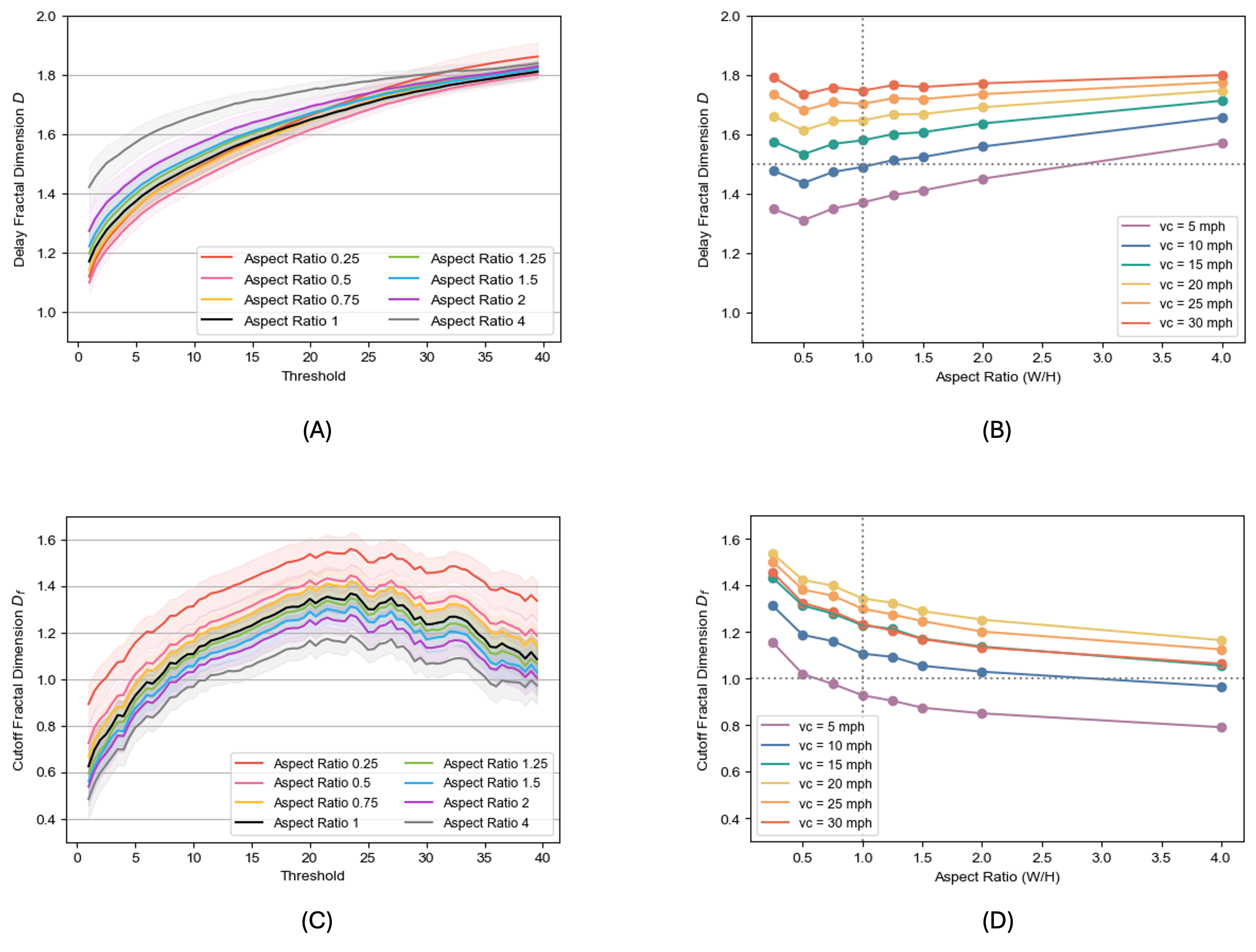}
\caption{\textbf{Sensitivity of estimated fractal dimensions to box aspect ratio.}
    \textbf{(A)} Estimated delay fractal dimension $D$ across critical speed thresholds for various aspect ratios (W/H). $D$ remains relatively stable for ratios between 0.5 and 1.5, though smaller aspect ratios produce slightly lower estimates.
    \textbf{(B)} Cross-sections of $D$ at selected thresholds ($v_c$), plotted against aspect ratio, confirming mild sensitivity within a moderate range.
    \textbf{(C)} Estimated cutoff fractal dimension $D_f$ across thresholds for various aspect ratios. $D_f$ shows stronger dependence, with lower aspect ratios yielding higher estimates.
    \textbf{(D)} Cross-sections of $D_f$ at selected $v_c$, plotted against aspect ratio. These show clear systematic trends, reinforcing the need to align box geometry with data resolution when estimating fractal dimensions.}
    \label{fig:aspectratio_sensitivity}
\end{figure}

Next, we revisit the impact of box aspect ratio on fractal dimension estimation. Figures~\ref{fig:aspectratio_sensitivity}A and C illustrate how the estimated delay fractal dimension $D$ and cutoff fractal dimension $D_f$ vary across different aspect ratios. The currently adopted ratio of 1 is motivated by the known anisotropic propagation of traffic congestion: with a space-time resolution of $\Delta x = 0.02$ miles and $\Delta t = 6$ seconds, square-shaped boxes (aspect ratio 1) align with the visual structure of traffic clusters in the discretized time-space diagram.

We find that $D$ is relatively insensitive to aspect ratio changes within the range of 0.5 to 1.5, whereas $D_f$ exhibits greater sensitivity. In particular, lower aspect ratios result in systematically lower estimates of $D$ and higher estimates of $D_f$. This suggests that under-sampling in time (\textit{i.e.}, tall narrow boxes) may compress delay structure, while over-sampling in space may fragment spatial clusters, inflating box counts.

Figures~\ref{fig:aspectratio_sensitivity}B and D provide cross-sectional views at selected threshold values, plotting $D$ and $D_f$ against aspect ratio. While $D$ curves tend to flatten out, $D_f$ declines steadily with increasing aspect ratio, underscoring its vulnerability to box geometry. These trends reinforce the notion that the optimal aspect ratio depends on the underlying resolution of the dataset and the anisotropic dynamics of the system.

Together, these findings emphasize that both the aspect ratio and the box size range must be tuned to the space-time resolution and structural characteristics of the system under study. While we propose intuitively justified settings for our dataset, broader applications may require further sensitivity analyses and theoretical support to ensure resolution-aware and scale-consistent fractal estimation.

 \subsubsection*{Critical exponents by lanes} \label{ax:all_lanes}

\begin{figure}[htbp!]
    \centering
    \includegraphics[width=\textwidth]{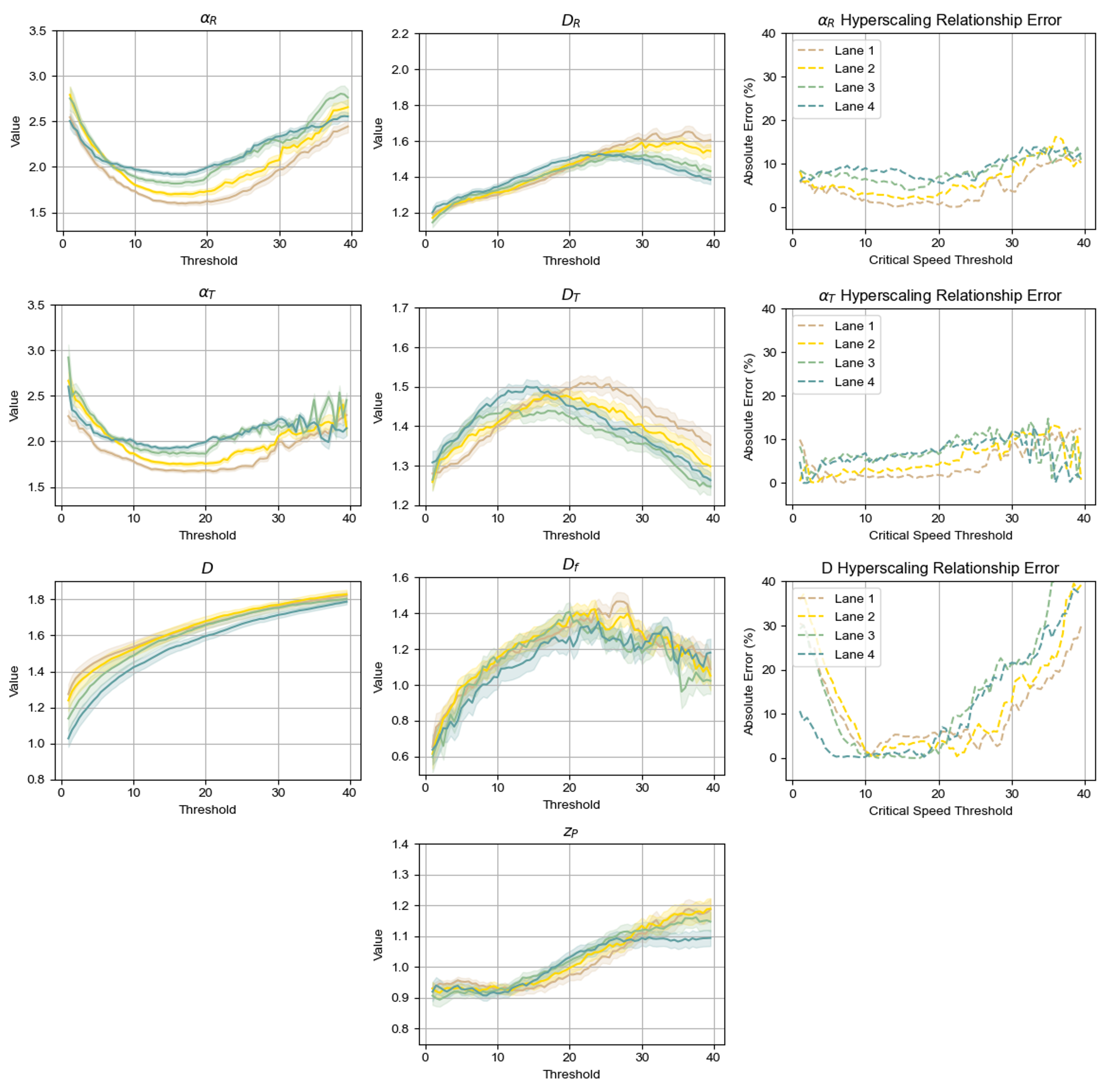}
    \caption{Lane-by-lane comparison of estimated exponents and hyperscaling errors using clusters with size $\geq 2$}
    \label{fig:perlane_filter2}
\end{figure}

\begin{figure}[htbp!]
    \centering
    \includegraphics[width=\textwidth]{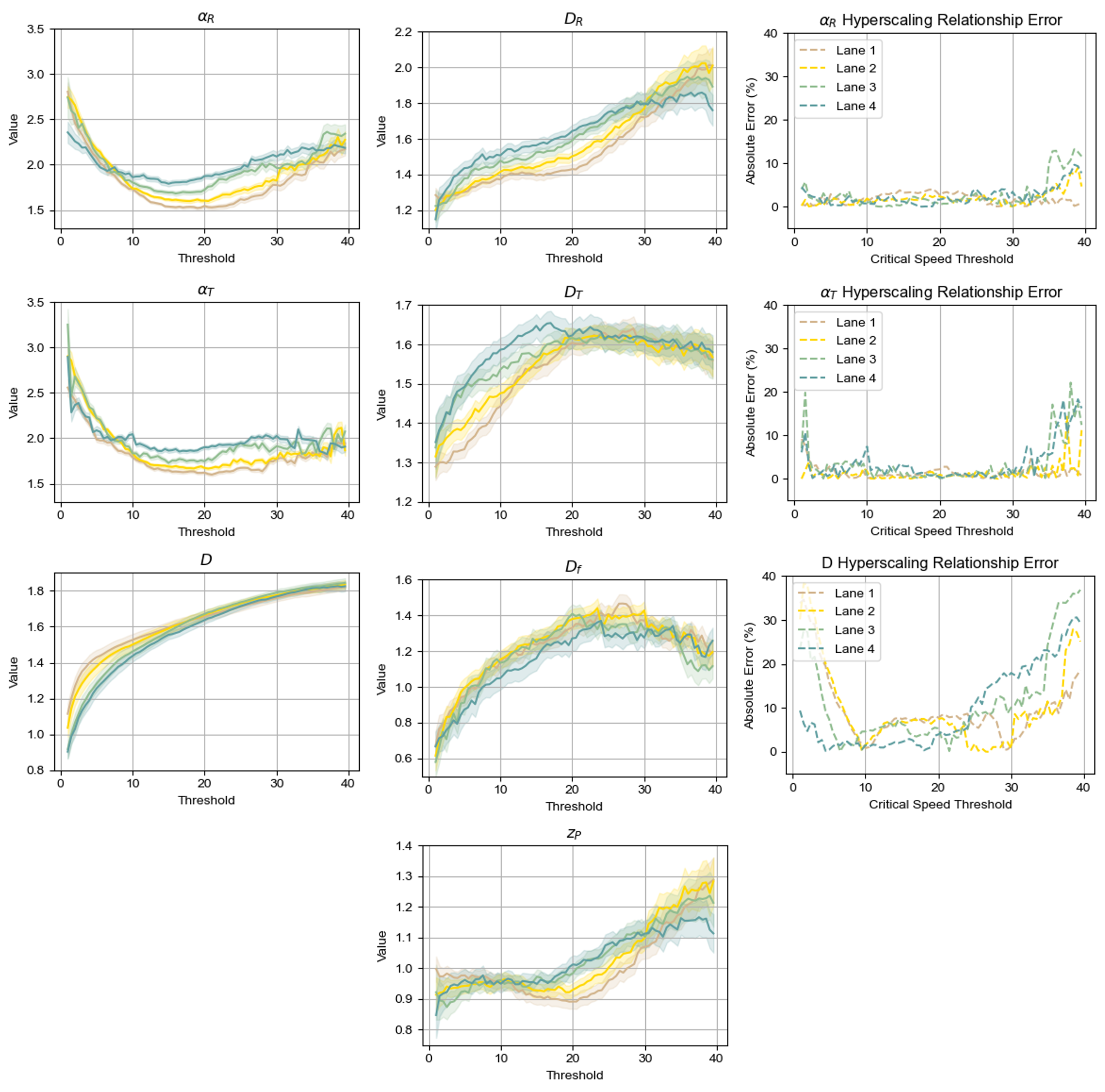}
    \caption{Lane-by-lane comparison of estimated exponents and hyperscaling errors using clusters with size $\geq 10$}
    \label{fig:perlane_filter10}
\end{figure}

Figures~\ref{fig:perlane_filter2} and \ref{fig:perlane_filter10} compare the estimated exponents across lanes, for cluster sizes $\geq 2$ and $\geq 10$, respectively. Similar to the Fisher exponent $\tau$, both $\alpha_R$ and $\alpha_T$ show a tendency for higher values in the outer lanes, with a more stable trend emerging once small clusters are excluded. While $D_R$ and $D_T$ do not display any clear lane-dependent pattern, the delay fractal dimension $D$ and cutoff fractal dimension $D_f$ show an opposite trend—higher values are observed in the inner lanes.

The hyperscaling relationships for $\alpha_R$ and $\alpha_T$ exhibit lower absolute error in the inner lanes. But when small clusters are excluded, where errors become uniformly low across all lanes. The delay fractal dimension hyperscaling shows a distinct shift; \textit{i.e.}, although the error curves have similar shapes, the inner lanes exhibit minimum error at higher critical thresholds than the outer lanes. This indicates that the scaling regime in inner lanes is delayed, possibly reflecting differing traffic dynamics or ramp influences.


\subsubsection*{Growth Exponent of Eastbound Data}
In the eastbound direction, where traffic remains in the free-flow regime throughout the observation window, one might expect a clearer observation of the early-time KPZ growth behavior, similar to Figure~\ref{fig:beta_width}. However, the same issues with missing data are present. As in the westbound case, incomplete trajectory coverage during the early morning hours suppresses the initial growth of the interface width $W(t)$, followed by an artificial surge once full data coverage resumes. This results in a temporary overestimation of the growth exponent and obscures the underlying dynamics in the eastbound direction as well. On days with fewer data artifacts, the expected scaling with $\beta \approx 1/3$ is still observed, but accurate estimation of the KPZ exponent requires uninterrupted, high-resolution measurements from the beginning of the observation period. When only free-flow is observed in the time–space diagram, it becomes difficult to infer whether the flow conditions are evolving. Therefore, the second row of the figure presents the evolution of average flow over time, confirming that the flow gradually increases, even under free-flow conditions.

\begin{figure}
    \centering
    \includegraphics[width=\linewidth]{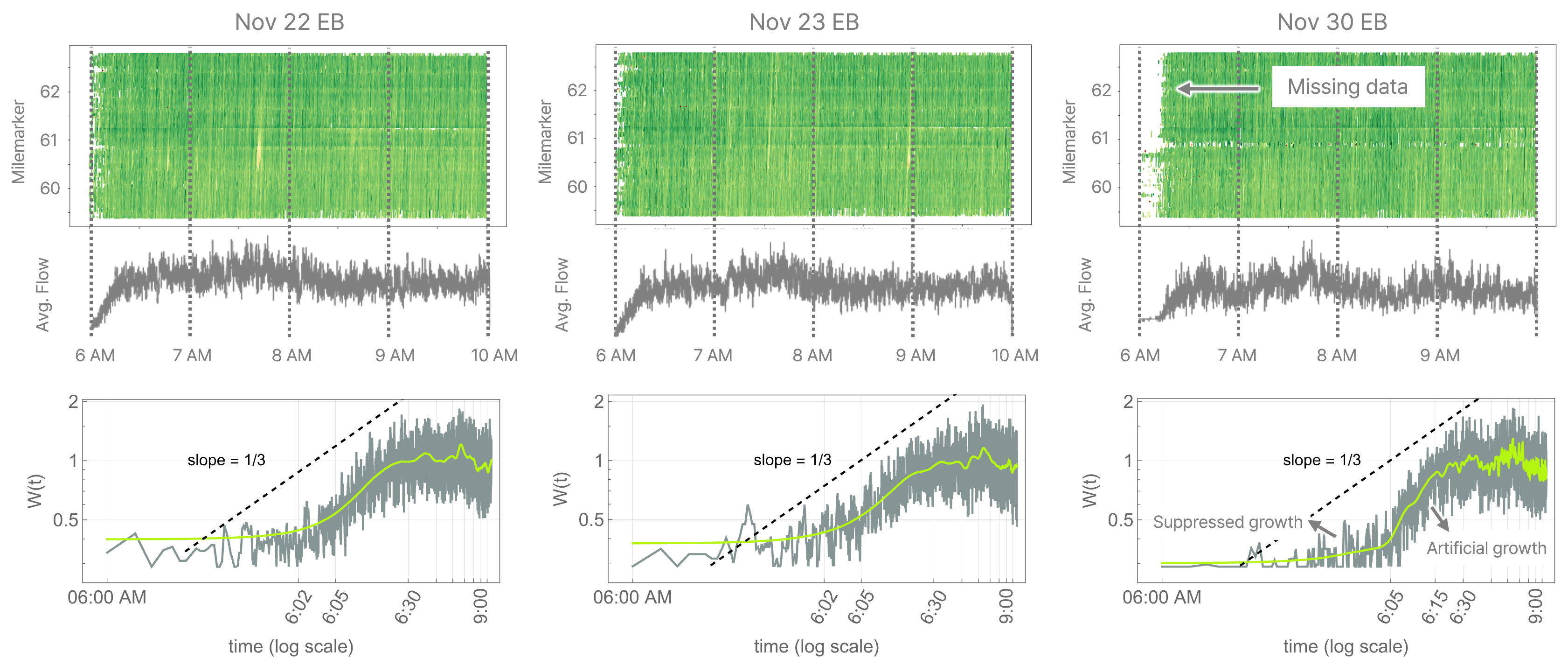}
    \caption{\textbf{Time evolution of the interface width $W(t)$ for three representative days of eastbound data, averaged across all four lanes at one-second resolution.} The results are consistent with those in Figure~\ref{fig:beta_width}. The accompanying flow evolution in the second row reveals a gradual increase in average flow during the early period, despite the time-space diagrams indicating persistent free-flow conditions throughout.}
    \label{fig:beta_width_eb}
\end{figure}

\subsubsection*{Dynamic correlation scaling}


Unlike the previous analysis in Figure~\ref{fig:corrlength0}, which averaged correlation functions across time windows, here we focus on the dynamic behavior of individual time slices within selected periods. This allows us to compare the empirical data directly with microscopic simulations from TASEP under similar finite-size constraints of $L=1000$, which shows an excellent agreement.

\begin{figure}[htbp!]
    \centering
       
\includegraphics[width=\textwidth]{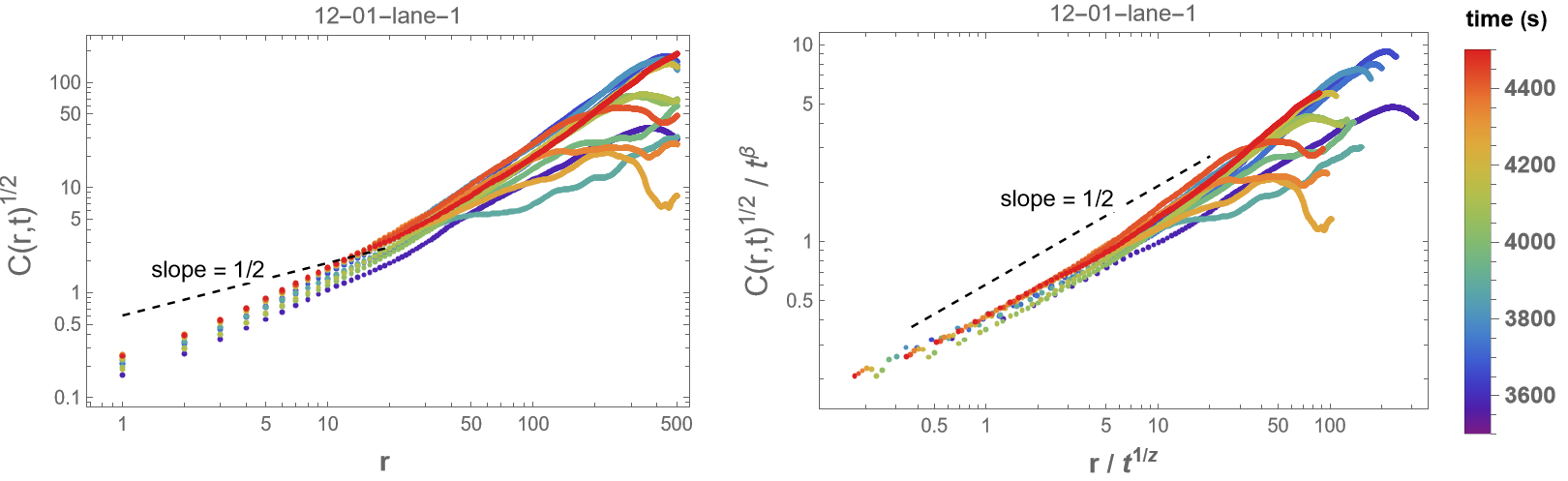}
\includegraphics[width=\textwidth]{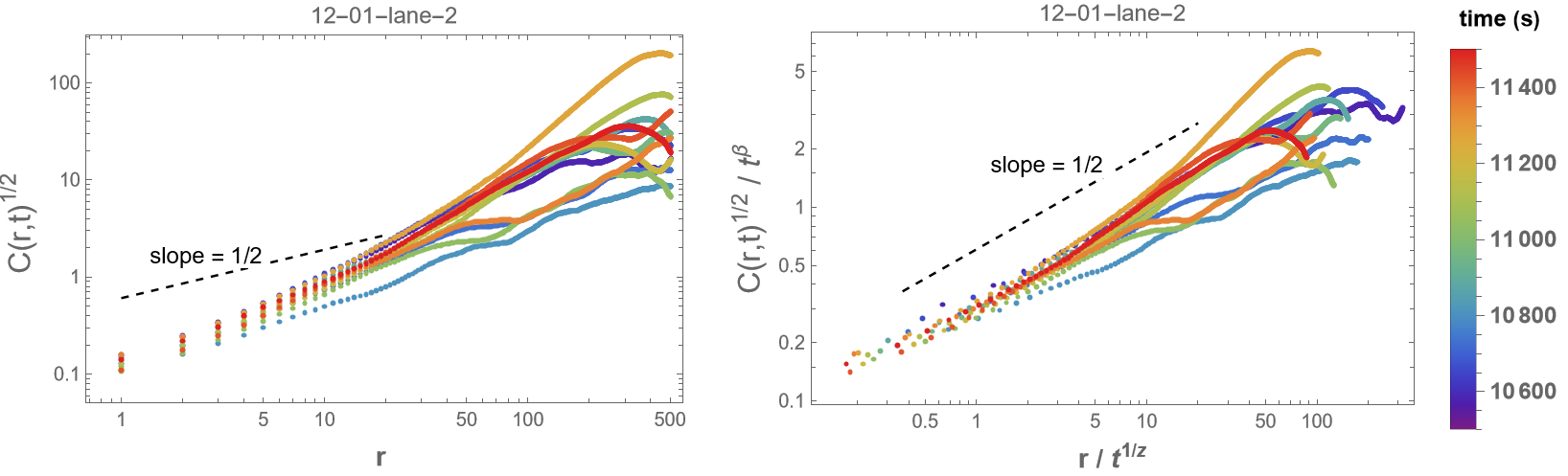}
\caption{\textbf{Height-height correlation and dynamic scaling collapse on data.} Data from two different time periods on December 1 lane 1 (one per row), shown both in original coordinates $ \sqrt{C(r,t)} $ as a function of $ r $ (left) and in rescaled coordinates$ \sqrt{C(r,t)}/t^{\beta} $ as a function of $ r/t^{1/z} $ (right) for several time points $ t $ and $ 1 \le r \le 500 = L/2 $. The collapsed curves follow the expected scaling behavior $ F(u) \sim u^{\alpha} $ at small $ u $, with a slope of $ \alpha = 1/2 $. Deviations at larger $u$, where a plateau is expected, likely reflect jam propagation effects.}

    \label{fig:corrlength}
\end{figure}

As shown in figure~\ref{fig:corrlength}, the rescaled data collapse plot exhibits the expected power-law trend $F(u) \sim u^{\alpha}$ for small scaled distances $r/t^{1/z}$. In this initial regime, the collapsed data points for $\sqrt C(r,t)/t^{\beta}$ closely follow the overlaid reference line of slope 1/2, indicating an effective $\alpha \approx 1/2$, which is reassuring. 
To get an idea of the rescaling effect, the left column of the figure shows the same data but in original coordinates, $\sqrt C(r,t)$ as a function of $r$. It can be seen that the effect on the data collapse appears to be mild compared to the original coordinates, and more effective for low values of $u$. However, notice that in original coordinates one expects $\sqrt C(r,t)\sim r^\alpha$, but this is not evident from the figure, which implies the usefulness and pertinence of the rescaling.

As $r/t^{1/z}$ increases towards the peak and plateau region in figure~\ref{fig:corrlength}, a noticeable spread among the curves for different times becomes apparent. We have verified that the spread can be explained by the propagation of jam clusters, which induce large dips in the correlation function when the separation distance $r$ is comparable to the jam wavelength of about 100-200 car lengths, or $r/t^{1/z}\approx 50-100$ in the figure. 

\begin{figure}[htbp!]
    \centering
       
\includegraphics[width=\textwidth]{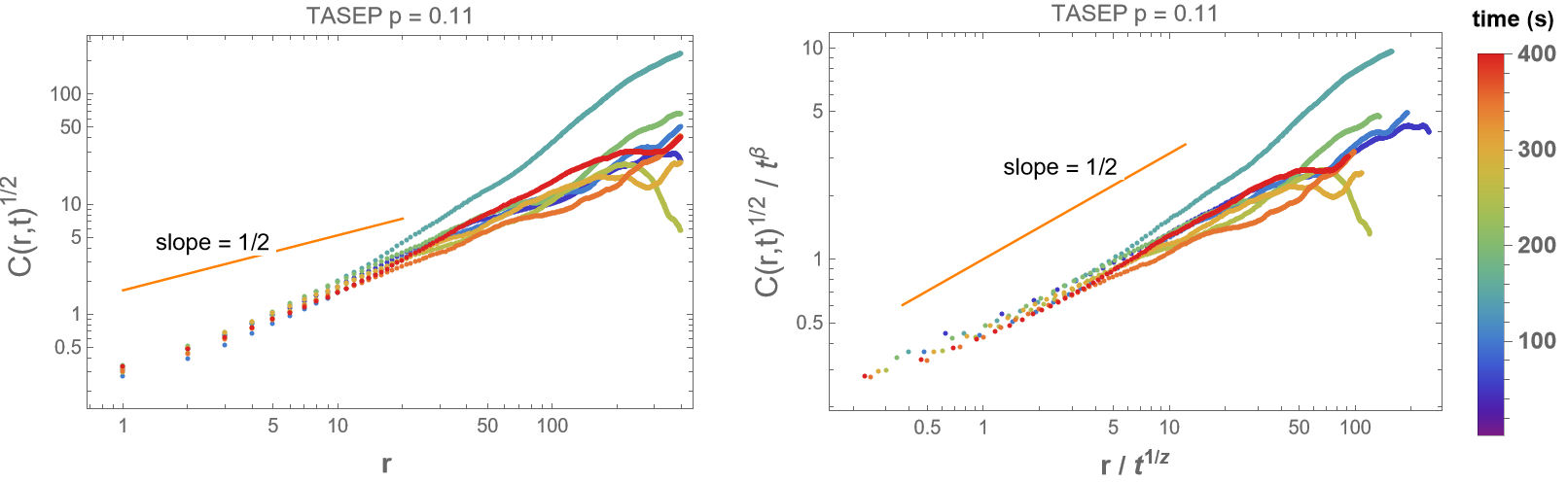}
\includegraphics[width=\textwidth]{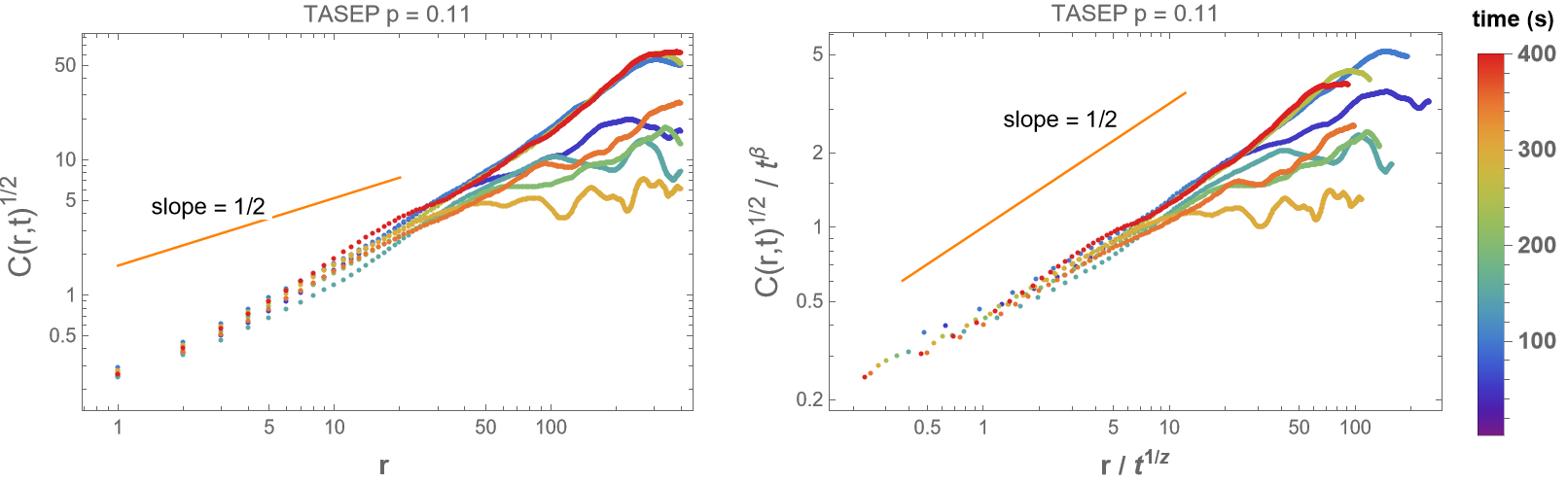}
       
\caption{\textbf{Height-height correlation and dynamic scaling collapse in TASEP simulations with random walk initial conditions and probability of 0.11.} Results from two simulations (one per row), showing the height-height correlation $ \sqrt{C(r,t)} $ in original coordinates as a function of $ r $ (left), and in rescaled coordinates $ \sqrt{C(r,t)}/t^{\beta} $ versus $ r/t^{1/z} $ (right), across multiple time points $ t $ and for $ 1 \le r \le 500 = L/2 $.}
    \label{fig:corrlengthTASEP}
\end{figure}

It turns out that strikingly similar results are obtained by simulating TASEP with random walk initial conditions; see figure~\ref{fig:corrlengthTASEP} for two realizations with a stubborn probability of $p=0.11$. That this figure matches qualitatively the results from the empirical data in figure~\ref{fig:corrlength} indicates that the scatter observed in $F(u)$ for large $u$ may be attributed to small system size, $L=1000$ vehicle car lengths, for both the 4.2 mi-segment and the finite-size simulation. 
Therefore, we are likely observing the finite-size version of the KPZ scaling function $F(u)$, where the deviations from $F(u)$  are not just random noise but systematic finite-size effects that may be predicted by the theory.

\subsubsection*{Sensitivity of Hurst Exponent Estimations}

\begin{figure}[htbp!]
    \centering
         \begin{subfigure}[b]{0.245
         \textwidth}
         \includegraphics[width=\textwidth]{Figs/Hurst/RS_r339Lane1_6Days_.pdf}
         \caption{Lane 1} \label{fig:l1_H}
     \end{subfigure} 
     \begin{subfigure}[b]{0.245\textwidth}
         \includegraphics[width=\textwidth]{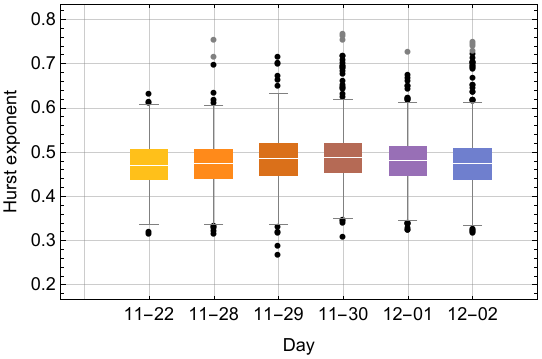}
         \caption{Lane 2} 
         \label{fig:l2_H}
     \end{subfigure} 
        \begin{subfigure}[b]{0.245\textwidth}
         \includegraphics[width=\textwidth]{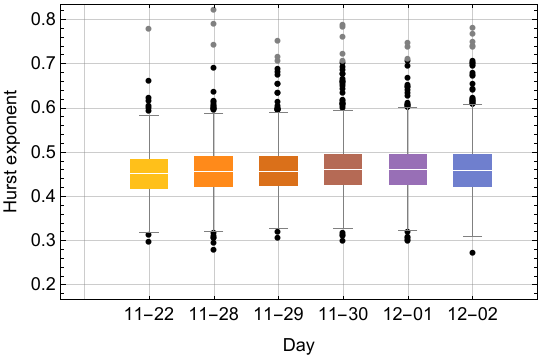}
         \caption{Lane 3} \label{fig:l3_H}
     \end{subfigure} 
     \begin{subfigure}[b]{0.245\textwidth}
         \includegraphics[width=\textwidth]{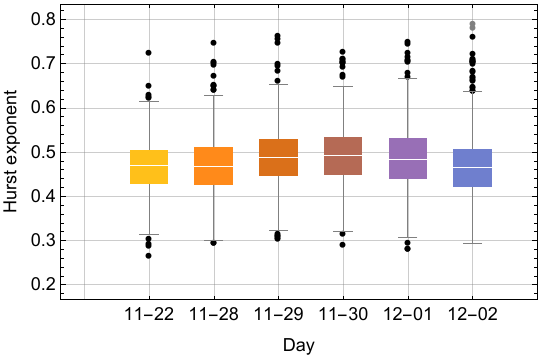}
         \caption{Lane 4} 
         \label{fig:l4_H}
     \end{subfigure} 

\caption{\textbf{Lane-by-lane comparison of Hurst exponents using the R/S method.} 
Box plots of Hurst exponent estimates computed from spatial cumulative profiles. }

    \label{fig:H_allrs}
\end{figure}

While the R/S method remains a widely used tool for estimating the Hurst exponent $H$, we report here its dependency on the choice of the window size parameter $r$ needed to eliminate deterministic trends in the data. As stated in Methods, we use a Gaussian filter with $r=L/3$, ensuring that the smoothing window spans the full scale of analysis. This choice balances the need to suppress low-frequency trends while preserving the stochastic structure relevant for Hurst exponent estimation. However, the selection of $r$ remains heuristic, and varying this parameter leads to slightly different estimates in our case. Figure~\ref{fig:H_allrs} reveals that the R/S estimates with $r=L/3$ exhibit minor but noticeable variations across different lanes. 

\begin{figure}[htbp!]
    \centering
         \begin{subfigure}[b]{0.45
         \textwidth}
         \includegraphics[width=\textwidth]{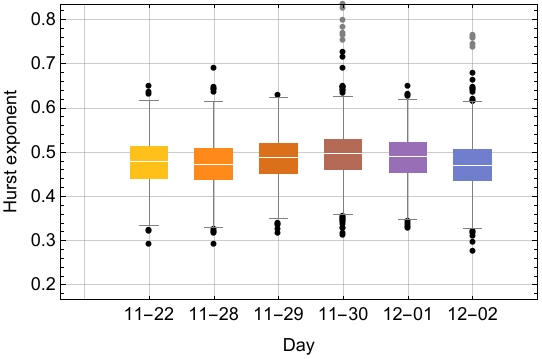}
         \caption{Window size $r=200$} \label{fig:l1_H}
     \end{subfigure} 
     \begin{subfigure}[b]{0.45\textwidth}
         \includegraphics[width=\textwidth]{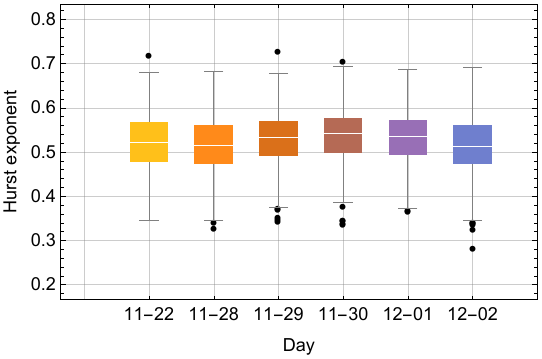}
         \caption{Window size $r=L=1000$} 
         \label{fig:l2_H}
     \end{subfigure} 

\caption{\textbf{Effects of the window size in the R/S method for lane 1.} 
Box plots of Hurst exponent estimates computed from spatial cumulative profiles. }

    \label{fig:H_window}
\end{figure}

Figure~\ref{fig:H_window} compares the R/S estimates with $r=L/5$ and $r=L$ for lane 1, where it can be seen that a lower window size corresponds to lower estimates and more outliers. Although variations are noticeable, it is reassuring to note that the evidence in favor of $H=1/2$ remains strong.

\noindent\textit{Detrended fluctuation analysis}

We have also considered detrended fluctuation analysis (DFA), which involves dividing the data into non-overlapping segments of size $ s $, fitting a polynomial of order $ m $ to each segment, and computing fluctuations after subtracting this local trend. Although the DFA is often preferred for its robustness to certain types of nonstationarity, in our experiments (not shown here), we found that the method was even more sensitive to the choice of the polynomial order $ m $ than R/S was to the smoothing parameter $ r $. Inappropriate choices of $ m $ can lead to over- or underfitting of trends, thereby distorting the estimated scaling behavior.

\noindent\textit{Maximum likelihood estimation}

\begin{figure}[htbp!]
    \centering
         \begin{subfigure}[b]{0.245
         \textwidth}
         \includegraphics[width=\textwidth]{Figs/Lane1_1AllDays.png}
         \caption{Lane 1} \label{fig:l1_H}
     \end{subfigure} 
     \begin{subfigure}[b]{0.245\textwidth}
         \includegraphics[width=\textwidth]{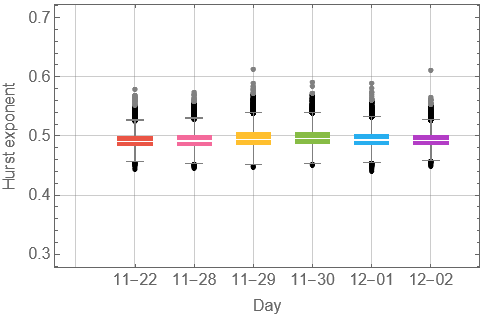}
         \caption{Lane 2} 
         \label{fig:l2_H}
     \end{subfigure} 
        \begin{subfigure}[b]{0.245\textwidth}
         \includegraphics[width=\textwidth]{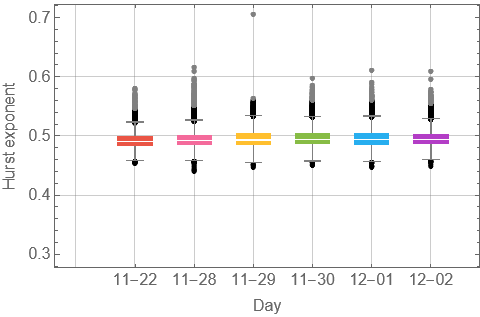}
         \caption{Lane 3} \label{fig:l3_H}
     \end{subfigure} 
     \begin{subfigure}[b]{0.245\textwidth}
         \includegraphics[width=\textwidth]{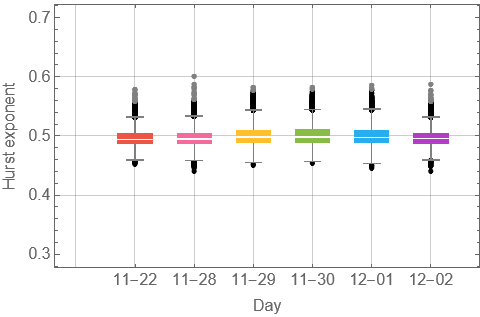}
         \caption{Lane 4} 
         \label{fig:l4_H}
     \end{subfigure} 

\caption{\textbf{Lane-by-lane distributions of Hurst exponents estimated via fBM-based MLE.} 
Box plots show the Hurst exponent estimates derived from fitting spatial cumulative traffic profiles to fractional Brownian motion using maximum likelihood estimation. 
Across all lanes, the distributions are narrowly centered around $H = 0.5$, indicating improved consistency compared to the R/S method.}
    \label{fig:H_all_fbm}
\end{figure}

Figure~\ref{fig:H_all_fbm} displays the distributions of Hurst exponent estimates for each lane across six days using the MLE approach based on fractional Brownian motion. Compared to the R/S method, the MLE estimates are consistently more robust across lanes and exhibit tighter concentration around the KPZ-theoretical value of $H = 1/2$. This reinforces the reliability of the MLE-based method, though it comes with the caveat of assuming an underlying fBM structure.

\subsubsection*{Data Artifacts} \label{ax:artifacts}

\begin{figure}[htbp!]
\centering
     \begin{subfigure}[b]{0.4\textwidth}
         \includegraphics[width=\textwidth]{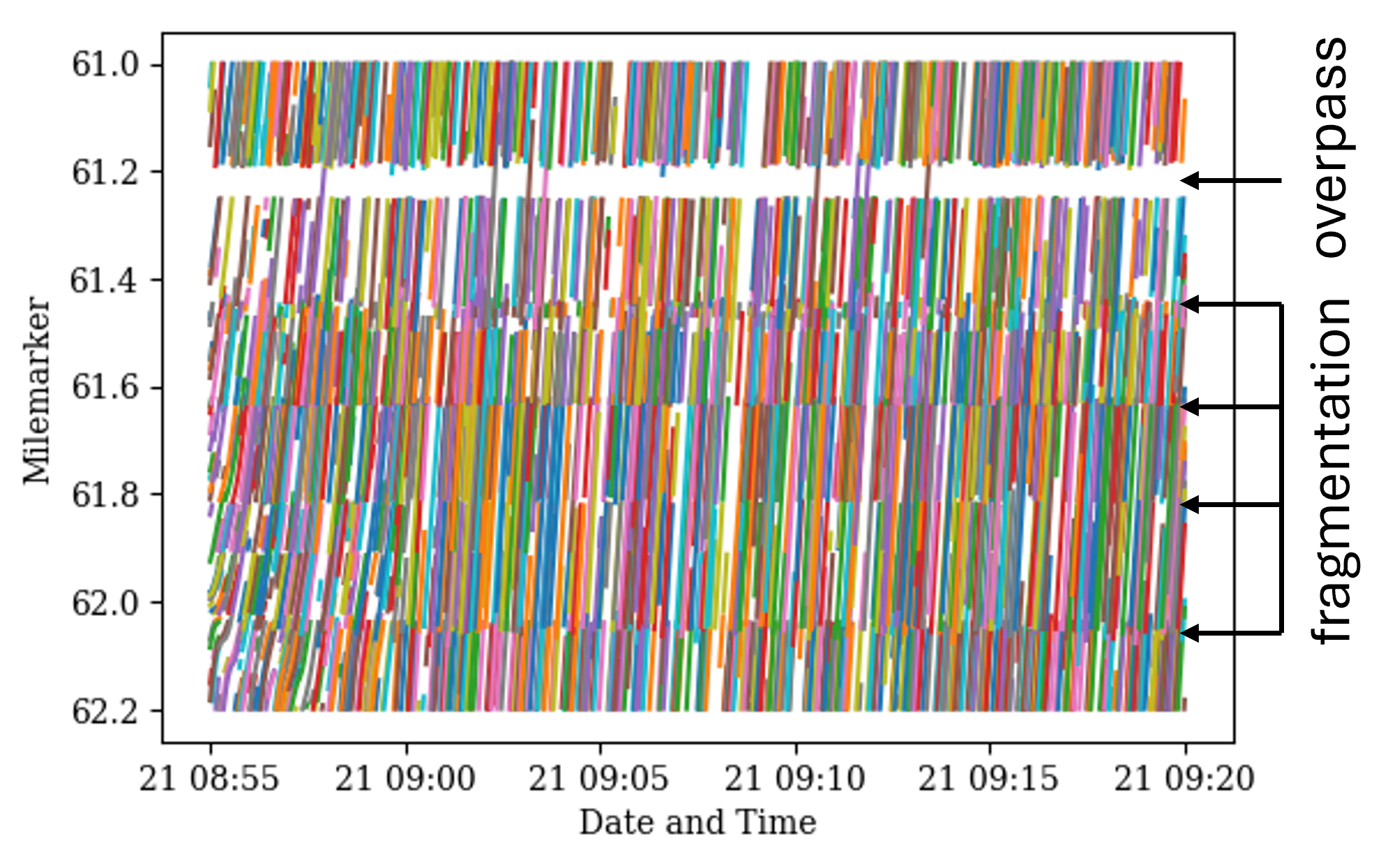}
         \caption{} \label{fig:artif1}
     \end{subfigure} \quad \quad \quad
          \begin{subfigure}[b]{0.4\textwidth}
         \includegraphics[width=\textwidth]{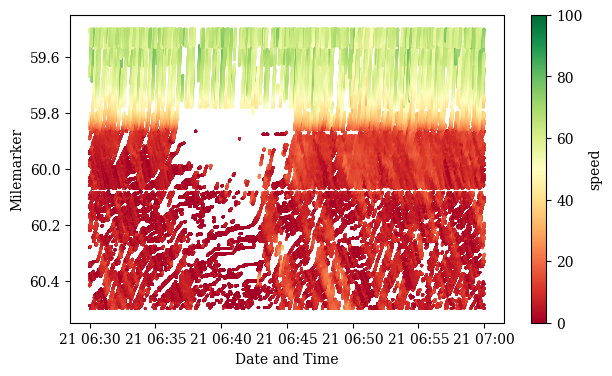}
         \caption{} \label{fig:artif2}
     \end{subfigure} \hfill
    \caption{\textbf{Visualizations of data artifacts}: Known artifacts in raw I-24 MOTION dataset \textbf{(A)} fragmentation and overpass, and \textbf{(B)} missing patches}
\end{figure}

The artifacts in the raw data, specifically tracking errors causing vehicle fragmentation and apparent missing parts, are likely to affect any utilization of this data. In figure \ref{fig:artif1}, each vehicle ID is colored distinctly to show how vehicle IDs are fragmented, although it has a possibility to be the same vehicle. Note that these fragmentation locations do not align with the on-off ramps, rather apparent in the overlapping areas of cameras. Additionally, large missing parts in certain spatiotemporal packets are non-trivially observed: see figure \ref{fig:artif2}. While analyses using this dataset should be conducted with caution due to these known artifacts, the dataset is unrivaled in its breadth and depth by capturing essential traffic jam properties such as post-accident discharge phenomena and consistent shock waves. This may make it a potential game-changer in the field of vehicle trajectory data. 

\newpage

\subsubsection*{Supplementary Figures } \label{fss_all}
\newcounter{tmpfig}
\setcounter{tmpfig}{\value{figure}}

\begin{figure}[H]
\centering
\begin{subfigure}[b]{0.24\linewidth}
    \includegraphics[width=\linewidth]{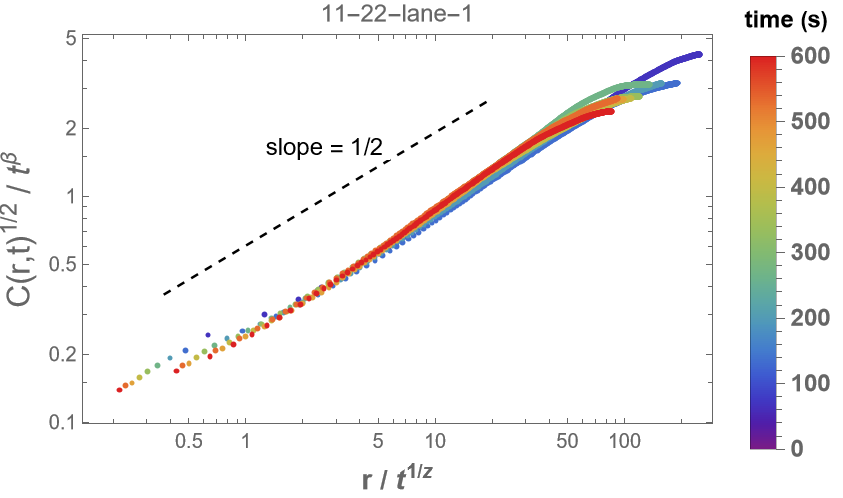}
    \caption{Lane 1 (11-22)}
\end{subfigure}
\begin{subfigure}[b]{0.24\linewidth}
    \includegraphics[width=\linewidth]{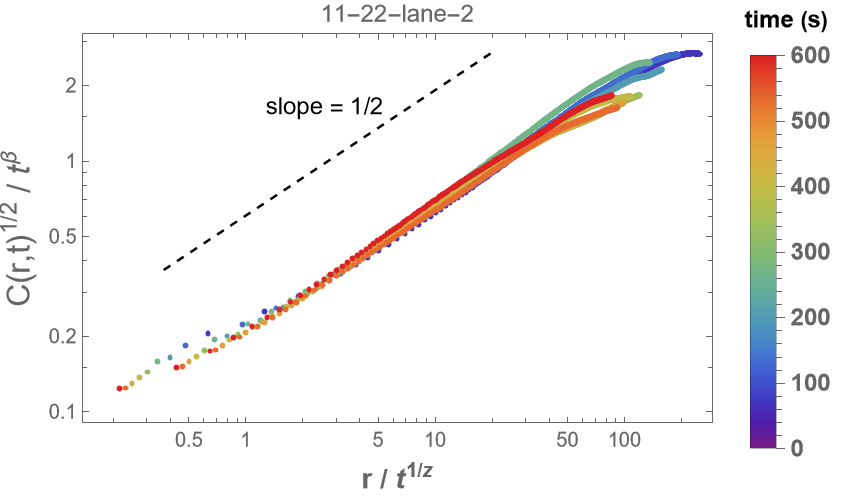}
    \caption{Lane 2 (11-22)}
\end{subfigure}
\begin{subfigure}[b]{0.24\linewidth}
    \includegraphics[width=\linewidth]{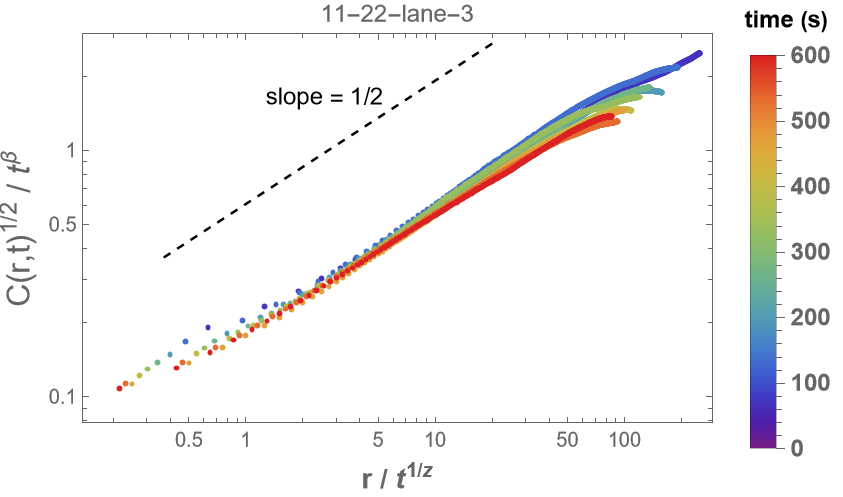}
    \caption{Lane 3 (11-22)}
\end{subfigure}
\begin{subfigure}[b]{0.24\linewidth}
    \includegraphics[width=\linewidth]{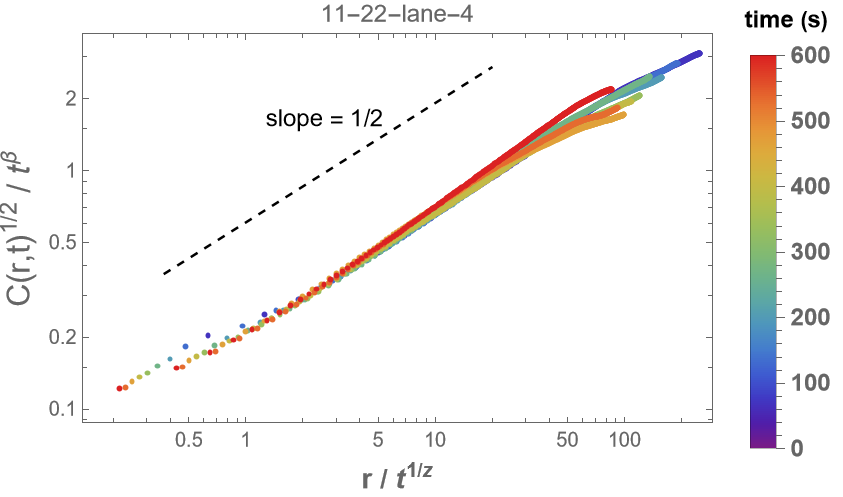}
    \caption{Lane 4 (11-22)}
\end{subfigure}
\\
\begin{subfigure}[b]{0.24\linewidth}
    \includegraphics[width=\linewidth]{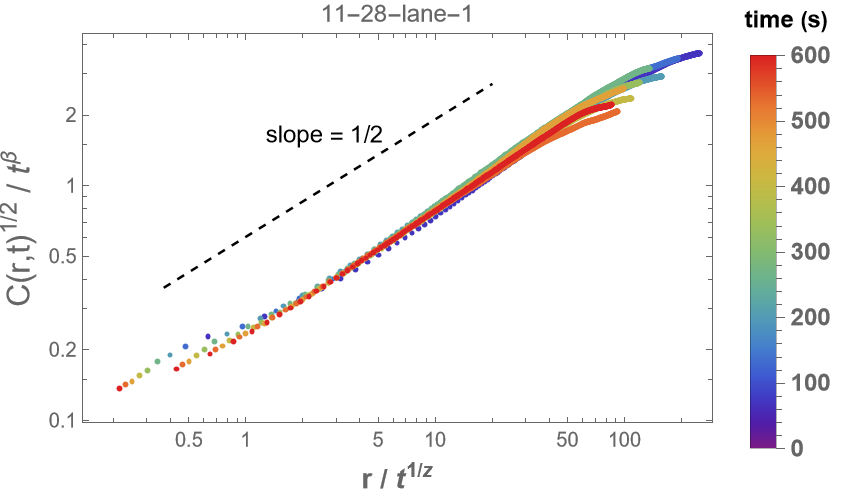}
    \caption{Lane 1 (11-28)}
\end{subfigure}
\begin{subfigure}[b]{0.24\linewidth}
    \includegraphics[width=\linewidth]{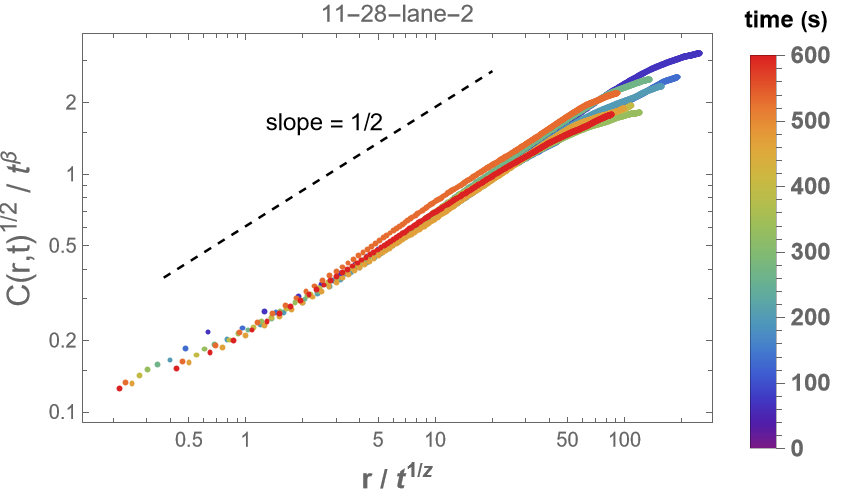}
    \caption{Lane 2 (11-28)}
\end{subfigure}
\begin{subfigure}[b]{0.24\linewidth}
    \includegraphics[width=\linewidth]{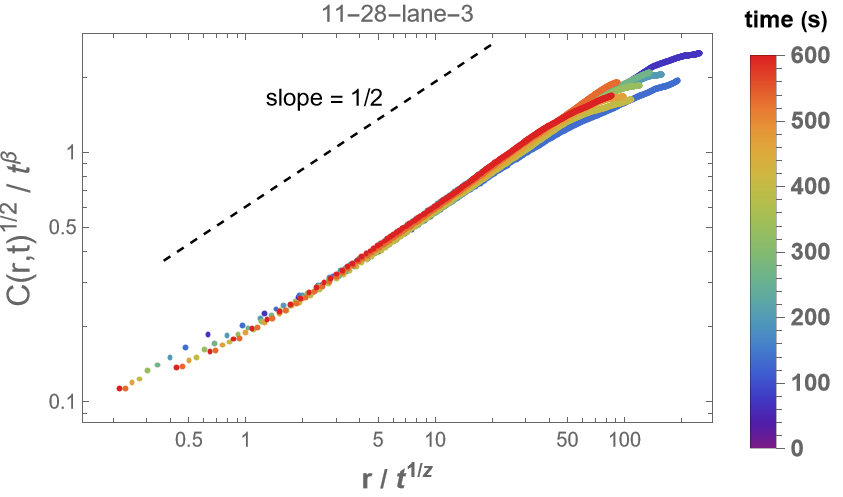}
    \caption{Lane 3 (11-28)}
\end{subfigure}
\begin{subfigure}[b]{0.24\linewidth}
    \includegraphics[width=\linewidth]{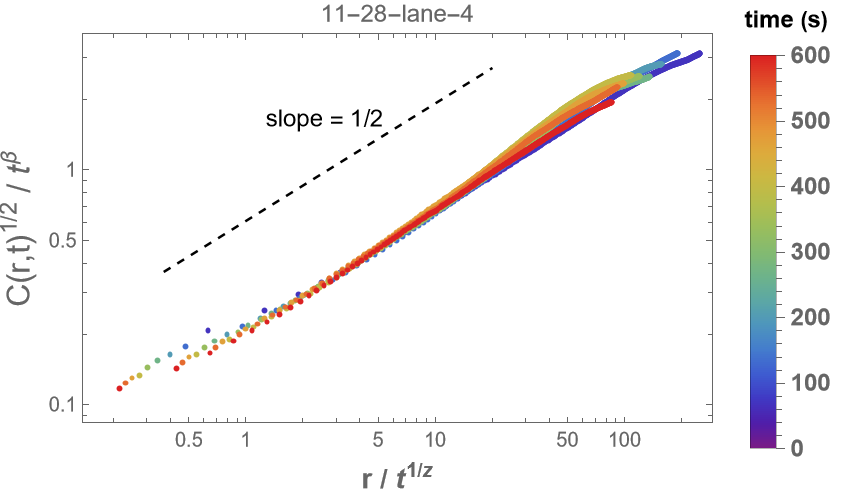}
    \caption{Lane 4 (11-28)}
\end{subfigure}
\\
\begin{subfigure}[b]{0.24\linewidth}
    \includegraphics[width=\linewidth]{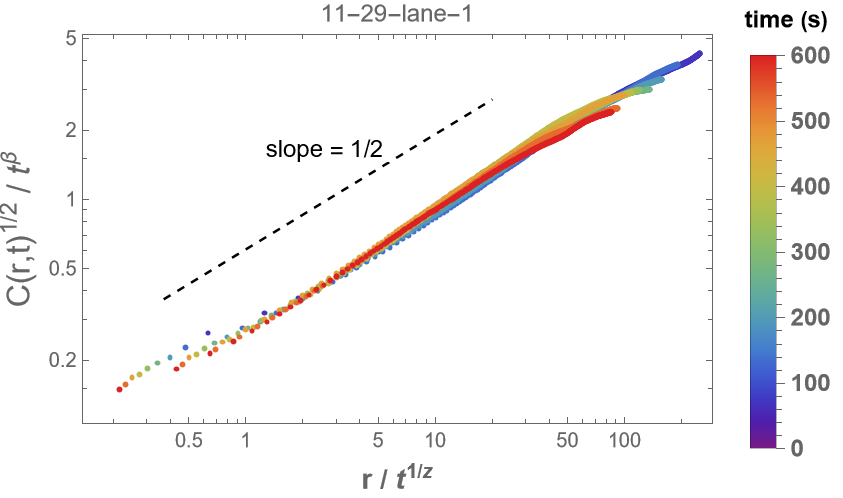}
    \caption{Lane 1 (11-29)}
\end{subfigure}
\begin{subfigure}[b]{0.24\linewidth}
    \includegraphics[width=\linewidth]{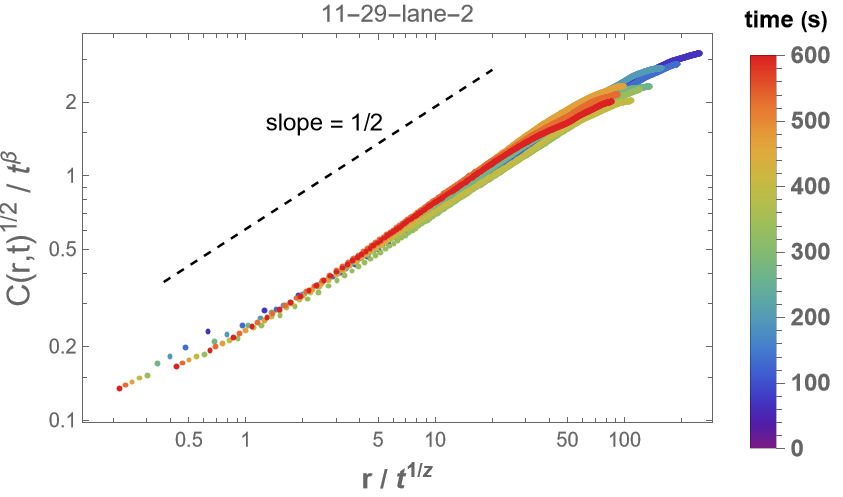}
    \caption{Lane 2 (11-29)}
\end{subfigure}
\begin{subfigure}[b]{0.24\linewidth}
    \includegraphics[width=\linewidth]{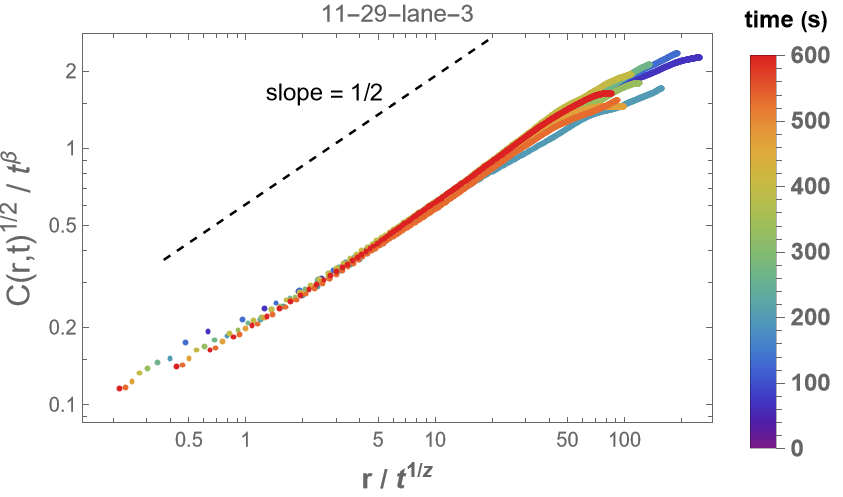}
    \caption{Lane 3 (11-29)}
\end{subfigure}
\begin{subfigure}[b]{0.24\linewidth}
    \includegraphics[width=\linewidth]{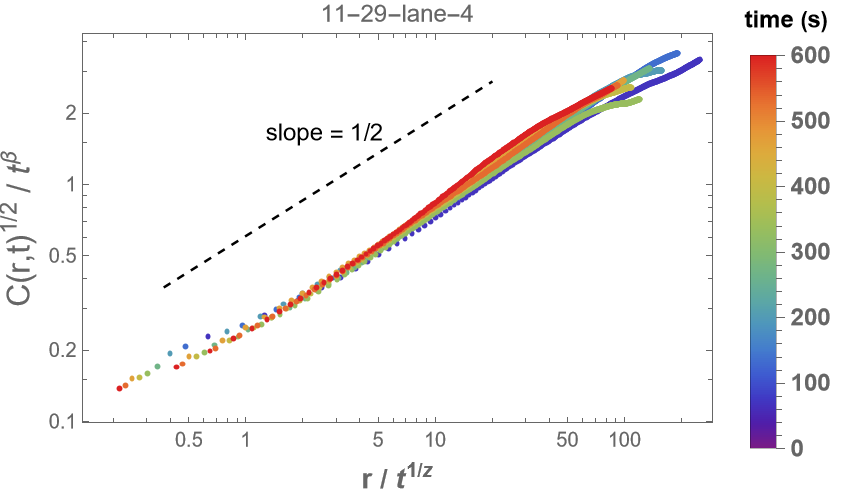}
    \caption{Lane 4 (11-29)}
\end{subfigure}
\\
\begin{subfigure}[b]{0.24\linewidth}
    \includegraphics[width=\linewidth]{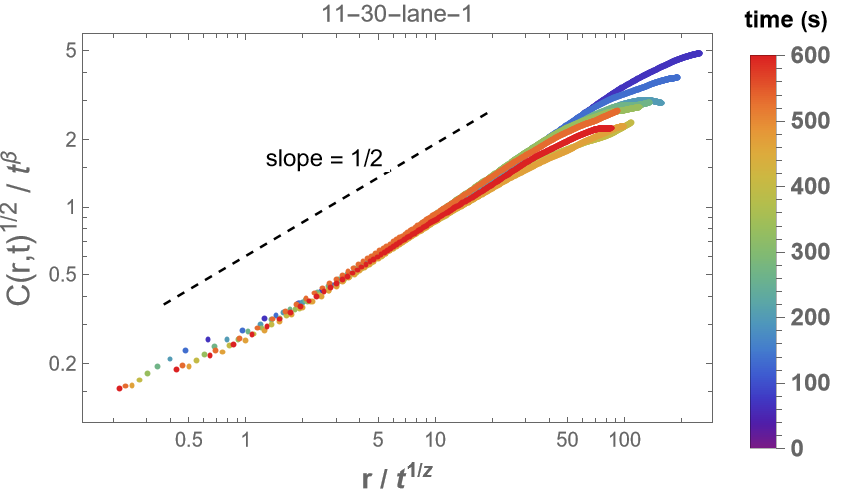}
    \caption{Lane 1 (11-30)}
\end{subfigure}
\begin{subfigure}[b]{0.24\linewidth}
    \includegraphics[width=\linewidth]{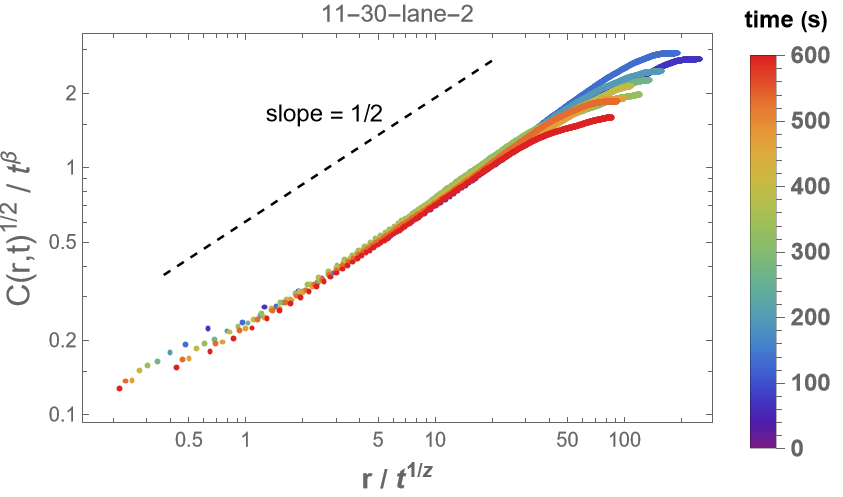}
    \caption{Lane 2 (11-30)}
\end{subfigure}
\begin{subfigure}[b]{0.24\linewidth}
    \includegraphics[width=\linewidth]{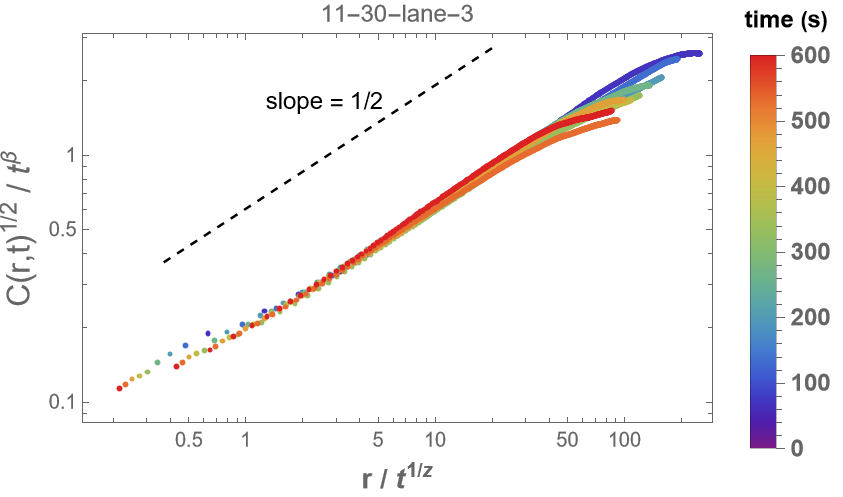}
    \caption{Lane 3 (11-30)}
\end{subfigure}
\begin{subfigure}[b]{0.24\linewidth}
    \includegraphics[width=\linewidth]{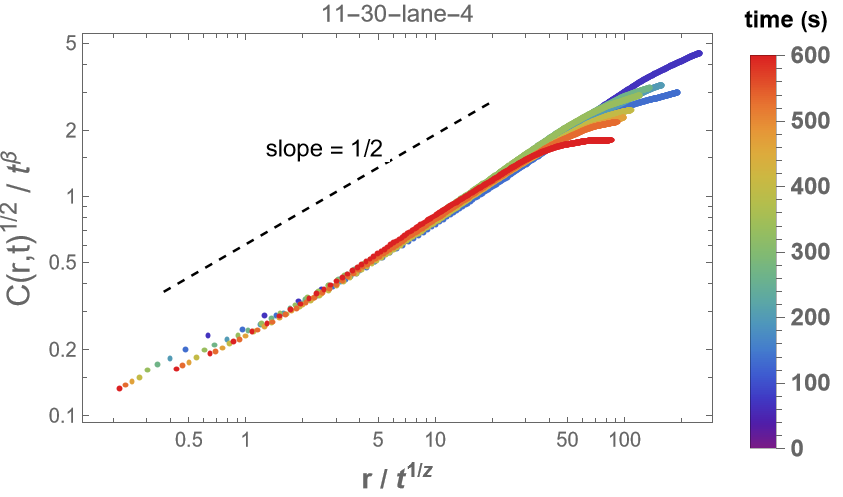}
    \caption{Lane 4 (11-30)}
\end{subfigure}
\\
\begin{subfigure}[b]{0.24\linewidth}
    \includegraphics[width=\linewidth]{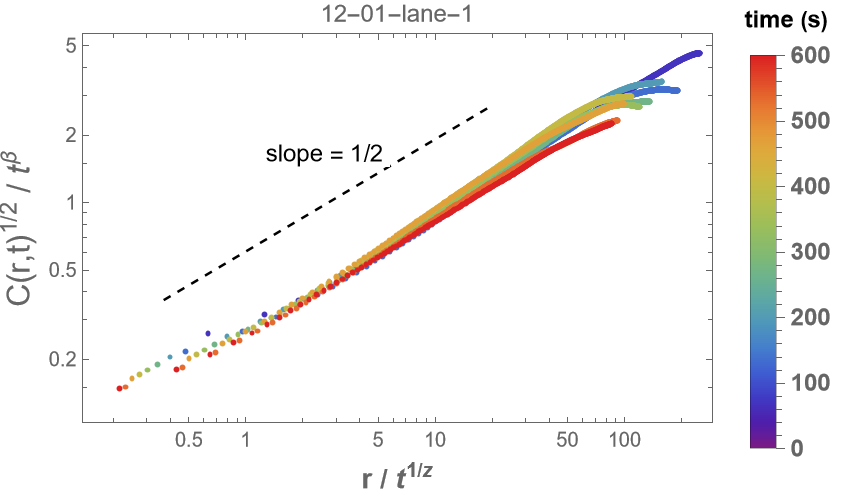}
    \caption{Lane 1 (12-01)}
\end{subfigure}
\begin{subfigure}[b]{0.24\linewidth}
    \includegraphics[width=\linewidth]{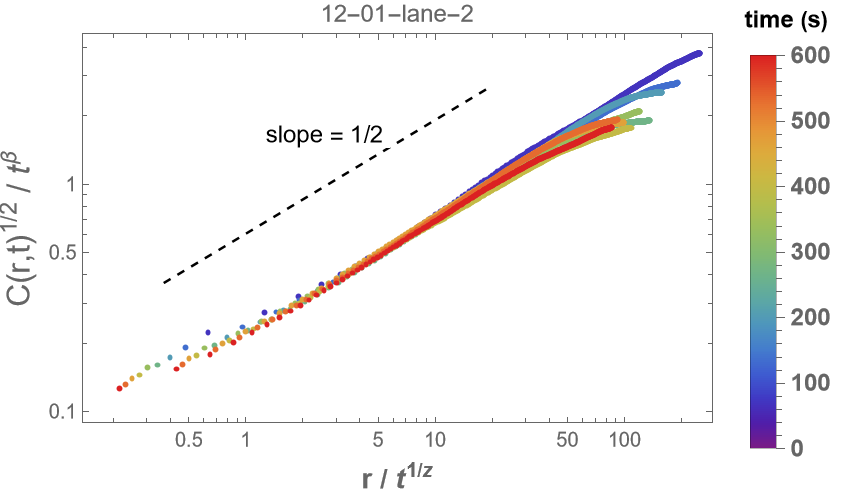}
    \caption{Lane 2 (12-01)}
\end{subfigure}
\begin{subfigure}[b]{0.24\linewidth}
    \includegraphics[width=\linewidth]{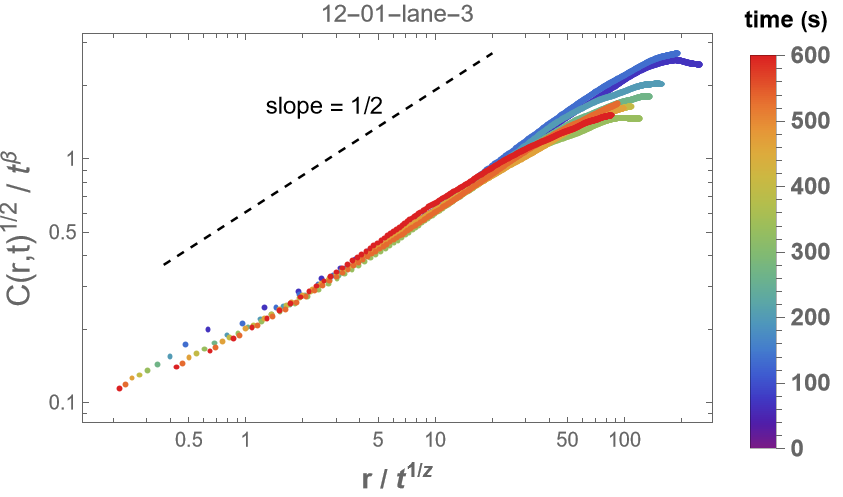}
    \caption{Lane 3 (12-01)}
\end{subfigure}
\begin{subfigure}[b]{0.24\linewidth}
    \includegraphics[width=\linewidth]{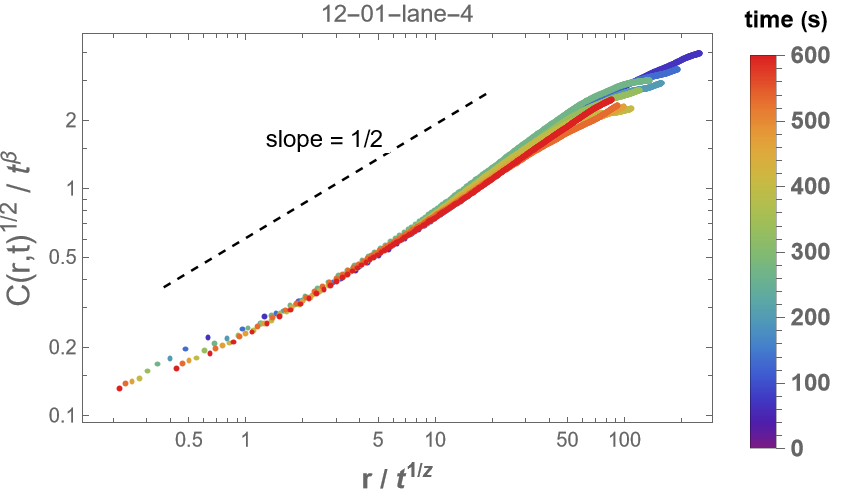}
    \caption{Lane 4 (12-01)}
\end{subfigure}
\\
\begin{subfigure}[b]{0.24\linewidth}
    \includegraphics[width=\linewidth]{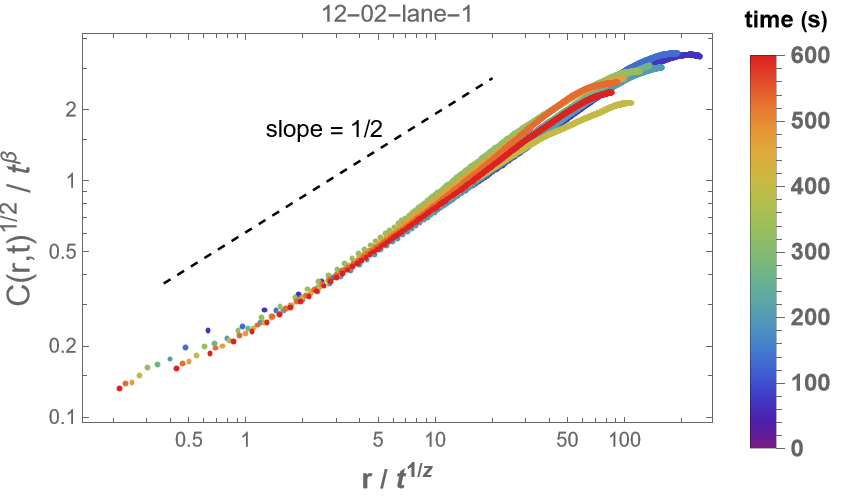}
    \caption{Lane 1 (12-02)}
\end{subfigure}
\begin{subfigure}[b]{0.24\linewidth}
    \includegraphics[width=\linewidth]{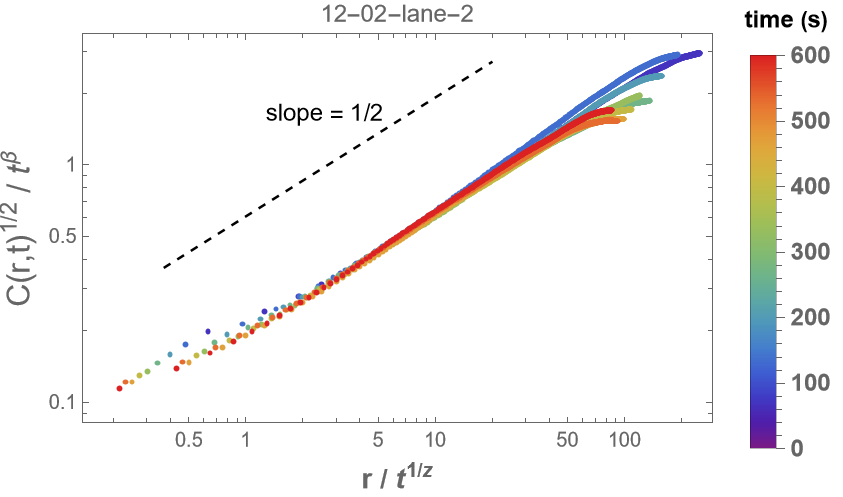}
    \caption{Lane 2 (12-02)}
\end{subfigure}
\begin{subfigure}[b]{0.24\linewidth}
    \includegraphics[width=\linewidth]{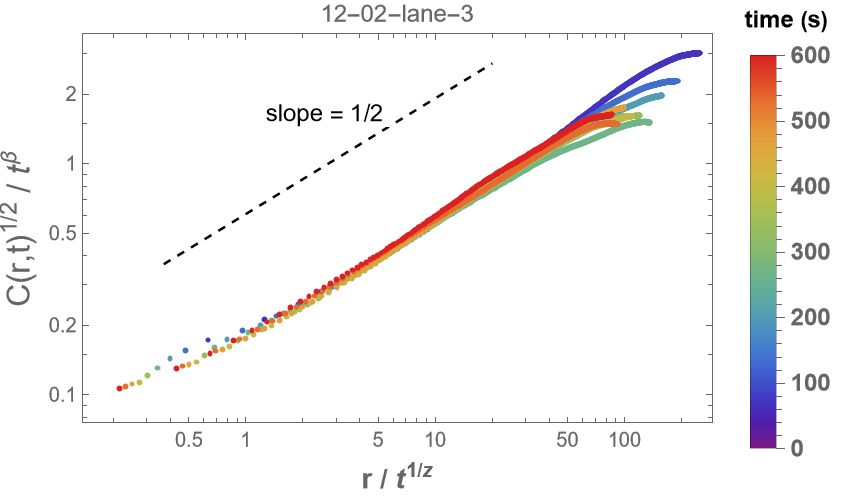}
    \caption{Lane 3 (12-02)}
\end{subfigure}
\begin{subfigure}[b]{0.24\linewidth}
    \includegraphics[width=\linewidth]{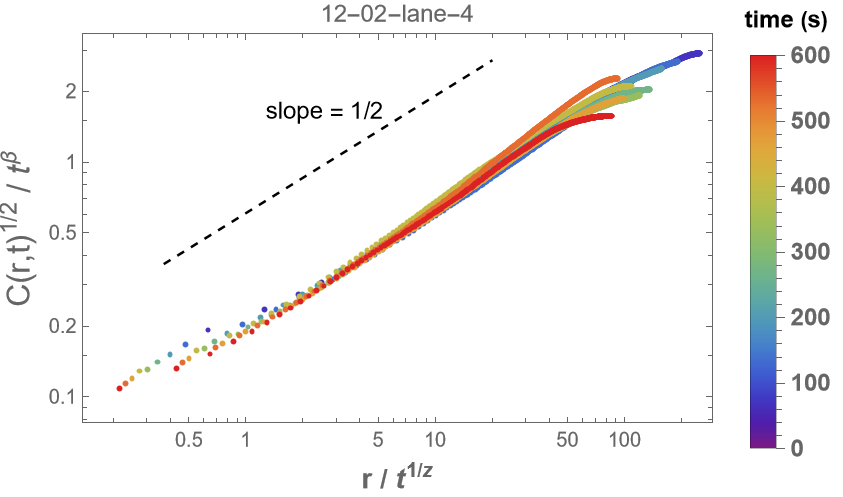}
    \caption{Lane 4 (12-02)}
\end{subfigure}
\\
\caption{\small\textbf{Rescaled height-height correlation functions for several times and $1\le r\le 400$ vehicle lengths for all days and lanes.} Remaining date–lane combinations not included in Figure~\ref{fig:corrlength0}.}

\label{fig:cxt_all}
\end{figure}

\begin{figure}[htbp!]
    \centering

    \begin{subfigure}[b]{0.77\linewidth}
        \includegraphics[width=\linewidth]{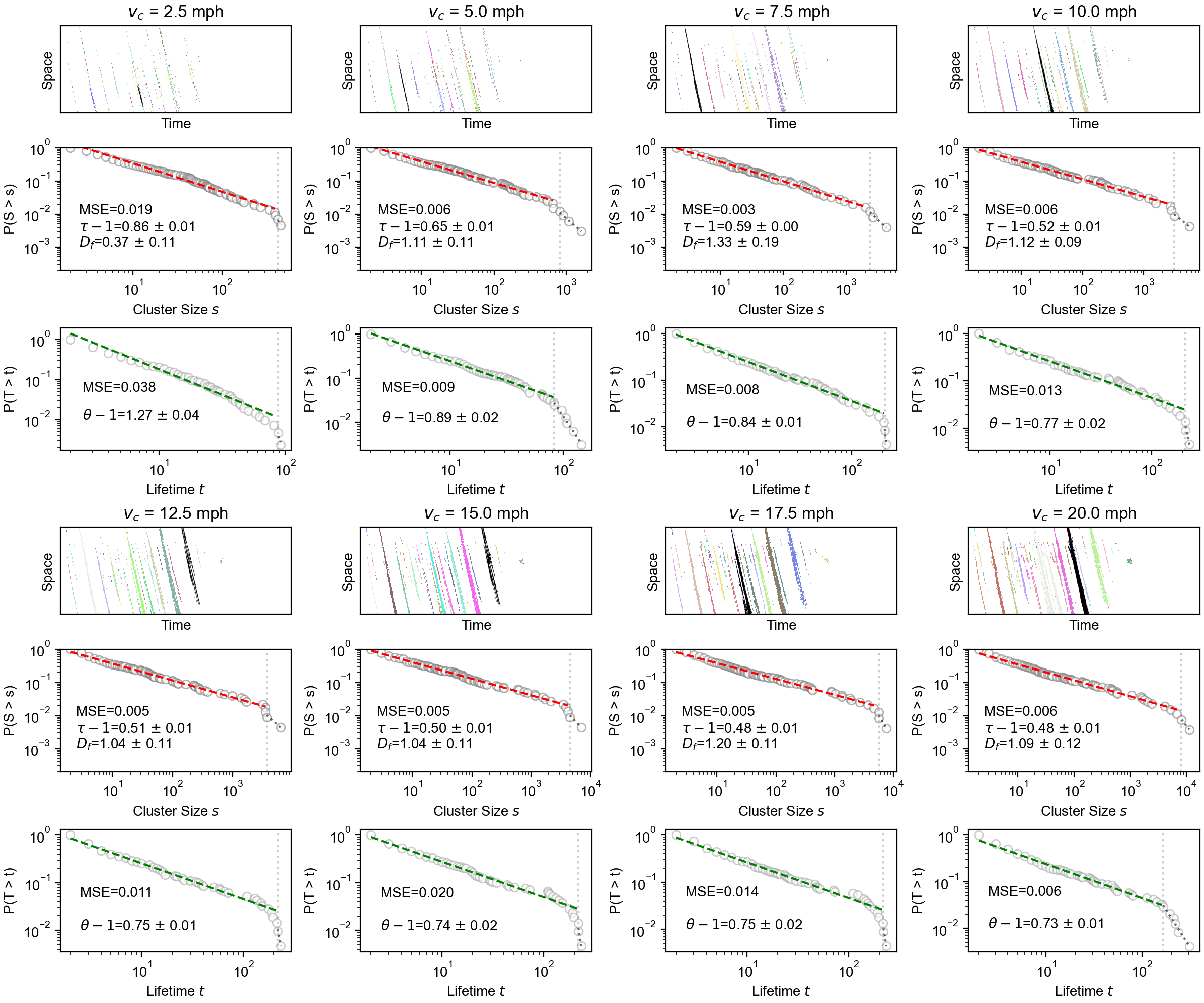}
        \caption{November 28}
        \label{fig:tau_1128}
    \end{subfigure}
    \hfill
    \begin{subfigure}[b]{0.77\linewidth}
        \includegraphics[width=\linewidth]{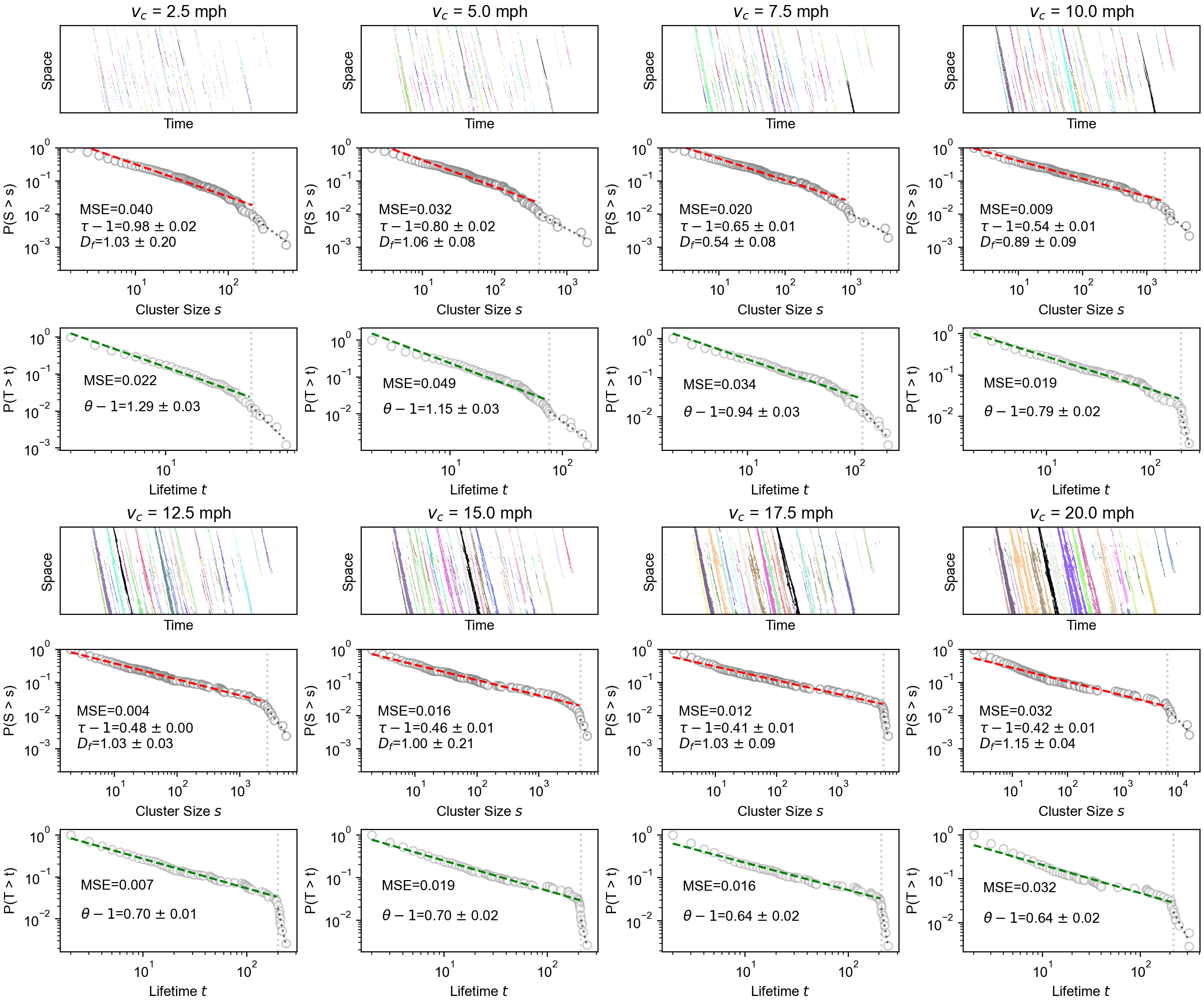}
        \caption{November 29}
        \label{fig:tau_1129}
    \end{subfigure}
\end{figure}

\begin{figure}[htbp!]
    \ContinuedFloat
    \centering
    
    \begin{subfigure}[b]{0.77\linewidth}
        \includegraphics[width=\linewidth]{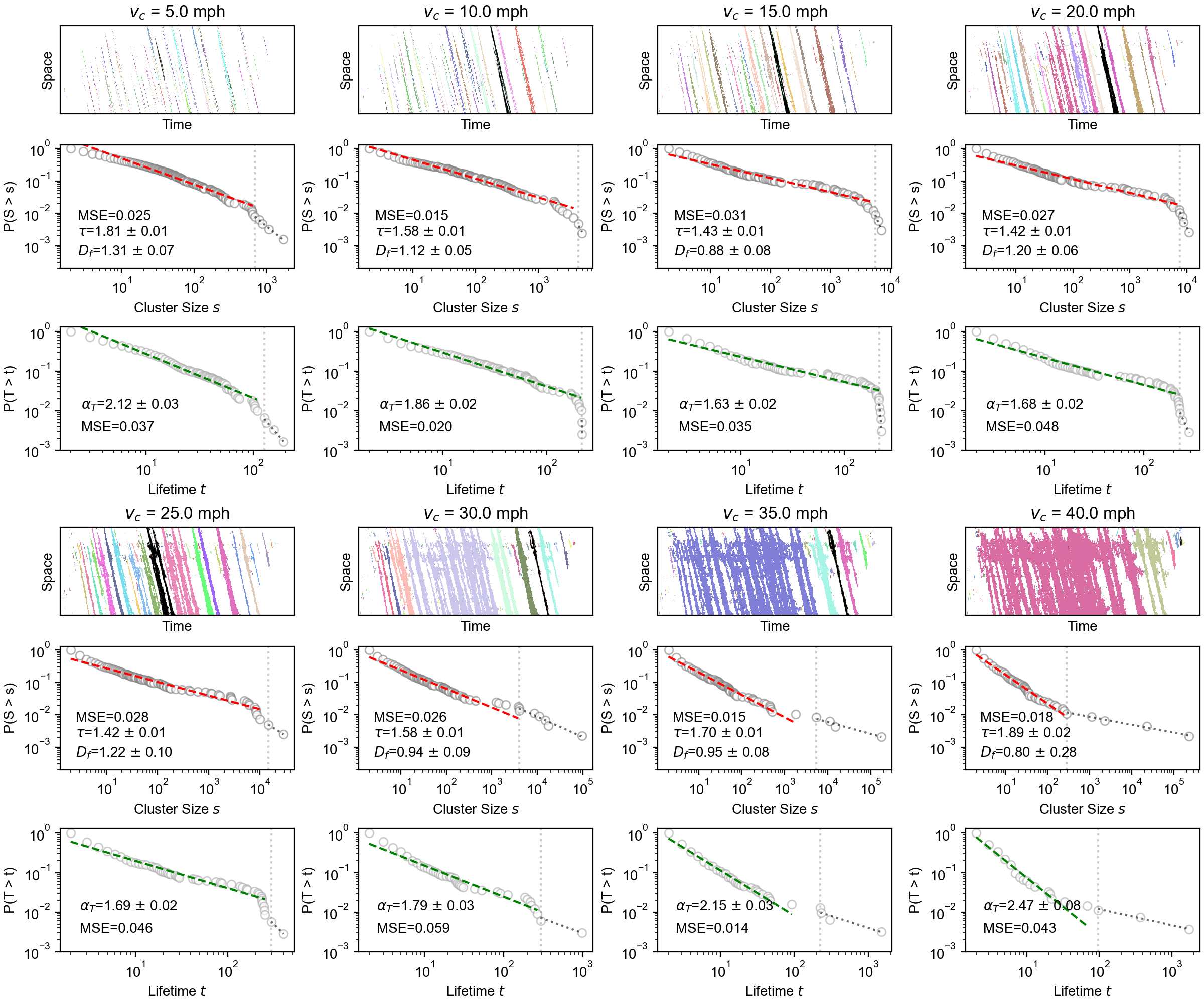}
        \caption{November 30}
        \label{fig:tau_1130}
    \end{subfigure}
    \hfill
    \begin{subfigure}[b]{0.77\linewidth}
        \includegraphics[width=\linewidth]{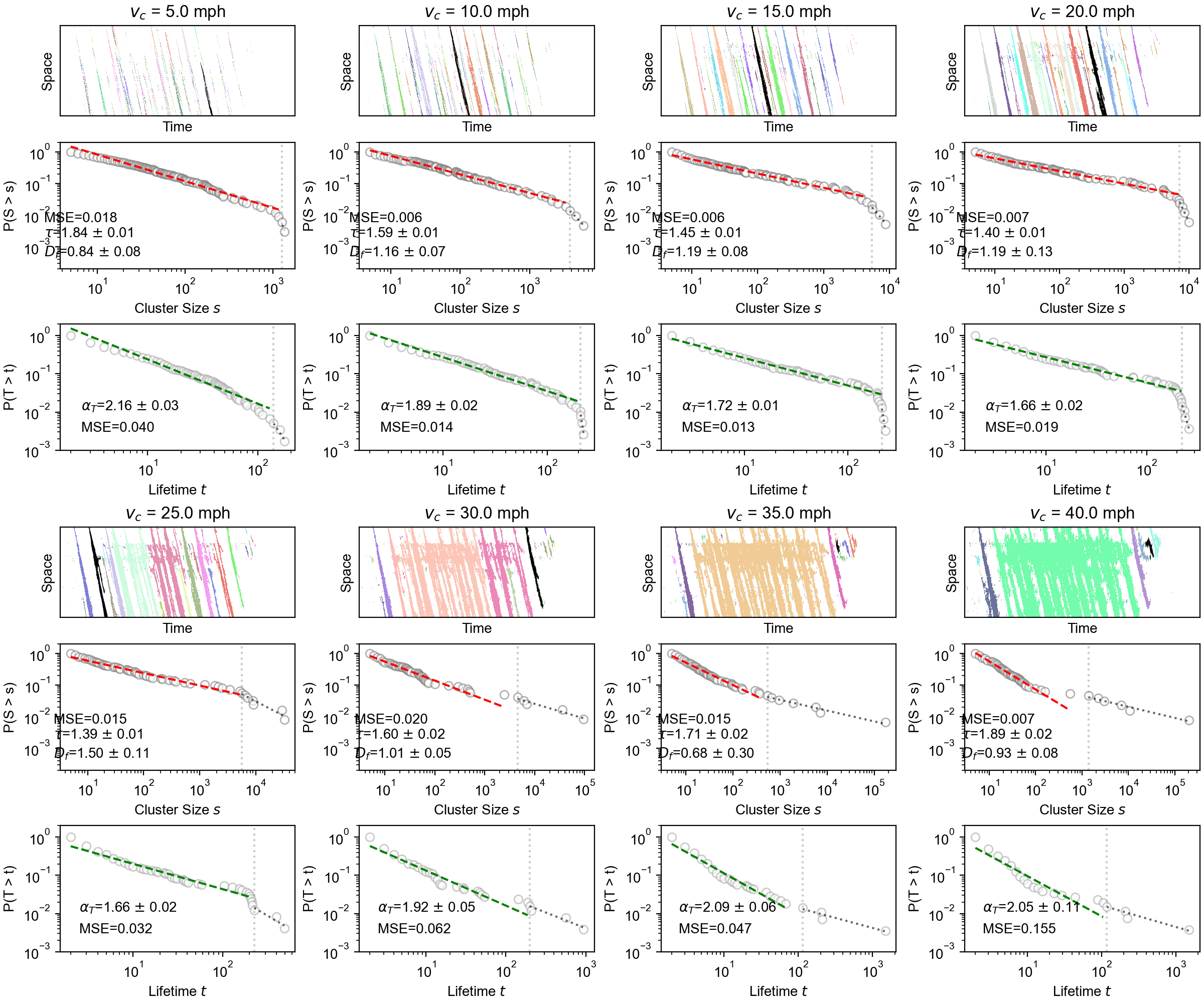}
        \caption{December 1}
        \label{fig:tau_1201}
    \end{subfigure}
\end{figure}

\begin{figure}[htbp!]
    \ContinuedFloat
    \centering
    \begin{subfigure}[b]{0.8\linewidth}
     \includegraphics[width=\linewidth]{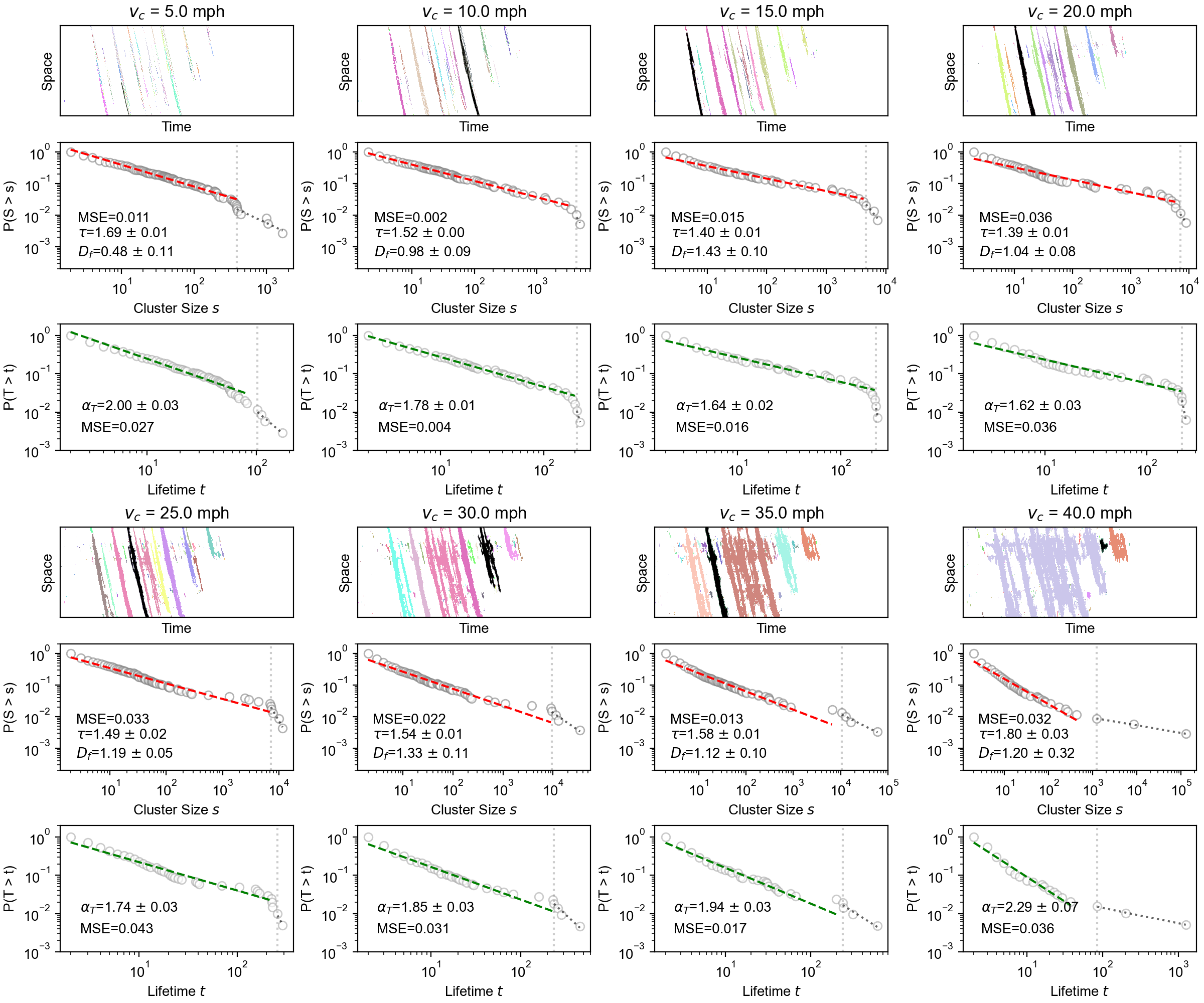}
        \caption{December 2}
        \label{fig:tau_1202}
    \end{subfigure}
    \caption{\textbf{Fisher exponent $\tau$ and $\alpha_T$ across threshold values.} 
    Estimated values for lane 1 (leftmost) across a range of critical speed thresholds, shown for the remaining five days}
    \label{fig:tau_all}
\end{figure}

\begin{figure}[htbp!]
    \centering

    \begin{subfigure}[b]{0.77\linewidth}
        \includegraphics[width=\linewidth]{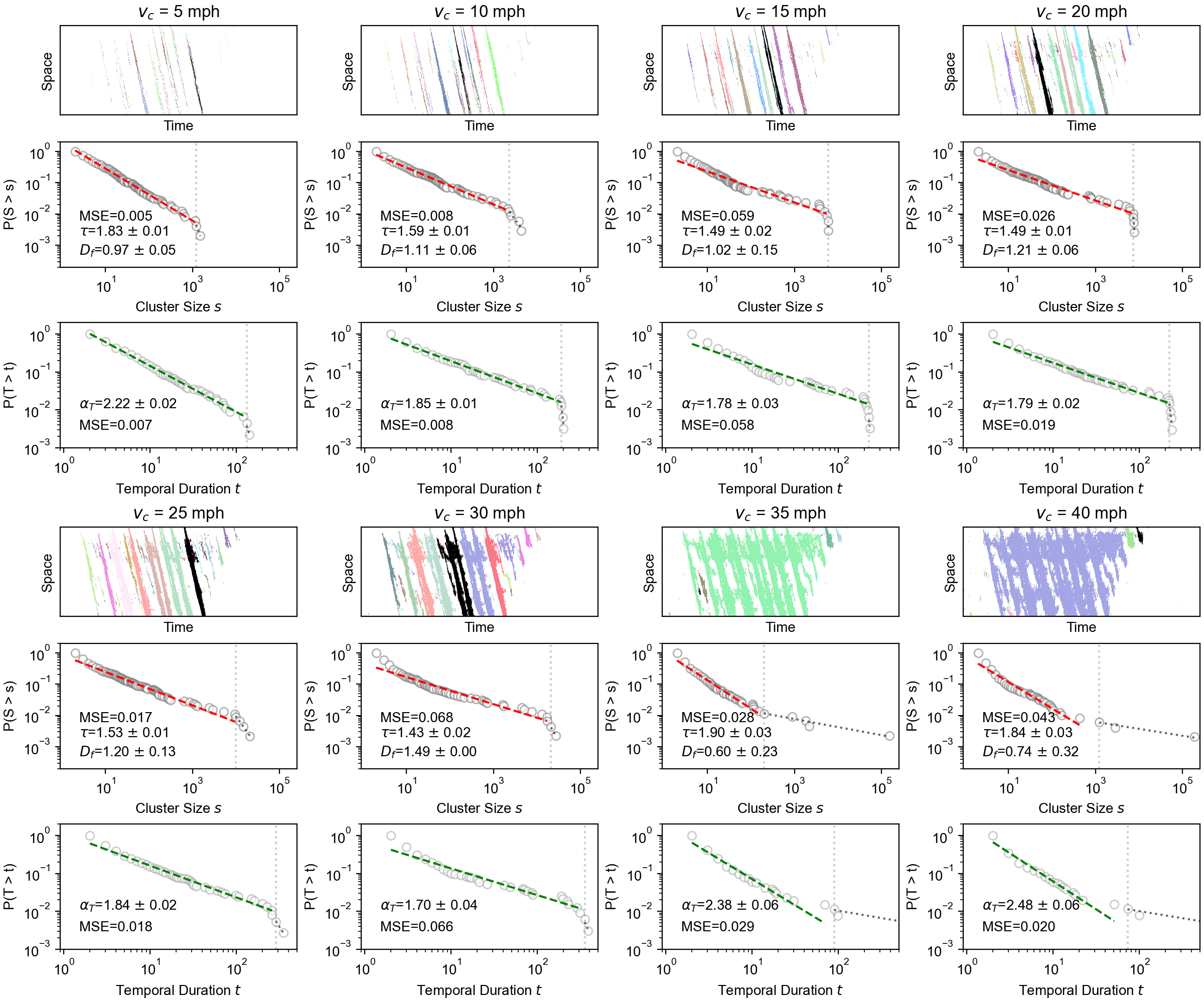}
        \caption{Lane 2}
        \label{fig:tau_1122_2}
    \end{subfigure}
    \hfill
    \begin{subfigure}[b]{0.77\linewidth}
        \includegraphics[width=\linewidth]{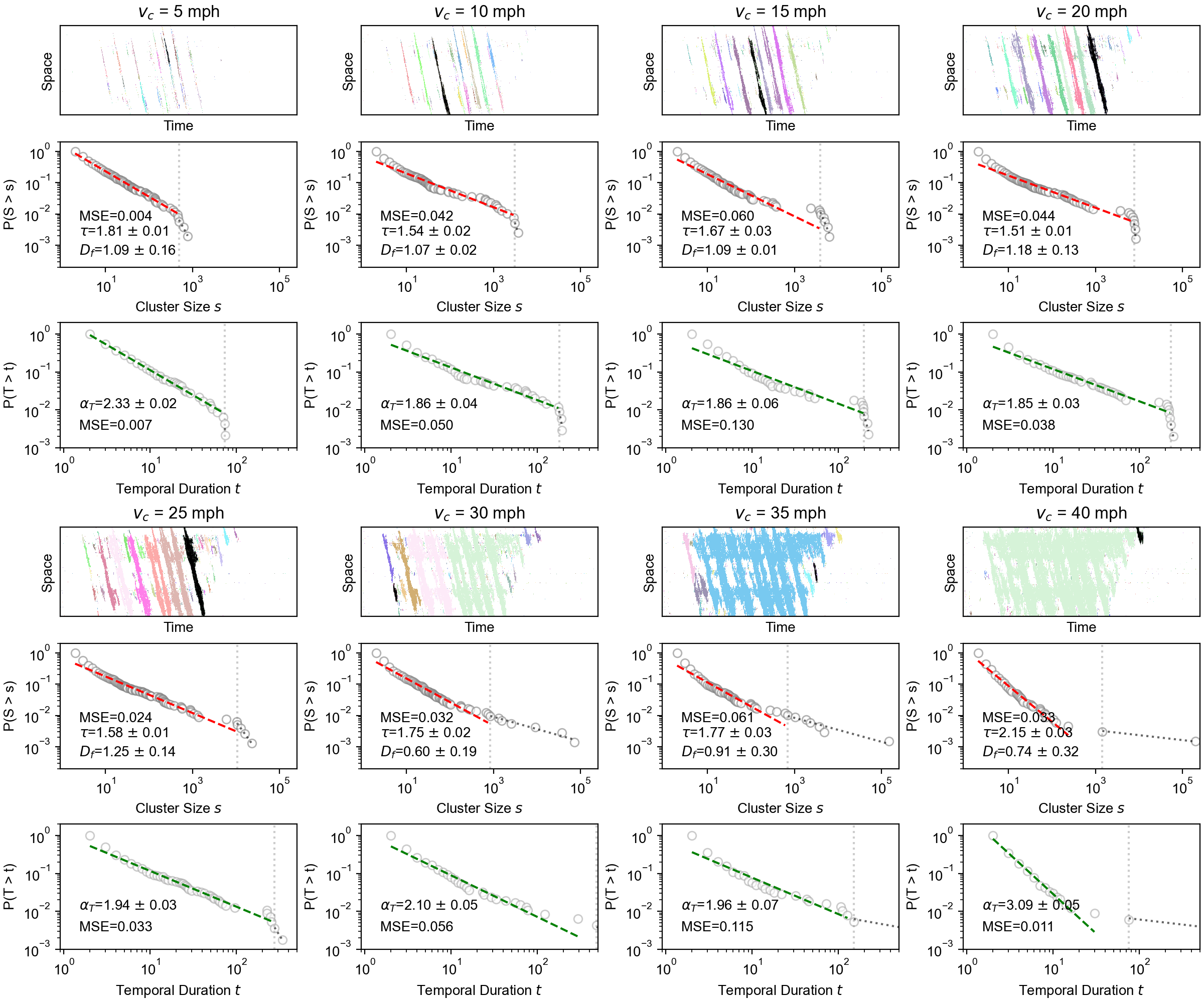}
        \caption{Lane 3}
        \label{fig:tau_1122_3}
    \end{subfigure}
\end{figure}

\begin{figure}[htbp!]
    \ContinuedFloat
    \centering
    \begin{subfigure}[b]{0.77\linewidth}
        \includegraphics[width=\linewidth]{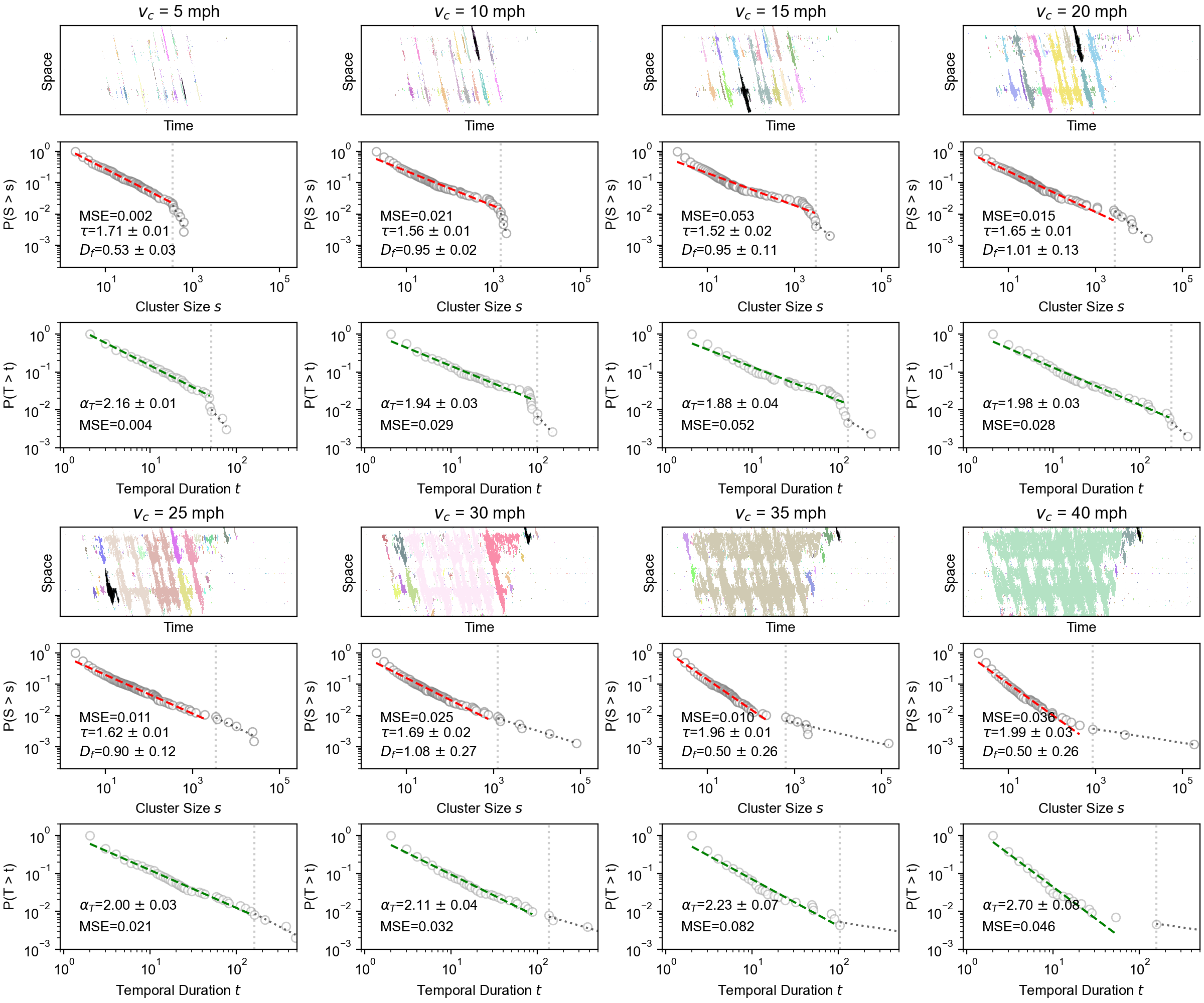}
        \caption{Lane 4}
        \label{fig:tau_1122_4}
    \end{subfigure}
    \caption{\textbf{Fisher exponent $\tau$ and $\alpha_T$ across threshold values.} 
    Estimated values for lanes 2, 3, and 4 across a range of critical speed thresholds, shown for Nov. 22. Lateral merging of clusters is more pronounced in the outer lanes.}
    \label{fig:tau_all_lane}
\end{figure}

\end{document}